\documentclass[aps]{revtex4}
\usepackage{amsmath,amssymb}
\usepackage{color,graphicx}
\usepackage{multirow}
\usepackage{amsmath,amssymb}
\usepackage{float}
\newcommand{\be}{\begin{equation}}
\newcommand{\ee}{\end{equation}}
\begin{document}
 \title{Brownian motion  in  trapping enclosures:  Steep potential wells, bistable wells and
  false bistability of induced Feynman-Kac (well) potentials.}
 \author{Piotr Garbaczewski and Mariusz \.{Z}aba }
 \affiliation{Institute of Physics, University of Opole, 45-052 Opole, Poland}
 \date{\today }
 \begin{abstract}

 We investigate  signatures  of convergence for a sequence  of   diffusion processes on a line,
    in  conservative force fields   stemming from  superharmonic  potentials   $U(x)\sim x^m$,  $m=2n \geq 2$.
    This is  paralleled  by  a   transformation of each   $m$-th  diffusion generator $L = D\Delta  + b(x)\nabla $,      and likewise the related   Fokker-Planck operator $L^*= D\Delta  - \nabla [b(x)\,  \cdot]$,  into  the affiliated  Schr\"{o}dinger  one   $\hat{H}= - D\Delta + {\cal{V}}(x)$. Upon a proper   adjustment  of  operator domains, the   dynamics is set   by  semigroups  $\exp(tL)$, $\exp(tL_*)$ and $\exp(-t\hat{H})$, with $t \geq 0$.    The Feynman-Kac integral kernel of  $\exp(-t\hat{H})$ is the major  building block  of  the relaxation process  transition probability density,   from which  $L$ and $L^*$ actually follow. The   spectral "closeness"   of  the pertinent  $\hat{H}$  and the Neumann Laplacian $-\Delta _{\cal{N}}$  in the interval  is  analyzed  for  $m$ even and  large.  As a byproduct of the discussion, we give a detailed
     description of an  analogous affinity,  in terms  of the $m$-family of  operators  $\hat{H}$ with  a priori chosen ${\cal{V}}(x) \sim x^m$, when  $ \hat{H}$  becomes    spectrally "close" to the Dirichlet Laplacian  $-\Delta _{\cal{D}}$  for  large $m$.
     For completness, a somewhat puzzling  issue of the absence of negative eigenvalues     for   $\hat{H}$  with  a   bistable-looking  potential ${\cal{V}}(x)= ax^{2m-2} - bx^{m-2}, a, b, >0, m>2$    has been addressed.
 \end{abstract}
 \maketitle

  \section{Motivation.}

  Our present investigation  is  motivated by that of  Ref. \cite{bounded1},    where a number of  somewhat  puzzling features  of  L\'{e}vy jump-type processes  in  bounded domains,   in particular these   with reflecting boundaries,  were reported.  The concept of reflected L\'{e}vy flights  has found a  coverage in the mathematical literature,  \cite{asmussen}.   However,  if the  fractional  dynamics (e.g.  L\'{e}vy-stable or   fractional  Brownian motion) is to be restricted to a bounded domain, in view of the nonlocality of the  process  and  its generator,   there are  many inequivalent proposals for   fractional analogs of the    domain-restricted Laplacian (Dirichlet,  Neumann  and the likes).

   This ambiguity  is particularly annoying  in  the reflective case. It   shows up in attempts to invent an appropriate process behavior at the boundary  and  extends  to its particular technical implementation, i.e. not only to identifying  a  proper  nonlocal version of the   Neumann condition,  but also to a detailed path-wise  definition of   the  reflection proper at the boundary.   Examples  are  provided by   computer-assisted procedures for  the  generation of sample paths and  varied realizations of the   reflection  or  quasi-reflection (happening in a vicinity of the boundary)   scenario, \cite{dybiec2,metzler,appr}).  The inferred jump-type processes are inequivalent as well, differing in their spectral and statistical characteristics. This  affects theoretical predictions for  relaxation properties and the near-equilibrium asymptotic behavior.

   On the other hand,  while departing from the Langevin picture of the L\'{e}vy  motion  that is confined by superharmonic potentials $\sim x^m/m, m=2n, n\gg 1$, one arrives at fractional Fokker-Planck equations, whose limiting properties   as $m\rightarrow \infty $  (the convergence issue is  mathematically disputable),   are ultimately interpreted in terms of reflected L\'{e}vy flights  in the  interval, \cite{denisov,dubkov,dubkov1}.
     The point is that   no link  has been established  with the  Neumann fractional Laplacian     (neither with   any   convincing   form of the Neumann condition),   which is  a valid   generator of  the  L\'{e}vy  process  in a bounded domain with reflecting boundaries.

   In our opinion,  the pertinent link can be consistently analyzed  by employing   the transformation  of  the  fractional Fokker-Planck equation to the    fractional Schr\"{o}dinger-type equation. It  is a properly tailored   version of  the  technical tool, often  used  in the study  case of the  standard (Brownian)  Fokker-Planck equation, but  seldom  addressed in the literature on  confined  L\'{e}vy processes.
 Since the superharmonic approximation seems  to provide a suggestive method to understand  what is   possibly  meant  by  the reflected L\'{e}vy process in the bounded domain (we restrict considerations to the interval on $R$), the  pertinent  transformation to the Schr\"{o}dinger-type  dynamics   should  in  principle   provide an approximation of this governed by the Neumann Laplacian.

 In passing, we   mention an active research on confined L\'{e}vy flights, with the path-wise attitude,  \cite{spectral}- \cite{dybiec1}.  We also   mention  publications,  \cite{lorinczi}-\cite{belik},  introducing   a concept of    the  "random motion in potential landscapes", albeit with alternative  notions of the "landscape".

 We emphasize that   our  intro  on    conditioned  L\'{e}vy processes  stands  merely for    a departure point  and actually   provides motivation    for the present  Brownian   endeavour.  From now on, we shall not refer to the L\'{e}vy framework anymore.   Instead, we   turn back  to the  extremally confined   Brownian motion,  where   the   sequential  limiting behavior (superharmonic approximation,  with a reflecting Brownian motion  introduced  as a  formal  limit) is taken for granted, or in the least unproblematic.

   A transformation of the Fokker-Planck operator to the Schr\"{o}dinger  one   (a respected metod of solution for the  Fokker-Planck equation, \cite{risken,pavl})  might  seem to be "obvious" as  well.
   The expected outcome  of the emergent sequential approximation  is  a possibly "smooth"   limiting behavior  of  Schr\"{o}dinger-type operators $-\Delta + {\cal{V}}(m)$ as $ m \gg 1$  and their spectral data, while approaching these normally attributed to the   reflected Brownian motion,  where the   motion generator is the  standard Neumann Laplacian  $(-\Delta )_{\cal{N}}$  with  the  well known  spectral solution, \cite{pavl}-\cite{gar1}.

   Links with the semigroup picture (Schr\"{o}dinger-type  dynamics) are    technically  much simpler  to analyze in  case of  the   confined Brownian motion, than in case of L\'{e}vy flights.  Therefore, we decided to  re-address this issue  in some detail, with  the aim of   detecting  any  hitherto  overlooked/neglected     subtleties  and obstacles/deficiencies   of the formalism  (including computer-assisted procedures) for:    (i)  superharmonic confining conditions, (ii)  permanent    confinement  between reflecting boundaries, (iii) mutual relationships between (i) and (ii), with an  emphasis on the issue of the "spectral  closeness" of the superharmonic Brownian  motion generator  to the   standard   Laplacian with Neumann boundary data and the reliability  of involved approximations.  And,  likewise on  the "closeness"  (whatever that means, but with the path-wise picture in mind) of the superharmonically confined Brownian motion to the  standard   reflected one.

One-dimensional  diffusion processes, which  are confined in the interior of the finite  interval in $R$,   are typically classified  to be reflecting (Neumann  boundaries) or  taboo (Dirichlet boundaries which  are not accessible), and  directly involve spectral solutions of    Laplacian operators  restricted to that     interval,  and  respecting the boundary data, i.e.   $\Delta _{\cal{D}}$   and $\Delta _{\cal{N}}$  respectively, \cite{gar,mazzolo,risken,pavl}.

    For each boundary data choice, the lowest  (interval-restricted) Laplacian   eigenvalues (strictly speaking,    the spectral gap)  determine the   time-rate of  the   so-called spectral relaxation to an invariant pdf (probability density function)    of the    Fokker-Planck equation  in question.  In the Dirichlet case, the drifted  diffusion in the interval   derives from the killed one via a proper conditioning, \cite{gar,pavl,pinsky,pinsky1},   and the  bottom   eigenfunction and the  lowest  eigenvalue  of  $\Delta _{\cal{D}}$ are explicit   constituents of  the  analytic formula for the transition probability density of the  process with inaccessible boundaries.

   In the literature there exist many proposals on how to  approximate Brownian  motion in the interval by means of
    strongly confining  random  model systems, both on the Langevin (thus Fokker-Planck)   and  Schr\"{o}dinger type
     (semigroups, generalized diffusion equations)  levels of description.
      Typically one  employs either a sequence of finite well potentials  with an increasing height of the barrier,
       or a sequence of strongly anharmonic potentials, alternatively $\sim x^{m}$, $\sim x^{m}/m$, $\sim m\, x^{m}$
       with $m=2n$ growing indefinitely (plus a number of other  options), \cite{gar1}-\cite{diaz}.

 These procedures tell us how  to   get  at a reliable   approximation of the Dirichlet (infinite well/finite interval-restricted)   operator $(-\Delta )_{\cal{D}}$, in terms of    Schr\"{o}dinger-type generators and semigroup evolutions      (with the Feynman-Kac paths redistribution involved, \cite{lorinczi}),  \cite{gar}-\cite{voros} see also \cite{robinett}. Although there  still remain   mathematical obstacles   that annoy  physicists, \cite{karw,diaz}.   In this case, a parallel    sequential approximation in terms of affiliated  Langevin-Fokker-Planck  evolutions is unavailable in the literature  and this topic, together with a number of illuminating explicit examples,  we  shall address in the Appendices.

 In comparison with the Dirichlet case,  the  Neumann one   has been left somewhat aside.
   In the current literature,  it is taken for granted (see  \cite{bounded1} and references there-in)
    that  the Brownian motion in  extremally anharmonic  force fields  (Langevin-driven case  with $b(x) \sim x^{2n}-1$, $n\gg 1$) not only   provides a  satisfactory  approximation of the reflected Brownian motion in the interval, but actually converges  to the latter.

    At this point we indicate  a  problem that will appear in our subsequent analysis, that   the convergence is not unifom,   and that the  standard Neumann condition is  actually  not reproduced  in the  controlled  point-wise limit.  This  statement needs to be reconciled with the fact that the     analytic description   of the reflected   motion   in the interval    is  given  in terms   of the spectral  solution  for the Laplacian with Neumann boundary conditions,     $\Delta _{\cal{N}}$,  c.f. \cite{linetsky}-\cite{pilipenko}.

     We point out  well established mathematical facts,   \cite{pinsky}:
    (i) the Schr\"{o}dinger operator with a constant potential and the Neumann boundary condition corresponds to
    the reflected Brownian motion, while (ii) the Schr\"{o}dinger operator with a constant potential and the
     Dirichlet boudary condition corresponds to what we name the taboo process, \cite{gar}, e.g.  a drifted
      Brownian motion that is conditioned never  to exit the open interval in question.  Moreover, (iii)
 there  is a one-to one correspondence between between the pertinent  Schr\"{o}dinger operators
   (thence  semigroup dynamics) and  drifted  diffusion generators  (thence  Langevin-driven  diffusion process).

  This motivates  the  basic aim of  the present paper, to re-examine   the theory of the   Brownian motion on the interval,    in   terms of  a sequence of  steep (superharmonic)  potential wells, with an emphasis  on the  familiar   and widely exploited  \cite{risken,pavl}  link  between the   Fokker-Planck dynamics  $\partial \rho = L^*\rho $    and that  governed by   the affiliated  Schr\"{o}dinger semigroup  $\exp(-t\hat{H})$, e.g. the generalized diffusion equation  $\partial _t \Psi = - \hat{H} \Psi $.    The latter  equation is known to  quantify the   random  motion   in terms of  the Feynman-Kac formula,  \cite{nonegative,non,streit,klauder,faris}  and  might be   interpreted  path-wise  as the Brownian     motion "in potential landscapes", c.f. \cite{lorinczi}, see also \cite{brockmann,geisel,belik}  for an alternative interpretation of the "landscape " notion.

    Our departure point will be the case of Brownian motion in  conservative force fields
      (explicitly  present in Langevin and Fokker-Planck equations), which    stem from confining  potentials of the form
    $U(x)= x^m/m, m=2n,  n\geq 1$.  Asymptotic stationary probability densities  (invariant pdfs of  Fokker-Planck equations),  to which the  random dynamical system relaxes,  have the Boltzmann form $\rho _*(x)\sim \exp (- U(x)/D)$.

     The affiliated  Schr\"{o}dinger operators  $\hat{H}=- D\Delta + {\cal{V}}$ include manifestly bistable (two-well)   potentials  $ {\cal{V}}(x)\equiv  {\cal{V}}_m(x)= ax^{2m-2} - bx^{m-2}$, $a,b >0, m>2$   , and somewhat surprisingly, there is  no bistability  impact upon  the functional  form of    manifestly unimodal   ground state functions  $\psi _0(x)\sim \exp(-U(x)/2D) \sim \rho _*^{1/2}(x)$.
 Moreover, these   eigenfunctions    correspond to the eigenvalue  zero  of the Schr\"{o}dinger operator
    (the problem  of spectrally  isolated zero energy
     bound states   continually   reappears in the literature, \cite{entropy}-\cite{makowski}).

       Although the  local  minima  of our  bistable   potentials  have values that are   below  zero,
         negative energy bound states are    conspicuously   absent.
   At the first glance, this  property  seems to contradict   standard intuitions    underlying  the concept of
     bistability and the related tunelling-through-the-barrier   conceptual imagery of  quantum theory
      (c.f. a    "canonical" discussion of  the negative eigenvalues splitting in case of   the familiar
        double-well potential,  \cite{landau}).  This slightly puzzling  point will be   addressed  in below  as well.

\section{Thermodynamics of Brownian motion and relaxation to equilibrium}

\subsection{Boltzmann-type  equilibria for Smoluchowski  diffusion  processes.}

  Let us consider a one-dimensional  diffusion process \cite{risken}, with  the Langevin representation
\begin{equation}
\dot{x} = b(x,t) + \sqrt{2D} B(t),
\end{equation}
 where $\langle B(s)\rangle =0$, $\langle
B(s)B(s')\rangle = \delta (s-s')$ and  $b(x)$ is a forward
drift of the process having the gradient form $b= 2D \nabla \Phi $,
$D$ being a  diffusion constant.
If an initial probability density $\rho_0(x)$ is given, then the
diffusion process  obeys  the Smoluchowski-Fokker-Planck
equation
\begin{equation}
 \partial _t\rho = D\Delta \rho - \nabla \, ( b \cdot
\rho )= L^*\,  \rho \, .
\end{equation}
where  $L^* = D\Delta - \nabla (b \, \cdot )$  is the Fokker-Planck operator  (we prefer the notation $L^*$ instead of the  often employed  $L_{FP}$).   We recall, \cite{risken,pavl},  that $L^*$  is a Hermitian   $L^2(R)$  adjoint of  the diffusion operator $L = D\Delta + b \nabla  $, which  in the mathematical  literature  actually stands for a legitimate generator of the pertinent Markovian stochastic process.

Let us  introduce an osmotic  velocity  field  $u = D\ln \rho $  and  the current velocity field  $v=b - u$. The latter
directly enters   the continuity equation $\partial _t \rho = - \nabla j$,
where  $j= v\cdot \rho $  has   a standard interpretation of a
probability current.

We restrict further discussion to  time-independent drifts,
 that are induced by   external (conservative,  Newtonian) force fields $f= - \nabla U$.
One arrives at Smoluchowski diffusion processes by  setting
 \begin{equation}
b = {\frac{f}{m\beta }} = - {\frac{1}{m\beta }}  \nabla U \, .
\end{equation}
This  expression accounts for a fully-fledged phase-space derivation of the spatial (Smoluchowski)  process, in the
 large  friction  $\beta $ regime.  It is  taken  for granted that  the fluctuation-dissipation  balance gives rise
  to the  standard form $D=k_BT/m\beta $ of the  diffusion coefficient,  \cite{risken,entropy}.

In the stationary asymptotic  regime we have $j\rightarrow j_*=0$.
We denote  $\rho _*= \rho _*(x) $   a strictly positive   probability   density,   to which $\rho (x,t)$  is presumed to  relax    as $t\rightarrow \infty $.
Accordingly in that regime  $v\rightarrow v_*=0$. Since   $b=f/m\beta  $  is time-independent,  there holds
\begin{equation}
b=u = D \nabla  \ln \rho _*  \, .
\end{equation}
 Consequently, we have  a stationary solution  of the Smoluchowski  (Fokker-Planck) equation  $L^*\, \rho _* =0$  in the  Gibbs-Boltzmann form
\be
  \rho _*(x) = (1/Z)\,  \exp[ - U(x)/k_BT]= \exp [(F_*  -  U(x))/k_BT],
\ee
 where $1/Z$ is a normalization constant.

 Here $F_*  = - k_BT \ln Z$     is the  minimal value of the time dependent Helmholtz free energy of the  random motion  $F=F(t)  = \left< U+ k_{B}T \ln \rho \right> $, to which $F(t)$ relaxes as  $t\rightarrow \infty $,   \cite{entropy},    $\left< . \right>$ denotes the mean value with respect to $\rho (x,t)$.

Note that we consider confining potentials $U(x)$ with the property    $\lim _{|x|\rightarrow \infty } U(x)   = +\infty $ and
  such that   $\exp(-U(x)/k_BT)$  is  $L^1(R)$ integrable (to secure the normalization).  Any bounded from below potential
   function $U(x)$ can be  redefined to become either  non-negative  or positive, since an additive  energy "renormalization"
   is irrelevant  for the   Langevin-Fokker-Planck dynamics and leaves the the functional form of $\rho _*(x)$  intact
    (this in view of the  $L^1(R)$ normalization).

\subsection{Pseudo-Schr\"{o}dinger    reformulation of the Fokker-Planck dynamics.}

  For computational convenience  it is useful to scale away the factor $m\beta $ in the formulas of the  previous subsection.
   This can be done by  changing the time scale in the Fokker-Planck equation   (2) from $t$ to $t/m\beta$,
     accordingly   replace  $D$ by $D \equiv k_BT$,  and thus
    $b(x)=-  \nabla U(x) /m\beta $ by $b(x)\equiv   -  \nabla U(x)$.

We shall intentionally  employ   the notation $D\equiv k_BT$, instead of  $D'= m\beta D= k_BT$.
From now on our $D$ in reality will  be a "graphical" replacement for  $D'$ (with time rescaling kept in memory).

Accordingly, if   we consider literally Eq. (2) with respect to the rescaled time, in the form
 $\partial _t \rho = D \Delta   - \nabla (b\rho)$, where $D\equiv k_BT$ and $b(x) = - \nabla U(x)$,
 we get an invariant pdf in the form:  $\rho _*(x) = (1/Z)\,  \exp[ - U(x)/D]$, often  employed in the literature.

 Although in principle $D>0$ is unrestricted, in  below we shall  mostly  refer to  the moderate noise range and specifically to
  $D=1$ or $D=1/2$   (that to stay in  conformity  with mathematically oriented approaches to diffusion problems).
  This, to some extent,  may  blur a   direct  comparison with   observable   effects of
 the specific   noise intensity choice. Indeed,  small $D$   narrows/compresses  the potential profile,
  while increasing its steepness. The choice of  larger $D$
 flattens/stretches  the potential profile).
    An issue of the small noise $D\ll 1$, and small versus  moderate or large noise,
  will briefly   return  back   in Section VI.

 Following a standard procedure \cite{risken,pavl}, given a stationary density $\rho _*(x)$,   one can  transform the
Fokker-Planck  dynamics, Eq. (2)  into an associated Hermitian (Schr\"{o}dinger-type) dynamical  problem in $L^2(R)$,  by means of a  factorisation:
\begin{equation}
\rho (x,t) = \Psi (x,t) \rho _*^{1/2}(x).
\end{equation}
Indeed, the Fokker-Planck evolution (2) of  $\rho (x,t)$   implies the validity  of  the   generalized diffusion (Schr\"{o}dinger-type)   equation
 \begin{equation}
 \partial _t\Psi= D \Delta \Psi - {\cal{V}} \Psi  = - \hat{H} \Psi ,
\end{equation}
for  $\Psi (x,t)$.  Note that  the   $\rho (x,t) \rightarrow \rho _*(x)$ as $t\rightarrow \infty $,  needs to be  paralleled by  $\Psi (x,t)  \rightarrow  \rho _*^{1/2}(x)$.

 We demand  that  $ \hat{H}\,  \rho _*^{1/2} = 0$, which implies that the admissible   functional form of the  potential function  ${\cal{V}}(x)$   derives  as a function  of
 $\rho _*^{1/2}(x)$:
\begin{equation}
{\cal{V}}(x) = D {\frac{\Delta \rho _*^{1/2}}{\rho _*^{1/2}}} =     \frac{1}{2} \left(\frac{b^2}{2D} + \nabla b\right),
\end{equation}
c.f.  Eq. (4), here  we have $b(x)= 2D \nabla \ln \rho _*^{1/2}(x)= -\nabla U(x)$.

Proceeding in reverse, the functional form  (8) of the potential function  ${\cal{V}}(x)$ is a guarantee for  the  existence
of the bottom  eigenvalue  zero of the  Hermitian operator  $\hat{H}= -D\Delta + {\cal{V}}$, associated  with a
  strictly positive  eigenfunction    $\rho _*^{1/2}(x)$.

If the  ($1/2mD$ rescaled) Schr\"{o}dinger-type  Hamiltonian
$\hat{H} = -D\Delta + {\cal{V}}$ is a  bounded from below,
self-adjoint operator in a suitable Hilbert space, then one arrives
at a  dynamical semigroup  $\exp(-t\hat{H})$,  which implies  $ \Psi(x,t) = [ \exp(-t\hat{H})\Psi](x,0)$.
 We note that, in our particular  F-P context, one is  bound to choose  $\Psi(x,0)= \rho _0(x)/\rho _*^{1/2}(x)$,  with   $\rho _0(x)$  being  an initial probability density  for  Eq.  (2).
  $\Psi (x,0)$ is  presumed to be the    $L^2(R)$  function,   hence for the   $\rho _0(x)$, the integrability condition $\int dx [ \rho ^2_0(x)/\rho _*(x)] < \infty $  must hold true. This imposes limitations on admissible initial data for the Fokker-Planck evolution.their decay at infinity must be significantly faster than this of $\rho _*^{1/2}$.

 The  separation  ansatz $\Psi (x,t)= \psi(x) \exp (-{\cal{E}}  t)$ converts Eq. (7) into  $\hat{H} \psi  = {\cal{E}} \psi $.
  It is clear that  with ${\cal{V}}$ given by (8),
 the square root   $\rho ^{1/2}_*(x)$ of the invariant density  (5)  is a particular  ($L^2(R)$-normalized ground state, \cite{zero}-\cite{vilela} solution of  the  eigenvalue equation  for $\hat{H}$:
\be
\hat{H} \psi (x) = [ - D\Delta   + {\cal{V}}] \psi (x)  =  0
\ee
corresponding  to the eigenvalue zero.  We recall that  $\hat{H}\rho_*^{1/2} =  0$ and  $L^*\, \rho _* =0$   hold true in parallel.

  The outlined procedure may be regarded as a reconstruction of the semigroup dynamics from an eigenstate (here,
  ground state function), \cite{vilela}. It is an alternative  to another, reverse engineering
  procedure \cite{klafter,stef}  (named  also a targeted stochasticity), whose main goal is to reconstruct
   the random motion (e.g. the   Langevin-driven motion and the Fokker-Planck equation) that  is
   compatible with a priori  given   equilibrium function (e.g. the stationary pdf) and a priori prescribed noise
    (here Brownian/Wiener   one).

\subsection{Brownian motion in   potential landscapes (Feynman-Kac potential spatial  profiles).}

   All previous observations actually derive  from  the Langevin-type equation (1) and thus from  the  traditional (microscopically motivated) picture of the
    Brownian motion in conservative force fields. Forces are of external nature,  while Wiener noise is  regarded as  an intrinsic property of the environment in which
    the  motion takes place.

   Concerning the generalised diffusion equation  equation  (7),  let us  set $D=1/2$  and  make an assumption that the  potential
   function ${\cal{V}}$ is a  continuous, bounded from below  function.  Then, we can introduce the  positive
     symmetric  integral kernel $k(t,x,y)=k(t,y,x)$  of the semigroup operator $\exp(- t \hat{H})$,
given  e.g.  by the Feynman-Kac formula, with an explicit ${\cal{V}}$   entry, c.f. \cite{gar,klauder,faris},
 see also \cite{gar,bounded1}.

The  Feynman-Kac formula  defines the transition density (integral kernel of the semigroup operator)  as  a weighted integral over sample paths  of the
  diffusion process (with the Wiener path measure being involved; here  paths  $\omega $ originate from $x$ at time $0$ and their destination is $y$ to be reached  at time $t>0$):
\be
  k(t,x,y) =  [\exp(- t \hat{H}](x,y)  =
       \int   \exp[-\int_0^t {\cal{V}}(\omega(\tau  )) d\tau ]\,   d\mu _{(x,0,y,t)}(\omega).
        \ee

In Ref. \cite{lorinczi}    Eq. (11)  is interpreted  in terms of a random mover in an {\it energy landscape}
  set on $R$   by the  potential   ${\cal{V}}(x)$ and  acting as a mechanism that reinforces  or penalizes the
  random mover tendency to  reside or  go into specific regions  of  space. A "responsibility" for a  weighted
  redistribution of random paths in  a given time interval,  is here transferred from external force fields
   of Eq. (1)   to  the spatial   variability  of    potentials  ${\cal{V}}(x)$  of Eq. (8), in particular
   their curvature and steepness.

The   Fokker-Planck-induced   dynamics of $\rho(x,t)$, Eq. (2),   in view of Eqs. (6) and (7)
 can be interpreted as follows:
\be
 \rho (x,t) = \rho _*^{1/2}(x)\, \Psi  (x,t) =\rho _*^{1/2}(x) \, \int k(t,y,x) \Psi (y)\, dy =
 \int \left[ k(t,y,x){\frac{\rho _*^{1/2}(x) }{\rho _*^{1/2}(y)}}\right] \rho  (y)\, dy = \int p(t,y,x) \rho (y)\, dy,
\ee
which in a self-defining manner introduces, \cite{gar}, the transition probability density  $p(t,y,x)$ of the
 (conditioned) Markovian  diffusion process   underlying   the temporal evolution of $\rho (x,t)$.
  We emphasise the relevance of the order in which variables $x$ and $y$ appear, because $p(t,x,y)\neq p(t,y,x)$.

   Indeed,  in contrast to the kernel function $k(t,x,y)$, transition pdfs are not symmetric functions of $x$ and $y$,  which  has  profound  consequences in the derivation of the explicit functional form for generators $L$ and $L^*$ of the diffusion process, while  evaluated  according to the familiar recipes. Namely the semigroup   $T_t = \exp(tL)$    has  a   generator
\be
 L f(x,t)= \lim _{t\downarrow 0}  {\frac{1}t} \left[   \int p(t,x,y) f(y)dy  - f(x)\right],
\ee
which reads  $L =  \Delta    + b \nabla $, with  $b(x) = -x $.  Clearly, we have  $\partial _t f(x,t) = L\, f(x,t)$.

The time evolution of  probability measures and associated probability density functions $\rho(x,t)$ is  governed by the  adjoint semigroup  $T^*_t = \exp (L^*t)$,  according to  $T_t^* \rho (x,t) = \int p (t,y,x) \rho (y) dz$, from which there follows   $\partial _t \rho (x,t) = L^* \rho (x,t)$, where $L^*  = \Delta  - \nabla (b\,  \cdot )$  comes from
\be
L^* \rho (x,t)= \lim _{t\downarrow 0}  {\frac{1}t} \left[   \int p(t,y,x) \rho (y) dy  -  \rho (x)\right].
\ee

While coming back to the unrestricted noise intensity  parameter $D$, we  realize that the pdf  dynamics  (2)  and (11),  as a direct consequence of Eq. (7),  can be represented   in  various  forms,
 \cite{zero,geisel,brockmann,belik}:
  \be
\partial _t \rho =   L^* \rho (x,t)=     \rho _*^{1/2}\, \partial _t \Psi  =  - \rho _*^{1/2} \hat{H} \rho _*^{-1/2}\, \rho (x,t)=
D\, [\,  \rho _*^{1/2}\, \Delta (\rho _*^{-1/2} \rho ) -   \rho
_*^{-1/2} (\Delta \rho _*^{1/2})\, \rho \, ]. \ee
Remembering that  $\rho _*(x)  \sim \exp[- U(x)/D]$,  we  uncover a  direct impact of the Boltzmann   weight  upon the dynamics
of $\rho (x,t)$. It suffices to set $\rho _*^{\pm 1/2} \rightarrow   \exp[\mp U(x)/2D]$ everywhere in Eq. (11).

In the Langevin modeling of the   Brownian motion, the choice of a  Newtonian
  potential $U(x)$,    determines  both   an invariant pdf
   $\rho _*= (1/Z) \exp (- U/D)$   of the   F-P equation and the  forward drift (e.g. the conservative force field)
     $b(x)= D \nabla  \ln \rho _* = - D \nabla U$,    entering    $\partial _t \rho = D \Delta \rho - \nabla (b \rho )$.
       The  relaxation process   $\rho (x,0) \rightarrow
   \rho (x,t) \rightarrow \rho _*(x)$    in question is well posed, once a suitable  $\rho (x,0)$ is selected as  the
    initial data.

  In the dual    picture provided by the
 Feynman-Kac  (Schr\"{o}dinger)  semigroup,  an   a priori chosen   $\rho _*(x)$   determines  the
  Feynman - Kac potential    ${\cal{V}} =  D [\Delta \rho _*^{1/2}]/ \rho _* ^{1/2}$ and
  thence the  "potential landscape"  delineated by the spatial profile of ${\cal{V}}$,
   \cite{lorinczi}.  The  relaxation process refers to the time rate at which  the eigenfunction $\rho _*^{1/2}$ is
    approached in the course  of the time  evolution  $\Psi (x,t) \rightarrow   \rho _*^{1/2}(x)$
      governed by  $\hat{H} = - D\Delta +  {\cal{V}}$,  with the spectral property  $\hat{H}\rho _*^{1/2} = 0$. To this end one needs to know (at least approximately)  the first positive eigenvalue and thus the spectral gap.\\

{\bf Remark:}  In the study of an  impact  of strongly inhomogeneous  (disordered)  media  upon the random motion (with an independent noise
 input), \cite{belik}, a complementary notion of   {\it salience (attractivity) field}
  has been associated with   the exponential (Boltzmann)  weight      $s(x) =\exp[-U(x)/D]  = Z\, \cdot \rho _*(x)$.
 A  curve delineated by  the salience field  $s(x)$  (actually its square root in Eq. (12)),   might  deserve  as well an  interpretation  (modulo dimensional issues) of  the  "energy (salience) landscape".  It is the relative weight between    the source and  target locations  that    enforces or penalizes (reduces)   the   rates of spatial displacements of the random process.

\subsection{Joint   spectral solution for  motion  generators $L$, $L^*$ and $\hat{H}$.}

As long as the operators $L, L^*, \hat{H}$ are considered in  a  common for them  function space  $L^2(R)$,
the transformation from $L^*$ to  the Hermitian  (symmetric, eventually self-adjoint)  operator  $\hat{H}$ seems to be
 the natural way to address  spectral properties  of generators of motion and related relaxation time rates to equilibrium
 (we need the real eigenvalues of $\hat{H}$)  at the bottom of the non-negative spectrum).
It is however not a must  and  we can   admit other function spaces.

 Given a  unique  stationary solution (5) of the Fokker-Planck equation, $L^*\,  \rho _*(x) =0$.
 The uniqueness is granted if the Fokker-Planck operator has an energy gap at the bottom of its spectrum,
  \cite{pavl,pinsky,pinsky1}.
  This gap in turn controls the (generically exponential) time rate of relaxation of $\rho (x,t)$
  towards  the equilibrium  pdf $\rho _*(x)$.  This is the scenario of so-called spectral relaxation.

Concerning the existence of the gap, a sufficient condition to this end can be given by demanding that the Gibbs potential $U(x)$ is a continuous and continuously differentiable (e.g. $C^2(R)$) function, and that the semigroup potential ${\cal{V}}(x)$, Eq.  (8)  shares with $U(x)$   the confining property: $\lim _{|x|\rightarrow \infty }  {\cal{V}}(x)  = + \infty $, see e,g, \cite{pavl}.

Independently of the previous factorization (6) of $\rho (x,t)$, given $\rho _*(x)$,  we can introduce  an  alternative
factorization:
\be
\rho (x,t) = h(x,t) \rho _*(x).
\ee
A comparison with Eq. (6) shows that $\Psi (x,t)= h(x,t) \rho _*^{1/2}(x)$.

A direct consequence of the  Fokker-Planck equation (often named the forward  Kolmogorov equation) is  that $h(x,t)$  is a solution of the backward Kolmogorov equation:
\be
\partial _t h(x,t) = L\,  h(x,t),
\ee
where $L= D\Delta + b \nabla  $  stands for  the diffusion generator and the initial condition reads  $h(x,0)= \rho _0(x)/ \rho _*(x)$.

It is not yet widely appreciated information that the generator $L$  is Hermitian (symmetric,  \cite{pinsky},
and eventually self-adjoint) in the Hilbert space $L^2(R, \rho _*)$, whose scalar product is weighted by the invariant
 pdf $\rho _*$:   $(f,g)_{\rho _*} = \int_R f(x) g(x) \rho _*(x) dx$, \cite{pavl}.
 We point out that $- L$ is a nonnegative operator, with a discrete spectrum beginning form the bottom
  eigenvalue $0$.

 We  can readily introduce   an operator product (warning: we admit  action upon functions in the domain in a
  definite order,  from right to left), reproducing the operator $\hat{H}$ of Eq. (11):
  \be
  \hat{H}  = -  \rho _*^{1/2} L \rho _*^{-1/2} = - \rho _*^{-1/2}  L^* \rho _*^{1/2}.
  \ee
  We point out   that the spectrum of $\hat{H}$, $-L$ and  $-L^*$   is nonnegative.

 This is accompanied by the transformations:
 \be
  - L^* = \rho _*^{1/2} \hat{H} \rho _*^{-1/2}
 \ee
 and   $ L = \rho _*^{-1}  L^* \rho _*$.

  Three operators $\hat{H}$, $L$ and $L^*$     are Hermitian (and eventually self-adjoint) in function spaces $L^2(R)$, $L^2(R,\rho _*)$ and  $L^2(R,\rho _*^{-1})$  respectively, \cite{pavl}.

 The transformations (14)-(16) allow to deduce eigenfunctions and eigenvalues of each related spectral problem, if one  of them has been  actually solved.

  Namely, let us assume that we know the spectral data of $\hat{H}$. We denote $\psi _k^{\hat{H}}$ the $k$-th
   eigenfunction in $L^2(R)$ and   $\lambda _k^{\hat{H}}$   the associated  eigenvalue.  This  eigenvalue (up to a sign, (15)-(17))
    is shared by all three generators, while the  corresponding  eigenfunctions are related accordingly:
 \be
 \psi _k^{\hat{H}} = \rho _*^{-1/2} \psi _k^{L}
 \ee
 and
 \be
 \psi _k^{L^*} =  \rho _*^{-1} \, \psi _k^{L}.
 \ee
  In particular we have: $\psi _k^{L^*}  = \rho _*^{-1/2} \psi _k^{\hat{H}} $. We recall that the  pertinent  eigenfunctions "live" in  different function spaces and any reduction to $L^2(R)$ (unweighted)  integrals needs to be executed with due  caution.

The above transformations  reduce  the issue  of the spectral analysis of the Brownian motion to that of  the
 computational convenience.  See e.g. \cite{pavl,pinsky,pinsky1, linetsky,bickel,pilipenko} for  eigenfunction expansions  associated with the diffusion generator
  $L$  of a    reflected  process on the interval.

  The main value  of the spectral analysis of motion generators is that we may  compare their spectral gaps,
  if in existence, under  various physically motivated circumstances. They  allow  to
   quantify  the  convergence to equilibrium for an  ergodic diffusion process and
    actually coincide with the  exponential time  rates of convergence.\\

\subsection{Spectral versus nonspectral relaxation in the OU process. A comment.}

 At this point  it is worthwhile to  invoke  a discussion of a possibility of the
 non-spectral relaxation in the  one-dimensional Ornstein-Uhlenbeck  process, \cite{sokolov}.
  The basic  argument  therein  can be rephrased as  follows (we employ our notation).  Equation (6) is supposed to
 define a similarity transformation between the Fokker-Planck operator $L^*$ and $\hat{H}$, c.f. Eqs. (16)-18)
  (remember about the operator action ordering from the right to the left).

   If we have the spectral solution for
  $\hat{H}$ in hands, in terms of eigenvalues $\lambda$ and $L^2(R)$  eigenfunctions $\Psi _{\lambda }(x)$,  then
   Eq. (17) tells us that the eigenvalues of $L^*$  are  $- \lambda$,  while the eigenfunctions appear in the
transformed form    $\phi _{\lambda }(x) = \Psi _{\lambda }(x) \rho _*^{1/2}(x)$.  The probability distribution $\rho(x,t)$ that {\it  can be expanded} into  $\phi _{\lambda }(x)$
  will relax at rates, determined by gaps in the eigenvalue spectrum of $\hat{H}$. This  is called a {\it spectral relaxation   pattern}.

   It is clear that the validity of the aforementioned expansion is guaranteed  only, if $L^*$ is
  a Hermitian operator. To this end an appropriate function space needs to be specified, and we know that  $\rho (x,t)$ (and likewise   the initial datum $\rho (x,0)$)    must be an element of  $L^2(R, \rho _*^{-1})$, as mentioned before in a paragraph below Eq. (8).
   This excludes from considerations such initial pdf as the Cauchy distribution, mentioned in Ref. \cite{sokolov} (the pertinent discussion appears between the formulas (3) and (4)), in connection with the conjectured  non-spectral  relaxation pattern.

    The point is that  to employ  the Fourier series  (eigenfunction)  expansion one   needs to respect  carefully adjusted   domain properties of transformed operators, see (15)-(20).   Our freedom of choice of   initial conditions for $\rho (x,t)$ is   thus limited  to these pdf $\rho (x,0)$  only,  which   decay sufficiently faster at infinity  than $\rho _*^{-1/2}(x)$ grows, and thus   transform via Eq. (6)   into square integrable functions   $\Psi (x,t)$, \cite{pavl}.   Told otherwise, the Cauchy pdf is not a legitimate candidate for an initial distribution for the Fokker-Planck equation of the Ornstein-Uhlenbeck process.

   Assume that $\rho _*(x)$  stands for the invariant pdf of the Ornstein-Uhlenbeck process  (c.f. the Appendices section for more details). Let $f_n(x)$ stand for the eigenfunctions of the OU process generator $L= \Delta  -x\nabla $  in $L^2(\rho _*)$, with eigenvalues $ -\lambda _n= n$, $n=0, 1, 2,...$. Then,  according to Eq. (20),   $L^* (\rho _* f_n) = - n \rho _* f_n$.   In terms of the eigenfunctions of $L^*$, we have the   the   expansion formula:  $\rho (x,t)= \rho _*(x) \sum_{n=0}^{+\infty } c_n  e ^{-\lambda _nt} f_n(x)$, with $c_n = \int_R f_N(x)\rho _0(x)dx$.   Consequently, beginning from an arbitrary initial distribution $\rho _0(x)  \in L^2(R, \rho _*^{-1})$, the probability law of the process converges exponentially fast to the invariant distribution. This is the spectral relaxation scenario and there is no room for its violation in the theoretical setting of the  standard Ornstein-Uhlenbeck process..

\section{Steep potential wells in  the Brownian motion: sequential approximations of the reflecting  well/interval.}

\subsection{Extremally steep (superharmonic)  potentials in the   Langevin-Fokker-Planck approach.}

It is   a folk wisdom, that  a sequence of    symmetric single well  (superharmonic) potentials
$ U_{2n}(x)= (x/L)^{2n}/2n$, with the growth of their steepness (i.e. for   large values of   $n$),
 can be used as an   approximation of the infinite  well  potential    with  {\it  reflecting}  boundaries
  located at $x=\pm L$.
 The statement has been  widely  adopted in the literature,  its   validity taken for granted and   possible limitations
have been ignored or  bypassed.  While some caution is here necessary.

For computational simplicity,   in below,   we shall  pass to  the  dimensionless notation $x/L  \rightarrow  x$,  so that
  $U(x)= x^{2n}/2n$, for large values of $n$ stands for an approximation of the  infinite well,   supported on $[-1,1]$).
  Closely related   potentials   of the form   $U(x)= 2n\,  x^{2n}$  and $U(x) =
  x^{2n}$ likewise can be employed  as  the infinite well approximations.

   We point out that  one needs to observe possibly annoying  boundary subtleties. Namely,  at $x=\pm 1$ the pertinent
    potentials  take values  $1/2n$, $2n$ and  $1$ respectively,  and   their  respective  limiting values
    (point-wise limits,   as $2n$ grows to $\infty $)    read  $0$, $\infty $ and $1$. That
    needs to be kept under scrutiny, while  invoking the standard  definition of the   infinite well enclosure,
     set on the interval $[-1,1] \subset R$
     (we impose the  exterior boundary data, instead of   locally defined data which are normally
       assigned to the interval  endpoints only), c.f. \cite{robinett,karw}:
 \be
U_{well} =
 \left\{
       \begin{array}{ll}
        0 , & \hbox{$|x| < 1$;} \\
         \infty , & \hbox{$|x| \geq 1$.}
       \end{array}
     \right\}
 \ee

This infinite well definition stays in conformity with the limiting   $n\rightarrow \infty $
  behavior of $U(x) = 2n\, x^{2n}$ at $x=\pm 1$, but is  incompatible with   the remaining  $U(x)$
   options, which take  boundary values  $1$ or $0$, instead of  $\infty $.

  \begin{figure}[h]
\begin{center}
\centering
\includegraphics[width=55mm,height=55mm]{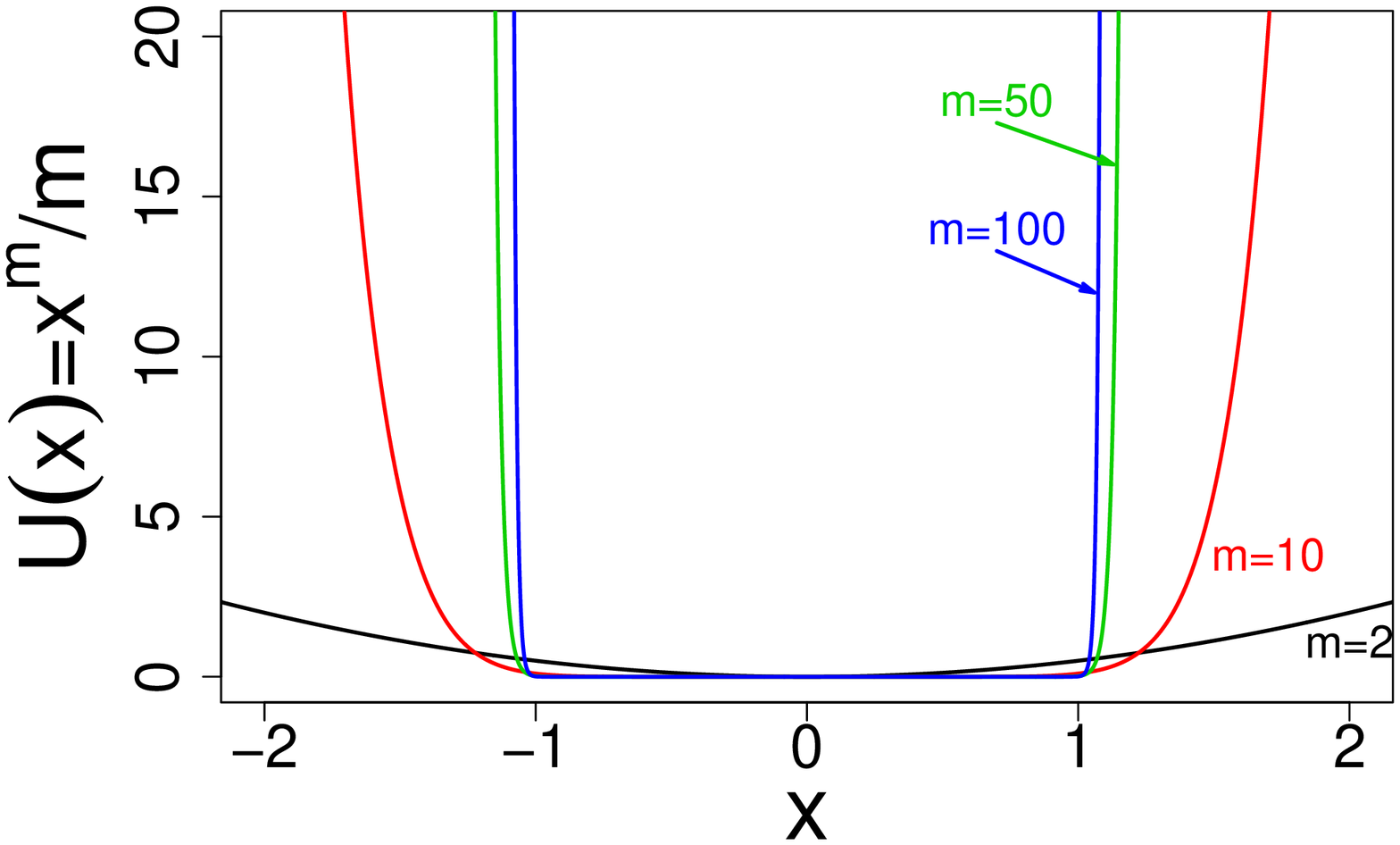}
\includegraphics[width=55mm,height=55mm]{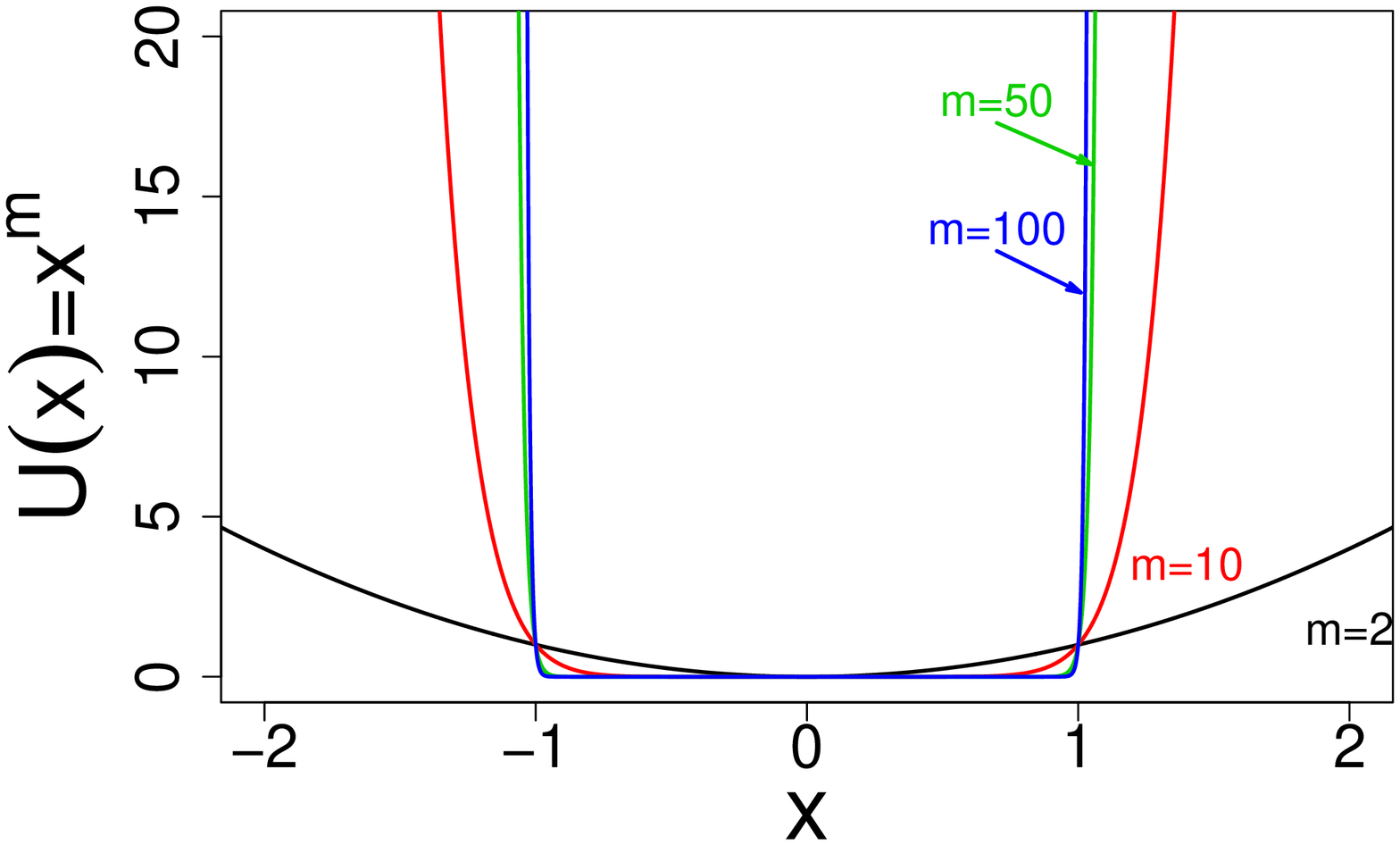}
\includegraphics[width=55mm,height=55mm]{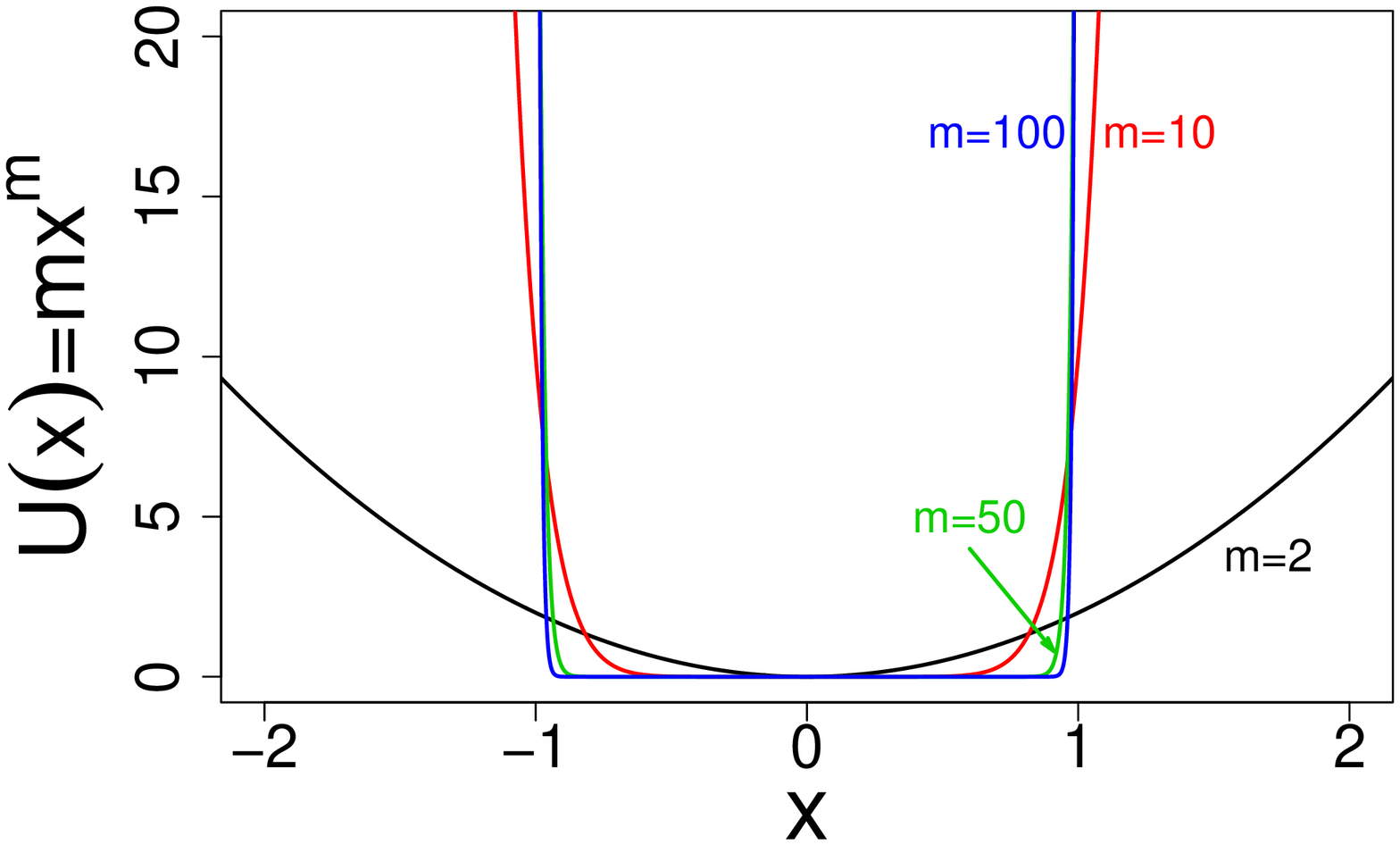}
\caption{Left panel:  $U(x)= x^{m}/m$  for  $m = 2, 10, 50, 100$. Note that $U(\pm 1) = 1/m$
for  $m<\infty $ and  sets at  $0$  as $m \rightarrow \infty $.   Middle panel:
 $U(x)= x^m$, we  have  $\lim_{m \rightarrow \infty } U(x) = U_{well}(x)= \infty $ for all $|x|>1$
 and $U_{well}(x)= 0 $   for all $|x|<1$, but  $U_{well}(\pm 1)=1$.   Right panel: $U(x) = m x^m$, we  have  $U_{well}(x)  =\lim_{m \rightarrow \infty } U(x)= \infty $
  for all $x\geq 1$ and    $U_{well}(x) = 0$   for all  $|x| < 1$.}
\end{center}
\end{figure}

Recently,  the $2n \rightarrow \infty $  sequential approximation   has  been explored   in  the   Langevin-Fokker-Planck
 description of the Brownian (and subsequently L\'{e}vy-stable)  motion  in the extremally anharmonic  (Newtonian)
   force  field $  b(x) \sim - x^{2n-1}$,
  to mimic and ultimately justify the  validity of {\it  reflecting boundary data} in  the emergent infinite well
   enclosure,  \cite{chechkin}-\cite{dybiec3}.  The  affiliated Schr\"{o}dinger-type  route has never  been analyzed.

On the other hand, an  analogous  limiting procedure,  while  employed on
 the Schr\"{o}dinger-type operator level of description,   has been found to  provide an  accurate
  approximation of the   traditional Dirichlet infinite well enclosure, \cite{froman}-\cite{robinett}.
  Surprisingly, in this case  the   Langevin-Fokker-Planck alternative has not been explored at all
   (this point we shall address in Section V).

Let us denote  $\rho(x)=A\exp[-U(x)]$, where  $U(x)\equiv U_m(x)=  x^{m}/m$,  $m=2n$.  The normalization condition  $
A\, \int_R  \exp[-x^{m}/m]\,dx= 1$,    upon a substitution  $y=x^{m}/m$, gives rise to
\be
1=2A (m)^{(1-m)/m} \int\limits_0^\infty \exp(-y)\,y^{1/m-1}dy,
\ee
where the integral expression is recognizable as the Euler Gamma function   $\Gamma(z)=\int\limits_0^\infty x^{z-1}e^{-x}dx$.
Accordingly, we have:
\be
A_m \equiv A_m  =   [2m^{(1-m)/m}  \Gamma(1/m)]^{-1}= [2m^{1/m}\Gamma(1+1/m)]^{-1}.
\ee

\begin{figure}[h]
\begin{center}
\centering
\includegraphics[width=55mm,height=55mm]{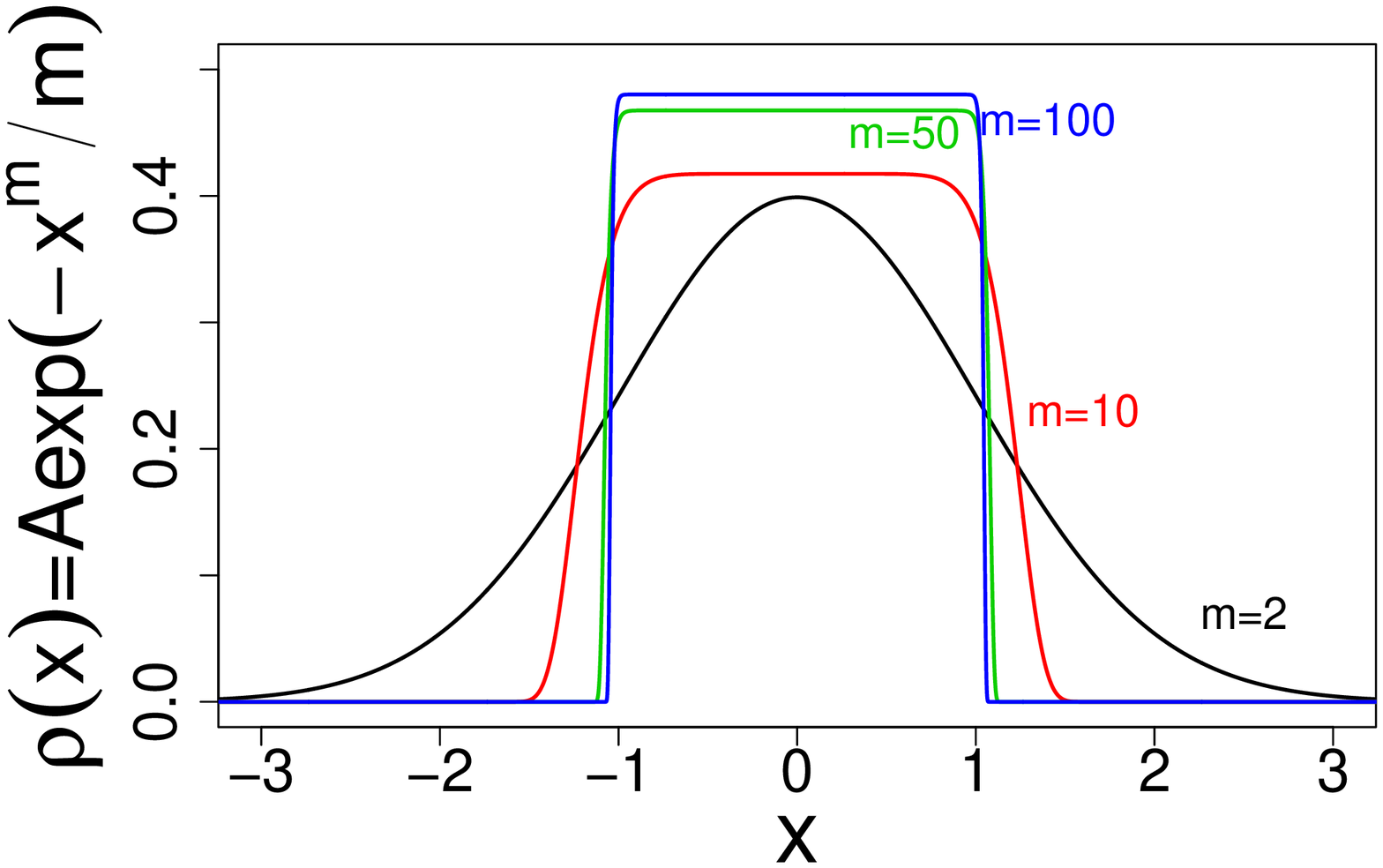}
\includegraphics[width=55mm,height=55mm]{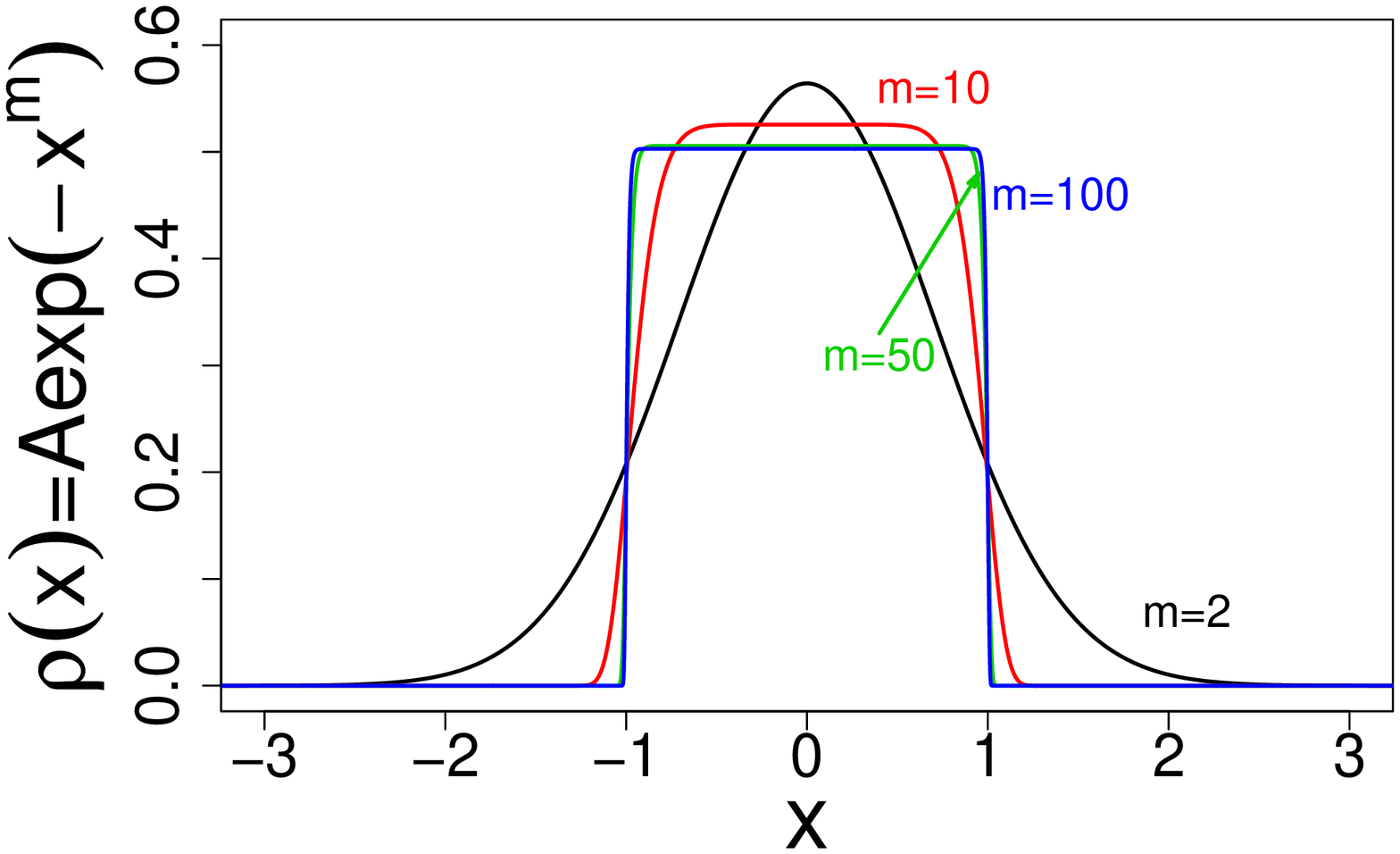}
\includegraphics[width=55mm,height=55mm]{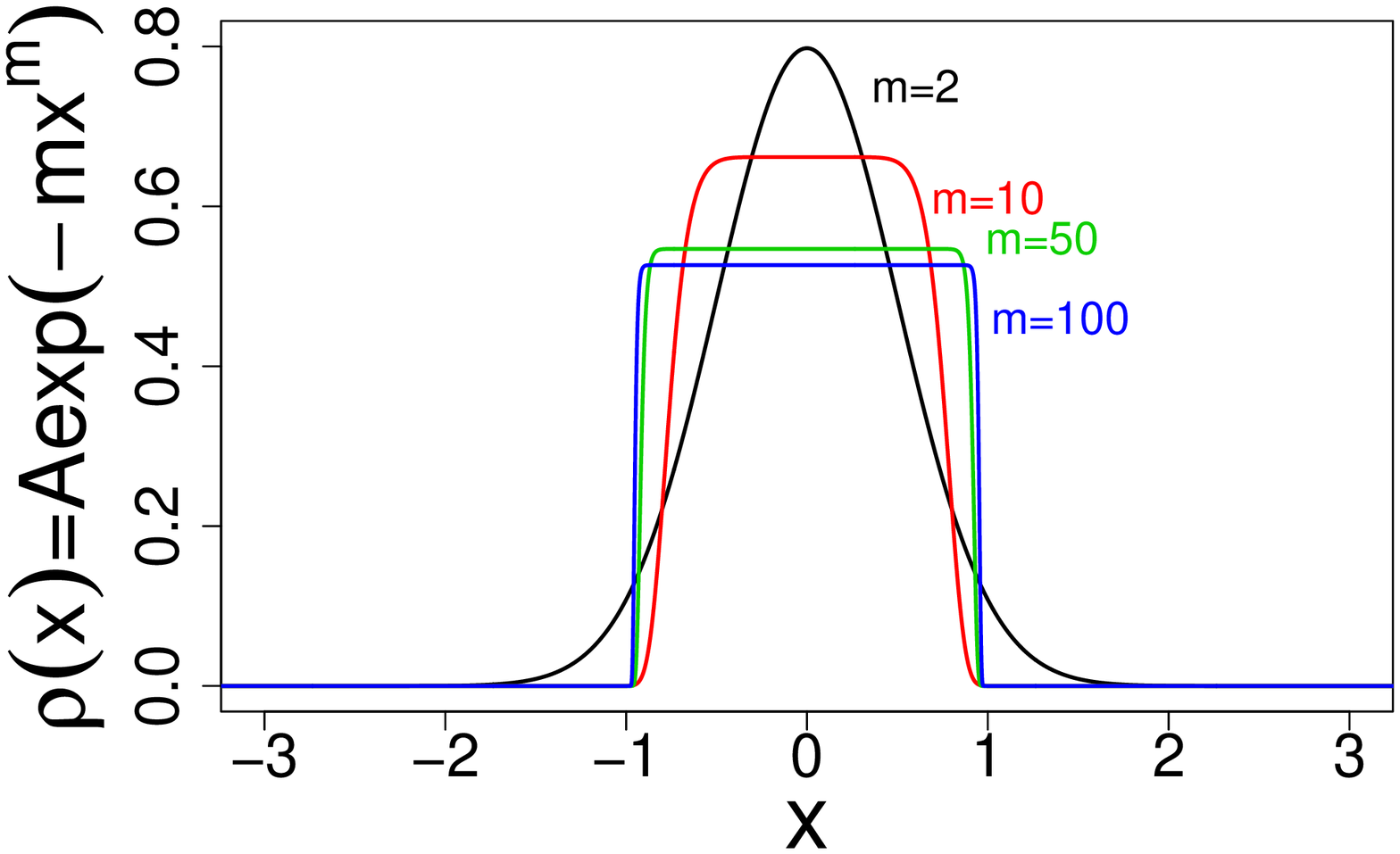}
\caption{We depict  (dimensionless, after  suitable  rescaling)  stationary Gibbs-Boltzmann densities of   the Langevin-driven
diffusion process (1)-(5),   $\rho _*(x)= A \,  \exp [ - U(x)]$,   $1/A  =
\int_{-\infty }^{+\infty } \exp[-U(x)]$, for different  versions  of strongly anharmonic potentials  $U(x)$
and selected values of  $m=2n$.    Note scale differences along the vertical axis. We have bypassed  subtleties of the
 boundary behavior of  $\rho _{well}(\pm 1) =   \rho _{*m\rightarrow \infty }(\pm 1)$, but one needs to keep in mind  that
  the pertinent  boundary  values read $1/2, 1/2e, \infty $ for $U(x)= x^m/m,\, x^m, \, mx^m$
  respectively.  }
\end{center}
\end{figure}

The  normalization coefficient $B$  for the case of  $U_m(x)= x^m$, $m=2n$,   is  the $ m^{1/m}$  multiple of that in
 Eq. (22):  $B = m^{1/m} A$.
Since $ m^{1/m} \rightarrow  1$ with $m\rightarrow \infty $, the limiting behavior (convergence rate) in both cases is similar.

 We note that $A\equiv A_m$ approaches
the limiting value $1/2$  from below    as a growing function, while $B\equiv B_m$ approaches $1/2$ from above as
a decreasing function.

\begin{figure}[h]
\begin{center}
\centering
\includegraphics[width=70mm,height=70mm]{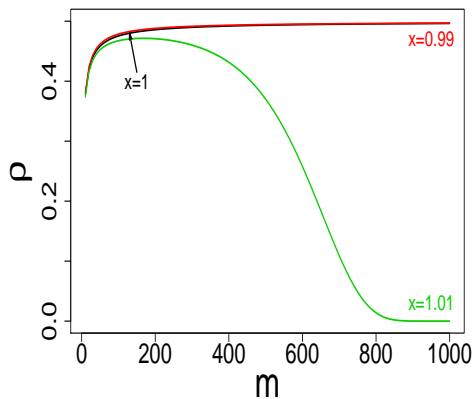}
\caption{The  (point-wise)  $m=2n$-dependence of the Gibbs-Boltzmann invariant pdf  $ \rho _*(x)= A_m \exp(- x^m/m)$
 at $x=0,99$ (red), $x=1$ (black) and $x=1.01$ (green).  In the figure resolution, we cannot distinguish the
 curve corresponding to $x=1$  from this for $x=0.99$, both have  a common limit $1/2$.}
\end{center}
\end{figure}

Since $\rho (0)= A$ or $B$ respectively, we  realize that  the   pertinent  normalization coefficients actually set the
 value of an asymptotic "plateau", with $1/2$ referring to a uniform probability distribution on the interval $[-1,1]$, formally regarded
as the $m\rightarrow \infty $ limit for the sequential approximation $x^m/m$.
However,  the uniform distribution is recovered  on an open interval
 $(-1,1)$  only, if one would  employ   $x^m$  and $m x^m$    approximating   sequences. See e.g. Figs. 1 to 3.

It seems  instructive to visualize  the dependence of $\rho _*(x)$   on  $m=2n$ in the close vicinity of boundary
 points $x=\pm 1$.    This  is  depicted in Fig. 4     for   $U(x) = x^m/m$, at points  $x=0.99, 1, 1.01$.

\subsection{The reference problem:   Reflected Brownian motion on $[-1,1]$.}

The uniform probability distribution on the interval $[-1,1]$  may be interpreted as  a signature of reflecting (Neumann)
 boundary data  and thence of of the reflecting Brownian motion in the interval, \cite{pavl,linetsky,bickel}.
  We emphasize  that nothing is  normally  said about the exterior   $R\backslash [-1,1]$  of the pertinent interval and  in the traditional  mathematical  lore,  the  Neumann  data  are defined  locally, \cite{pavl,pinsky}.

  An exact solution  of the  latter  problem refers to  the standard Laplacian, while restricted to the  interval
  and subject to  reflecting (e.g.  Neumann)  boundary conditions.   Solutions of the diffusion  equation
$\partial _t \Psi (x,t) = \Delta _{\cal{N}} \Psi (x,t)$  in  $ [-1,1]$,     need to respect   $(\partial _x \Psi ) (-1,t)= 0 =   (\partial _x \Psi )(+1,t)$ for all $t$.    The pertinent transition   density   reads, \cite{bickel}:
  \be
k_{\cal{N}} (t,x,y)  =     {\frac{1}{2}}  +   \sum_{n=1}^{\infty } \cos\left( {\frac{n\pi
}2}(x +1)\right) \cos\left( {\frac{n \pi }2} (y +1)\right)  \exp
\left(- {\frac{n^2\pi ^2}{4}} \, t  \right)
\ee
and is an integral kernel of the reflecting semigroup  $\exp( t\Delta _{\cal{N}})$.
The  Neumann  operator $  - \Delta _{\cal{N}}$ admits the eigenvalue $0$ at the bottom of its spectrum,
the corresponding eigenfunction being a constant  $1/\sqrt{2}$,  whose square  actually  stands for  a uniform
probability distribution on the interval of length $2$,  see \cite{pavl,bickel,pinsky}.

 Solutions of the diffusion equation with reflection at the boundaries of $D=[-1,1]$  can be modeled  by setting
 $p(x,t)= k_{\cal{N}}(t,x,x_0)$, while remembering that $p(x,0)= \delta(x-x_0)$.    We can as well resort to  $\Psi (x,t)= \int_D k_{\cal{N}} (t,x,y) \Psi (y) dy$.
  Note that all $n\geq 1$ eigenvalues coincide with these attributed to  the absorbing   case, \cite{gar}, and
   (up to dimensional constants)  coincide with eigenvalues $(n\pi /2)^2$ of the  standard  (quantum mechanical)
     infinite well problem,  with Dirichlet boundary data. The eigenfunctions respect Neumann conditions, and do
      not  {\it necessarily}  vanish at the boundary points (that would be the Dirichlet case).

We note that the exponential rate of convergence to equilibrium in the present case (D=1) reads $\pi ^2/4$  and is three
 times smaller than the rate  $E_2- E_1= 3\pi ^2/4$     established in Section III.D  as an asymptotic signature of the
  taboo process (conditioned  never to exit  the interval with inaccessible  boundaries, \cite{gar,pinsky}).

  The direct path-wise description of the reflected Brownian motion  belongs to a non-standard inventory, if compared with the standard Langevin modeling. It
    involves the so-called Skorokhod problem and  a class of stochastic differential equations with reflection,  \cite{pilipenko,bickel}.  This problem is   avoided  (or circumvented)
    in the pragmatic approach to the reflection issue via computer simulations of sample paths, where it is
     the boundary behavior  (proper  handling of the instantaneous reflection)  which appropriately alters
      the   statistical features of  propagation  of the otherwise   free  Brownian motion, \cite{denisov,metzler,dybiec}.

\subsection{Superharmonic  approximations of the  reflecting  well: (in)validity of  the  sharp
Neumann boundary condition.}

\begin{figure}[h]
\begin{center}
\centering
\includegraphics[width=40mm,height=40mm]{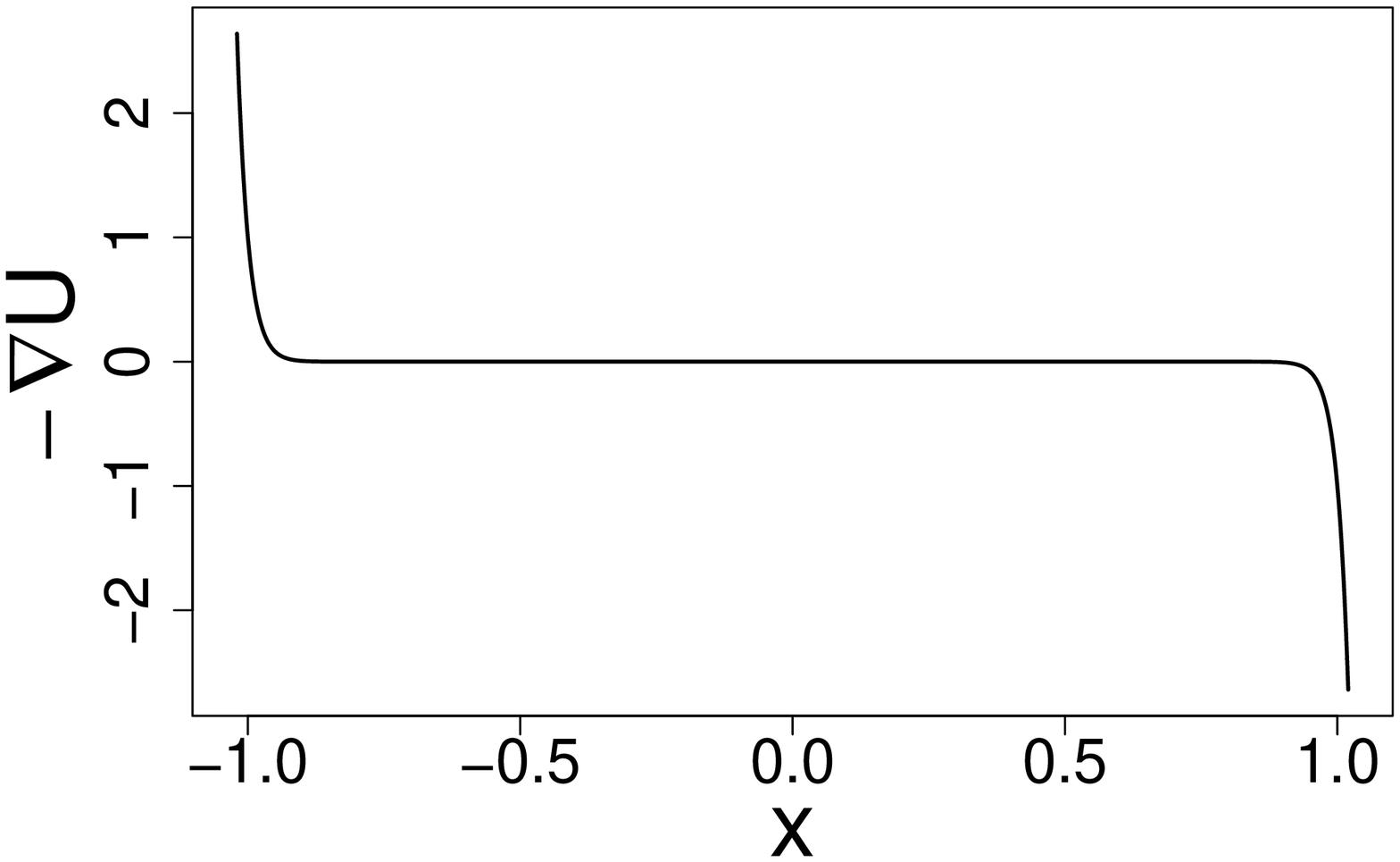}
\includegraphics[width=45mm,height=40mm]{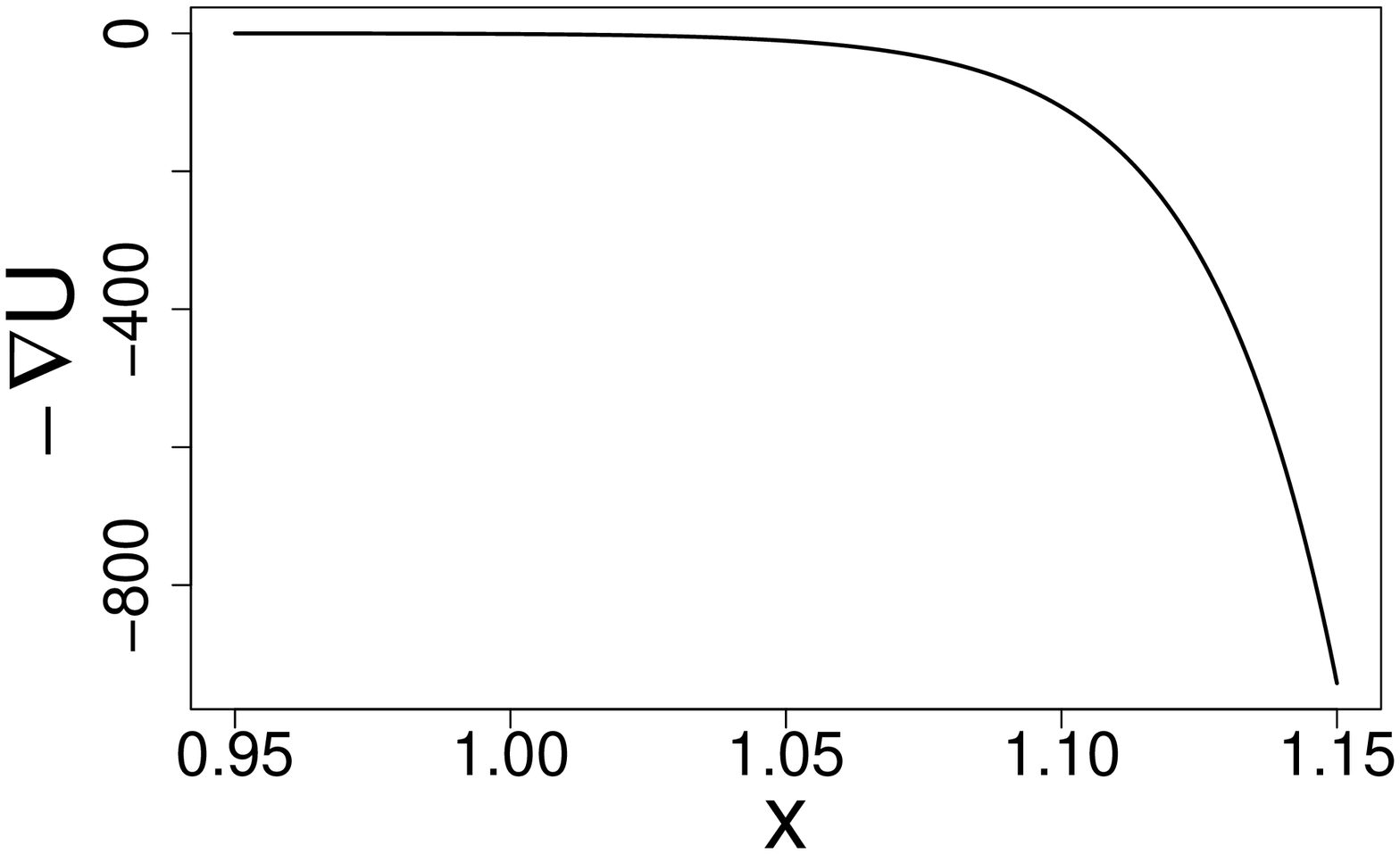}
\includegraphics[width=40mm,height=40mm]{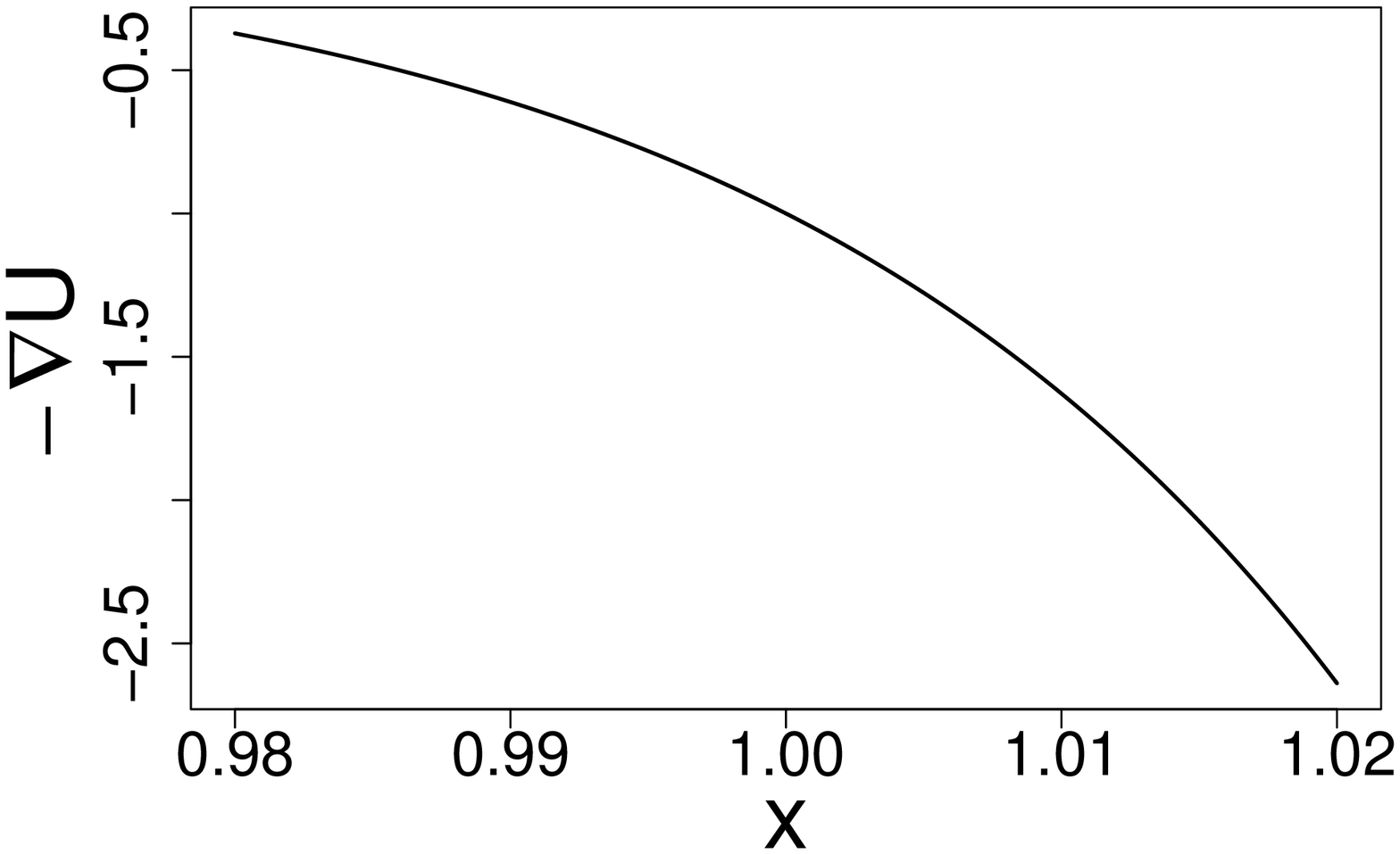}
\includegraphics[width=40mm,height=40mm]{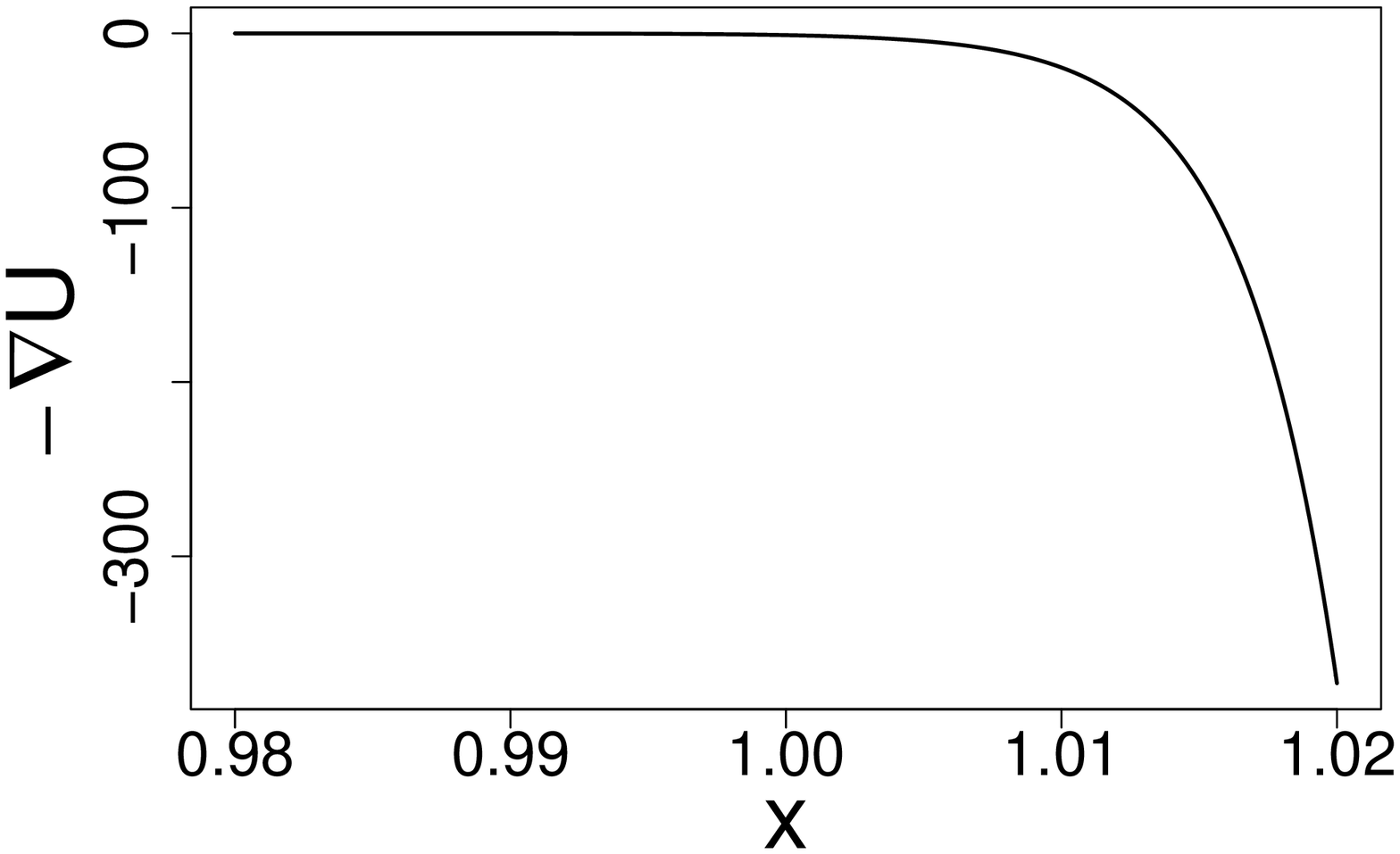}
\caption{$U(x)= x^m/m, m=2n$.  Graphical insight into where (on R) and when (large $m$)
 the forward drift of a diffusion process
  $ b(x) = - \nabla U(x)$ may be  considered as  vanishing. First three   $- \nabla U(x)= - x^{m-1}$   figures from the left  are
  depicted for $m=50$,  $x\in R$   (for any finite $m$ we have a residual tail
  of $ -x^{m-1}$ which   extends  to infinity).   Since  the  behavior of  $-\nabla U(x) $  in the vicinity of $x=1$  is  somewhat  blurred (for any finite $m$, we have
     $-\nabla U(1)= -1 $), the real state of affairs is depicted in the enlarged segment located in the interval
       $[0.98,1.02]$, for m=50 and next for  m=300.   Note scale indications on the vertical axis.}
\end{center}
\end{figure}

\begin{figure}[h]
\begin{center}
\centering
 \includegraphics[width=40mm,height=40mm]{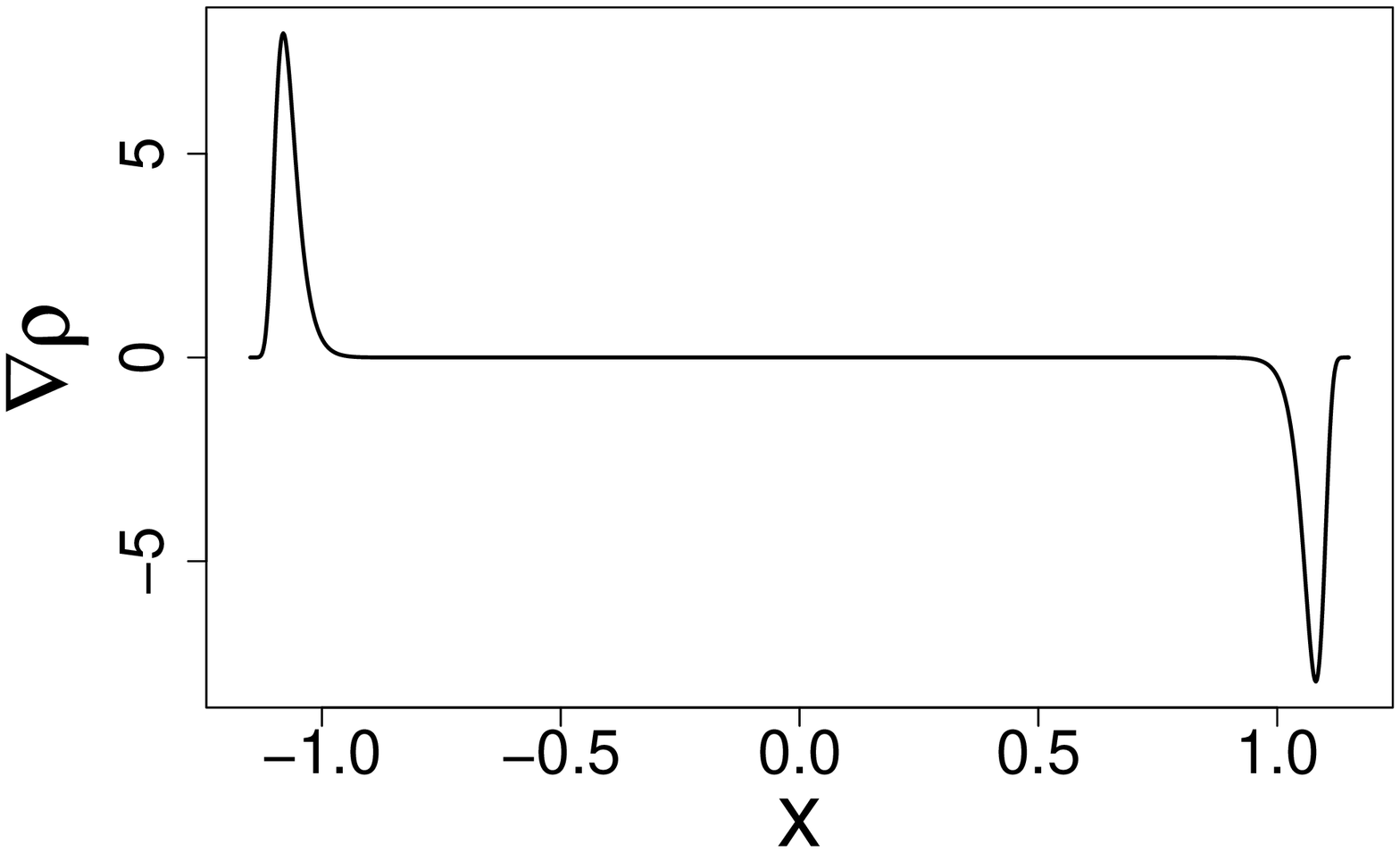}
\includegraphics[width=40mm,height=40mm]{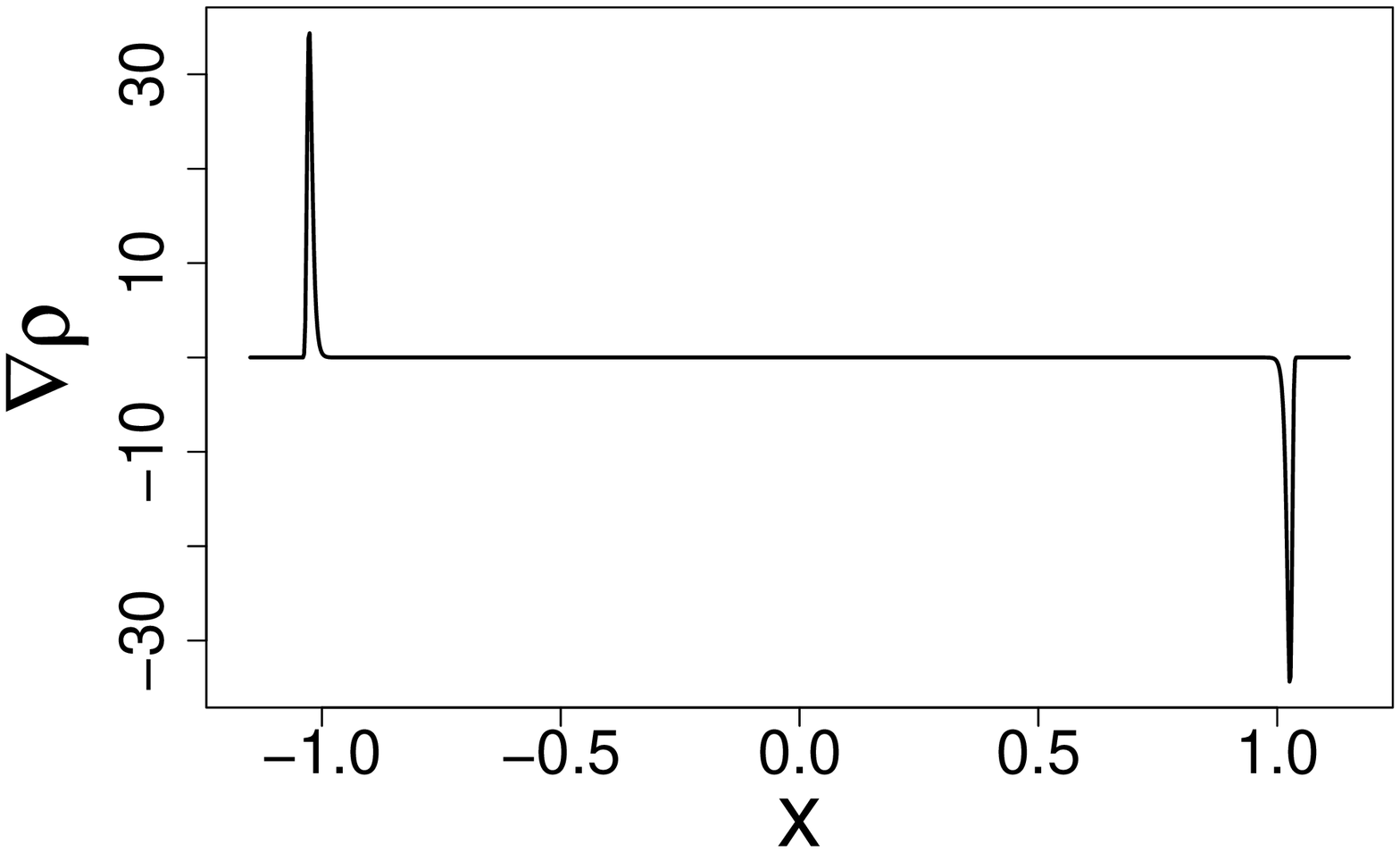}
\includegraphics[width=40mm,height=40mm]{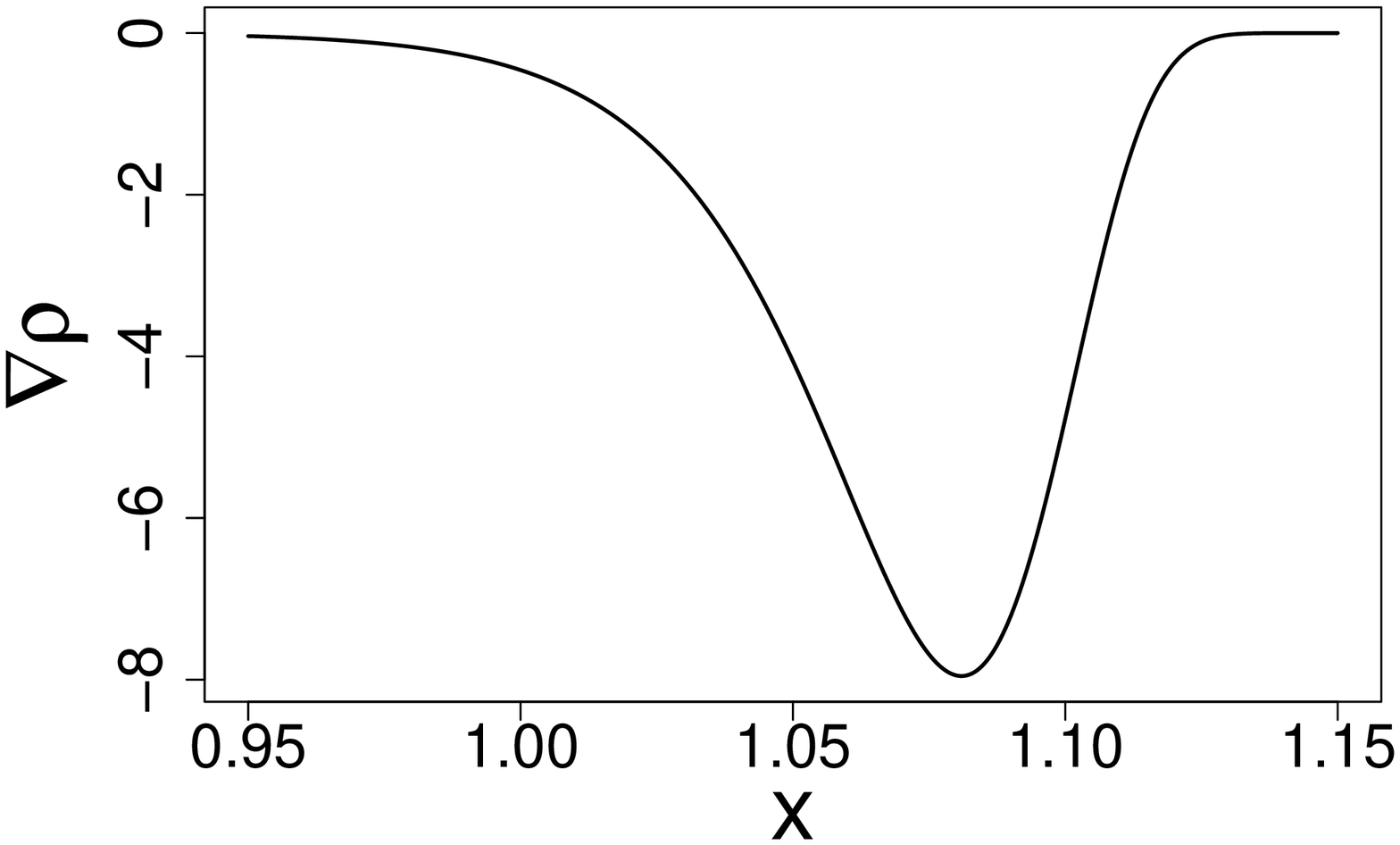}
\includegraphics[width=40mm,height=40mm]{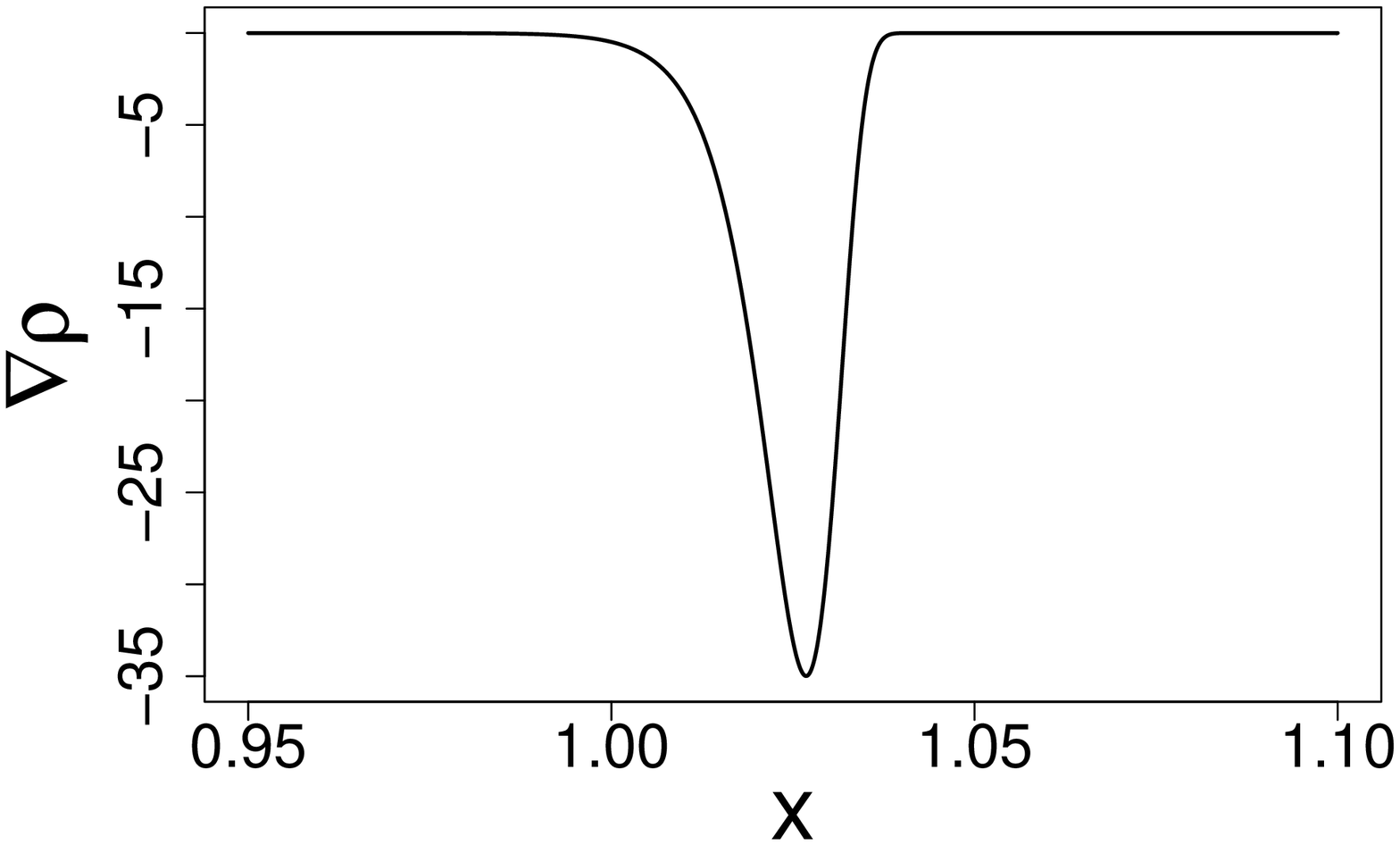}
\includegraphics[width=40mm,height=40mm]{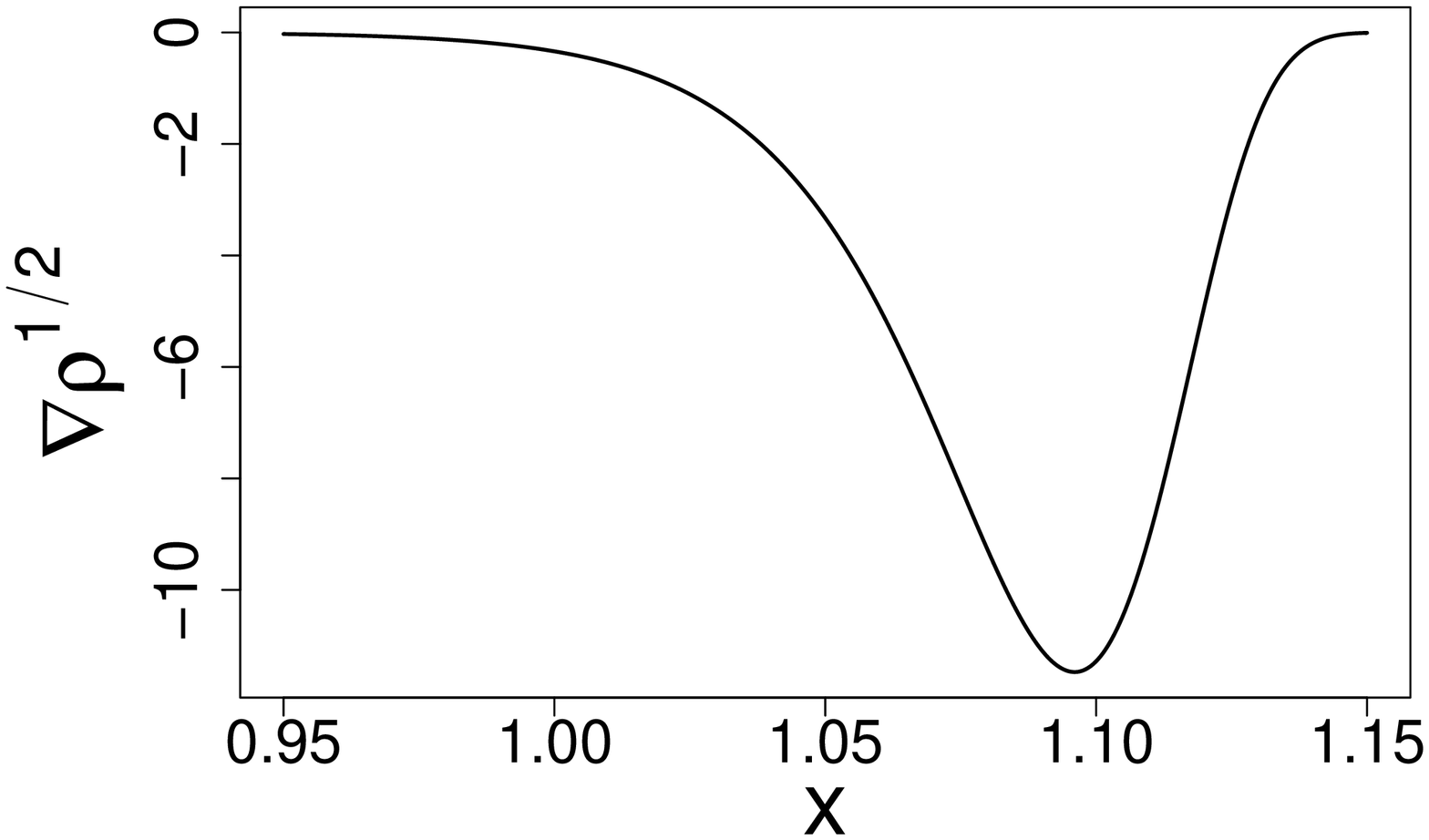}
\includegraphics[width=40mm,height=40mm]{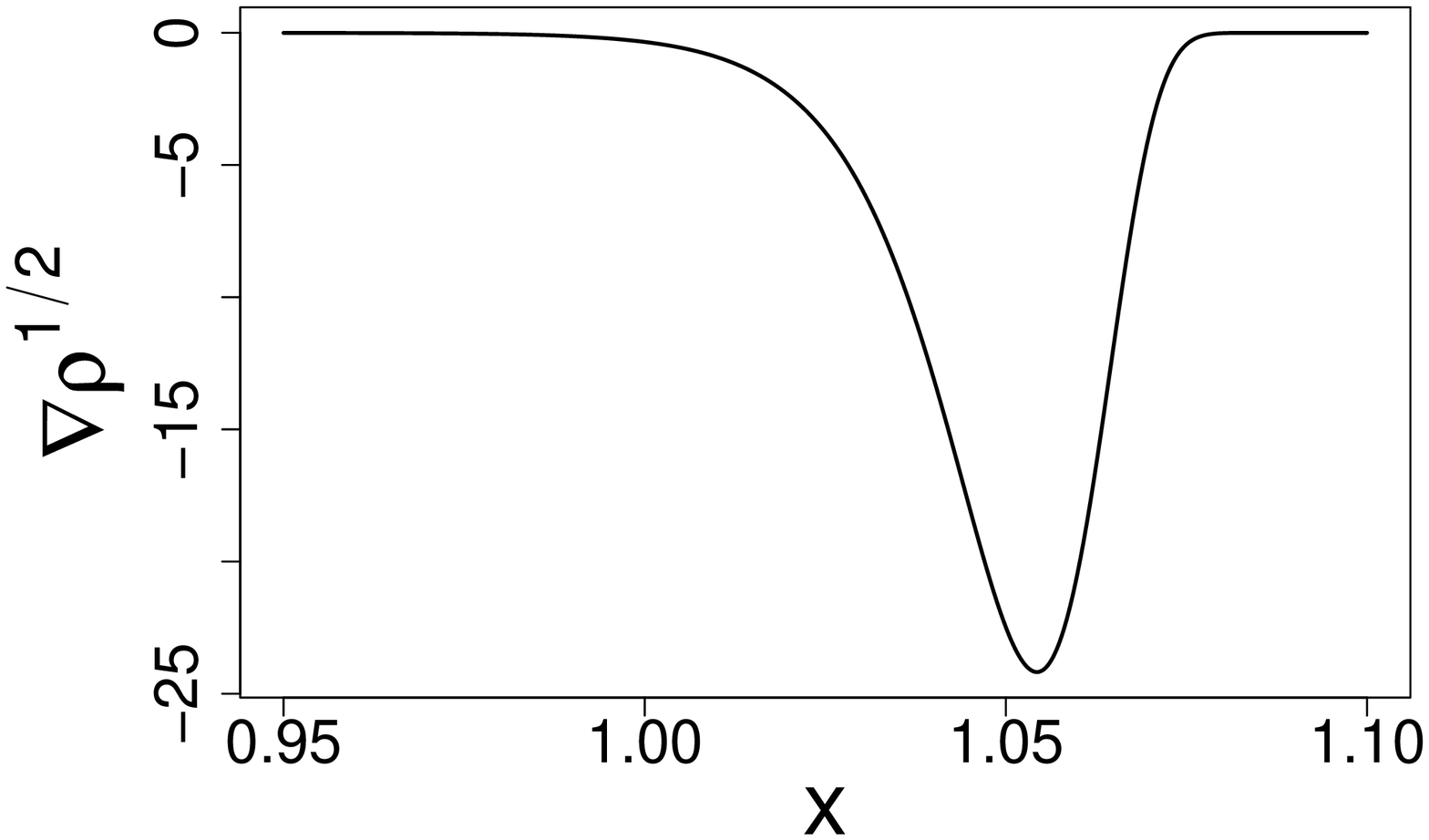}
\includegraphics[width=40mm,height=40mm]{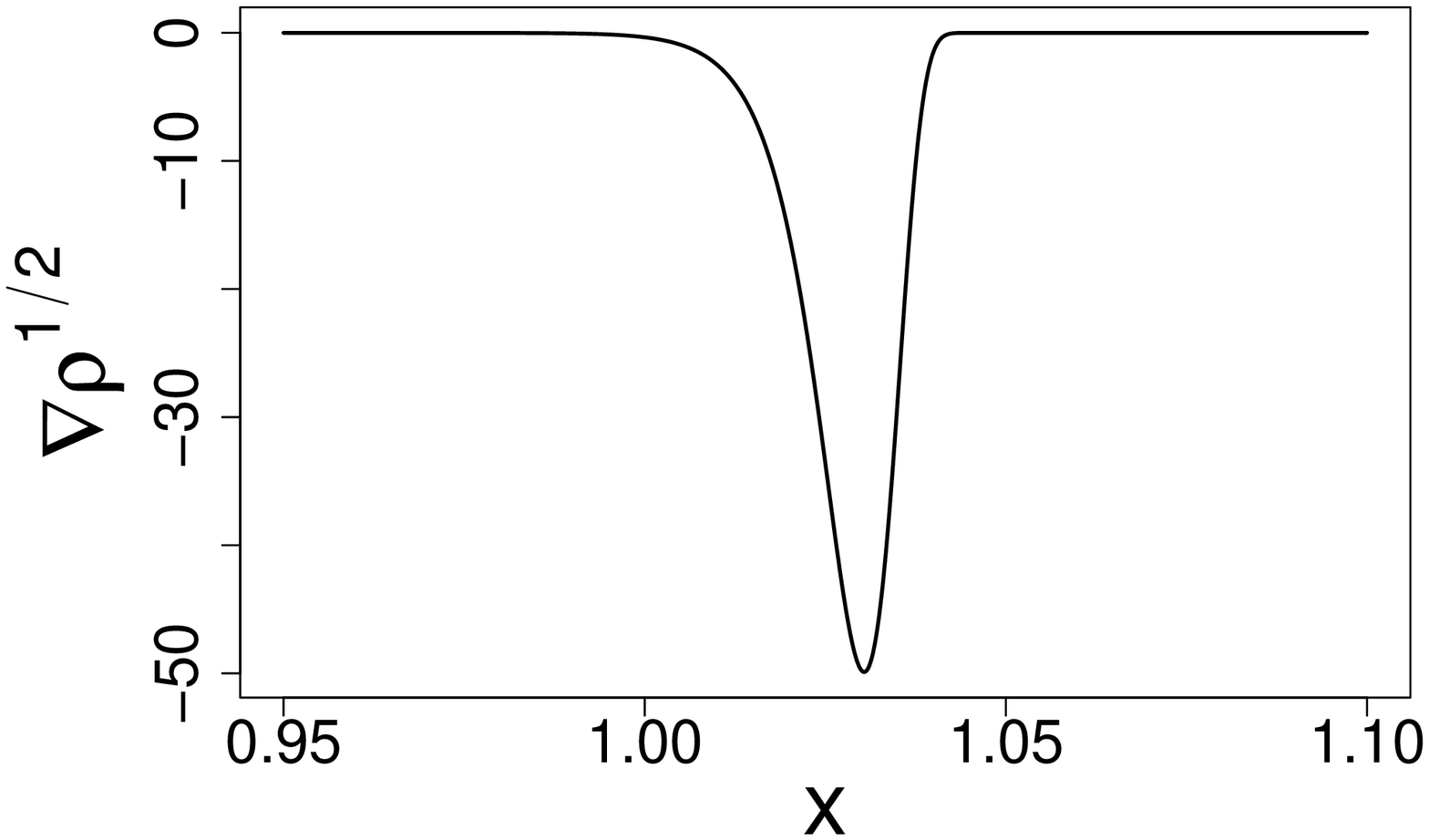}
\includegraphics[width=40mm,height=40mm]{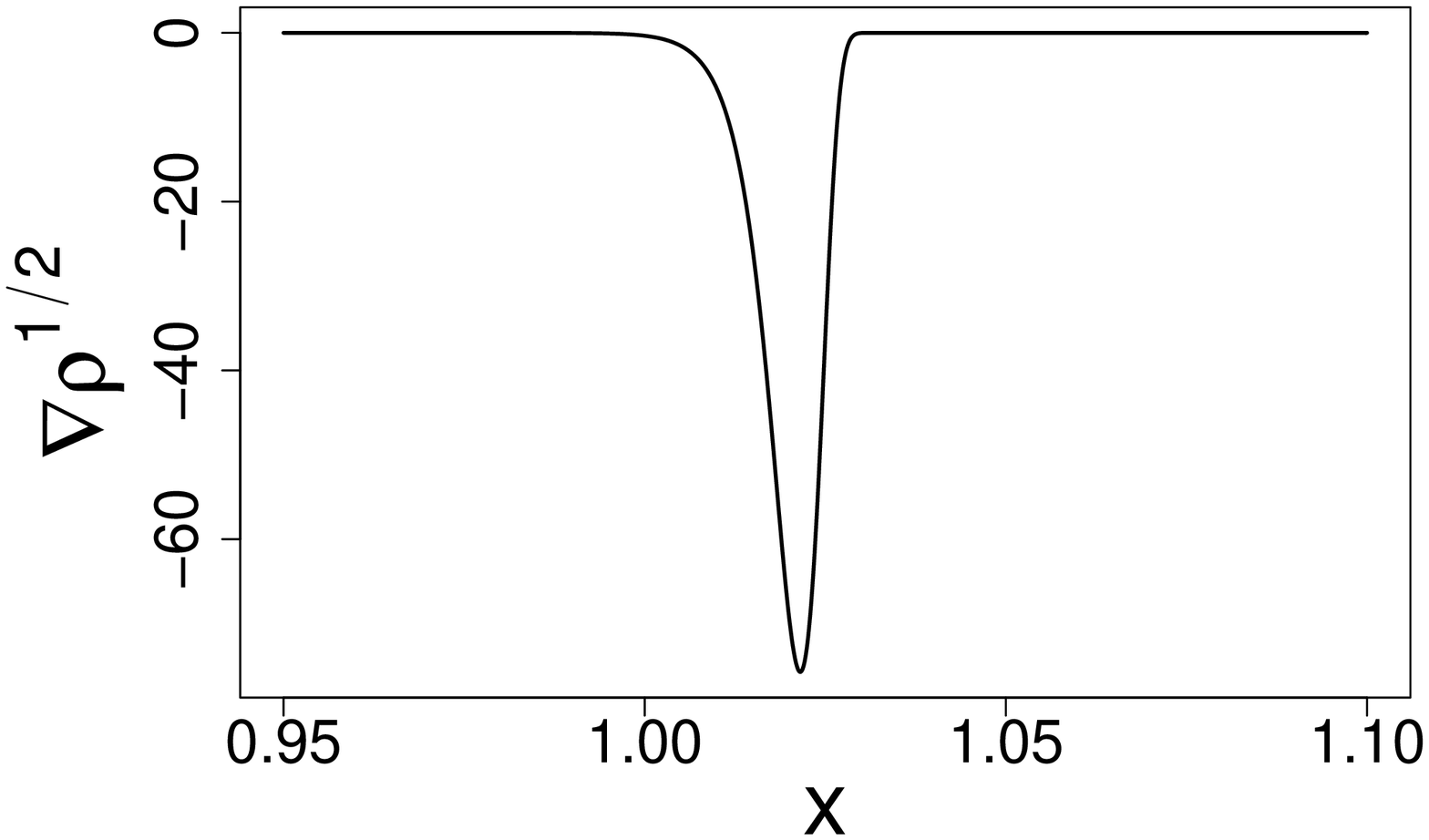}
\caption{$U(x)= x^m/m, m=2n$. Graphical insight into when (large $m$)  and where (vicinity  of $\pm 1$) we can interpret
$\nabla \rho _*(x)$ as vanishing (we recall that  $b(x)= -\nabla U(x) = [\nabla \rho _*(x)]/\rho _*(x)$).
 Upper row from the   left:  $\nabla \rho _* (x)=  - A\, x^{m-1} \exp (-x^m/m)$   for  $m = 50, 200$, followed
by  details   of $\nabla  \rho _*(x)$  in the vicinity of the
  point $+1$  for $m=50, 200$.  Second row reports  comparatively, for
  $m=50, 100, 200, 300$,   the behavior of  $\nabla  \rho _*^{1/2}(x)$ which  asymptotically ($m \rightarrow \infty $)
   is expected to show up  the signatures of Neumann boundary data.  C.f. Table I in below for a quantitative analysis.}
\end{center}
\end{figure}

As long as we prefer to deal with traditional Langevin-type  methods of analysis,  it is  of some   pragmatic interest  to know, how reliable  is  an approximation of the reflected Brownian motion in $[-1,1]$ by means of  the   attractive  Langevin driving (and thence the  Fokker-Planck equation),  with force terms (e.g. drifts)  coming from  extremally anharmonic (steep)  potential wells.

The main obstacle, we encounter here is that a "naive" $m=2n \rightarrow \infty $ limit is singular and cannot be
 safely executed on the level  of  diffusions  proper.
 We note that  for any finite $m$, irrespective of how large $m$ actually is,  we deal with a continuous and
 infinitely differentiable potential and likewise,  the Boltzmann-Gibbs pdf as  a consequence of (1)-(5).

On the informal, graphical level (Figs.  2 and 3) we  can  anticipate  that  the limiting pdf, is a constant  $1/2$ in the
closed interval (uniform distribution on the interval) and vanishes identically for $|x|>1$:
\be
\rho _{*,well}(x) = \lim_{m\rightarrow \infty }  \rho _{*m}(x)  =
\left\{
       \begin{array}{ll}
        {\frac{1}2}, & \hbox{$|x|\leq 1$;} \\
         0, & \hbox{$|x|> 1$.}
       \end{array}
     \right\}
\ee
This   is consistent with the   formal  limiting behavior  of $U(x) =  U_m(x)=  x^m/m$  (with $U_m$ explicitly   present   in the exponent of the BG density), as
$m \rightarrow \infty $,  and  properly  mimics the  infinite well potential enclosure, \cite{dubkov1}  whose boundary $\pm 1$  can be continuously
 reached from the interval interior:
 \be
U_{well} =  \lim_{m\rightarrow \infty } U_m(x)=
 \left\{
       \begin{array}{ll}
        0 , & \hbox{$|x|\leq 1$;} \\
         \infty , & \hbox{$|x|> 1$.}
       \end{array}
     \right\}
 \ee
 We note that the    infinite-valuedness  everywhere  beyond $[-1,1]$ (e.g. an exterior restriction)
  is an essential  complement to  potential  shapes depicted in  Figs. 2 and 3.

If   (25) is taken literally  as   a (more or less  mathematically  legal)   limit  of a sequence of  superharmonic potentials,  there is an obstacle  to be  overcome  or bypassed.    Namely, differentiability properties of   resultant (limiting) functions   $U$ and $\rho _*$  are   lost at the interval   $[-1,1]$  endpoints.   We emphasize that
 an infinite differentiability  is a valid  property of   $U(x)$,   $\rho  _*(x) $   and $\rho _*^{1/2}(x)$ for  all  $2\leq  m<\infty $, irrespective of how large $m$ actually is.

 Let us consider in more detail  the properties of $-\nabla U$ and $\nabla \rho _{*}$   in $R$,  for large $m=2n$,
  in the vicinity of  $\pm 1$.
Our  focus is on the  seemingly  obvious  reflecting behavior and the possible  (in)validity  of  Neumann boundary
conditions,  which  actually define the  reflecting Brownian motion in terms  of the
  Neumann Laplacian  $\Delta _{\cal{N}}$  in the interval, c.f.  Subsection  IV.B.

In the superharmonic  Langevin-type regime we have   $b(x)= -\nabla U(x) = \nabla \ln \rho _*(x) =  - x^{m-1}$.   Since $ \nabla \rho _*(x) =- A x^{m-1}\, \exp (- x^m/m)$,
we readily infer the location of its minimum  in the vicinity of $x=1$:
\be
x^m =  m-1  \Rightarrow   x= (m-1)^{1/m}
\ee
which, for large $m$, can be safely replaced   by  $x= m^{1/m}$.
One  may infer, c.f. \cite{kuczma}, a useful estimate for the actual location of the considered  minimum for finite but   large values of $m$:
\be
1 < {\frac{m}{m-1}} \leq   m^{1/m} \leq 1 +{\frac{2}{\sqrt{m}}}
\ee
with a limiting property $\lim m^{1/m}  =1$ as $ m\rightarrow \infty $.

     Accordingly for all finite values of $m$, irrespective of how large $m$ is,
 there is a lot to happen (in the lore of  turned over sample paths) in the narrow zone  of thickness $4/\sqrt{m}$,  beyond the interval  boundaries set at $\pm 1$ on $R$.

 \begin{table} [tbh]
\begin{tabular}{|c||c|c|c||c|c|c||c|c|c||}
\hline
& \multicolumn{3}{|c||}{$b(x)= \nabla \ln \rho _*(x) $}   & \multicolumn{3}{|c||}{$\nabla \rho _*(x)$} & \multicolumn{3}{|c||}{$\nabla \rho _*^{1/2}(x)$} \\
\hline
$m$ & $x=0.99$ & $x=1$ & $x=1.01$ & $x=0.99$ & $x=1$ & $x=1.01$ & $x=0.99$ & $x=1$ & $x=1.01$ \\
\hline
$50$ & -0.6111 & -1 & -1.6283 & -0.2823 & -0.4583 & -0.7368&  -0.2077& -0.3385& -0.5476\\
\hline
$100$   & -0.3697 & -1 & -2.6780 & -0.1769 & -0.4754 & -1.2517 & -0.1279& -0.3448 & -0.9154 \\
\hline
$ 200$ & -0.1353 & -1 & -7.2436 & -0.0660 & -0.4859 & -3.4102 & -0.0473& -0.3485& -2.4850 \\
\hline
$300 $  & -0.04954 & -1 & -19.5925 & -0.0243 & -0.4899 & -9.0155 & -0.0174& -0.3500 &-6.6452   \\
\hline
$600$ & -0.0024  & -1 &  -387.706& -0.0012 & -0.4943 & -99.9590&  -0.0009 & -0.3515& -98.4311 \\
\hline
$800$  & -0.000325 & -1 & -2836.47& -0.0002 & -0.4956 & -39.1928& -0.0001&-0.3520 &-166.71 \\
\hline
$\infty $ & 0 & -1 &   $-\infty $  & 0 & -0.5 & 0 & 0 & $- 1/2\sqrt{2} \simeq - 0.3535$&$ 0$\\
\hline
\end{tabular}
\caption{The vicinity $[0.99,1.01]$ of $x=1$.  The  approximation accuracy  of the   value   zero,  for
three gradient functions,  may be regarded satisfactory  ({\it fapp} - for all practical purposes) in the half-open
interval $[0.99, 1)$, but not at $1$, for  $m\geq 300$.  We note that the
 limiting behavior  of these  gradient functions  at $\pm 1$ is inconsistent with the formal definition of the
  reflecting Brownian motion, for which  a solution $\rho _*^{1/2}(x)$  of  $\Delta _{\cal{N}} \rho _*^{1/2}=0$
   needs to obey  $\nabla \rho _*^{1/2}(\pm 1) =0$.}
\end{table}

The smoothness properties of $\nabla \rho _*^{1/2}(x)$    (and the limiting behavior for $m\gg 1$)
can be read out from the formula   $(\nabla \rho_*^{1/2})(1) =  -  (A^{1/2}/2) \exp(-1/2m)$, where  $A=A_m$,
  c.f.  Eq. (35).   We note that the minimum of $\nabla \rho _*^{1/2}(x)$ is located a $x= [(2(m-1)]^{1/m}
  > (m-1)^{1/m} >1$.

    A detailed insight into the approximation accuracy of the  infinite well enclosure
   with Neumann boundary conditions, while in terms of superharmonic traps,  is provided in Table I.
     Beginning from $m=300$,  the  considered gradient functions are {\it fapp} (for  all practical purposes)
      equal  zero  in the interior of the well (e.g. for  $x<1$).
       Nonetheless,  irrespective of how large $m$ is, we have $-(\nabla U)( 1)= -1$.
        This gradient function rapidly varies for $|x|>1$,  in a narrowing "window"   (of thickness $\sim 1/\sqrt{m}$)
          close to $|x|=1$, c.f.  Figs. 5  and 6.

Both  $\rho _*(x)$ and $ \rho _*^{1/2}(x)$ have nonvanishing gradients  at $\pm 1$. We note that
 $(\nabla \rho_*^{1/2})(1) =  -  (A^{1/2}/2) \exp(-1/2m)$  is nonvanishing  for all $m$,   including the
 $m\rightarrow \infty $ limit. Indeed, since  $A= A_m$   converges to $ 1/2$  with
 $m\rightarrow \infty $, we get  the (point-wise)  limit  $\nabla \rho ^{1/2}_*(\pm 1) \rightarrow
  -1/2\sqrt{2}$.

    This shows that  the (expected  to hold true)  Neumann  boundary  condition is violated. Albeit  for
     sufficiently  large $m$ we can achieve  $\nabla \rho ^{1/2}_*(x) \equiv 0$  in  an arbitrarily close
      (interior, $|x|<1$)  vicinity of   $\pm 1$,  see e.g. Table I.

\section{Schr\"{o}dinger semigroup relative  of the Fokker-Planck dynamics in steep potential wells.}

\subsection{Reconstruction of the  Schr\"{o}dinger semigroup  from an eigenstate of the  motion generator.}

\begin{figure}[h]
\begin{center}
\centering
\includegraphics[width=55mm,height=55mm]{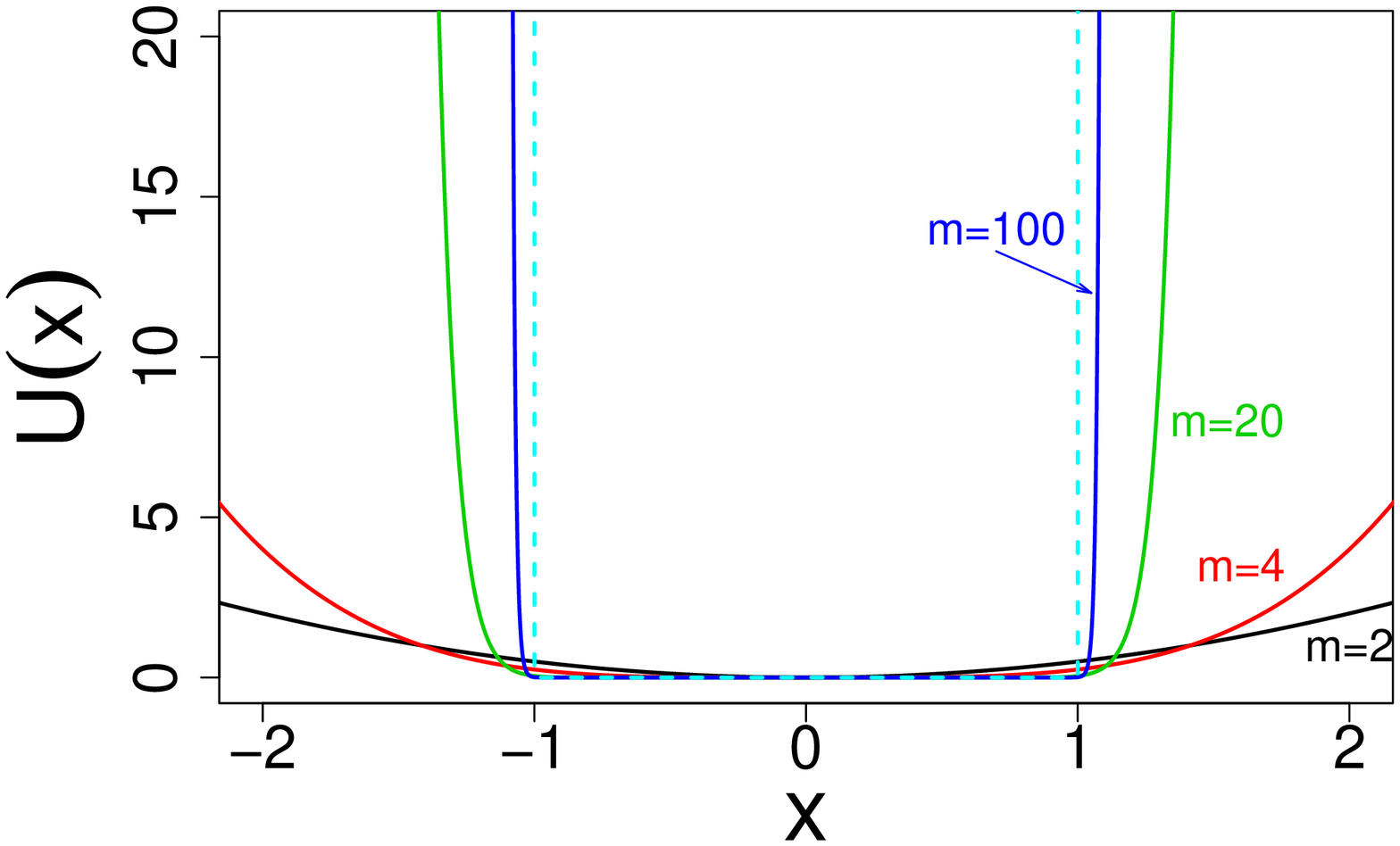}
\includegraphics[width=55mm,height=55mm]{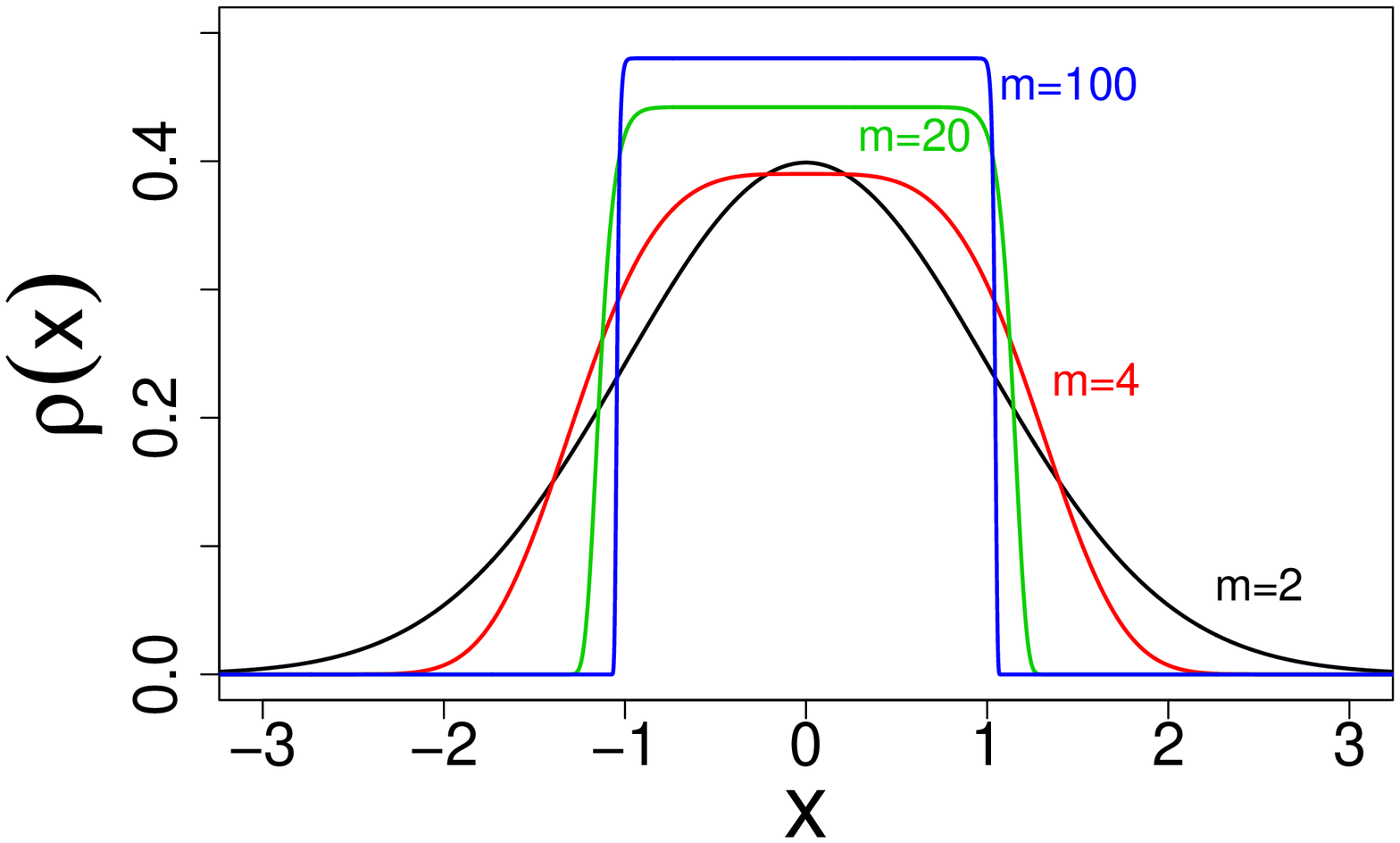}
\includegraphics[width=55mm,height=55mm]{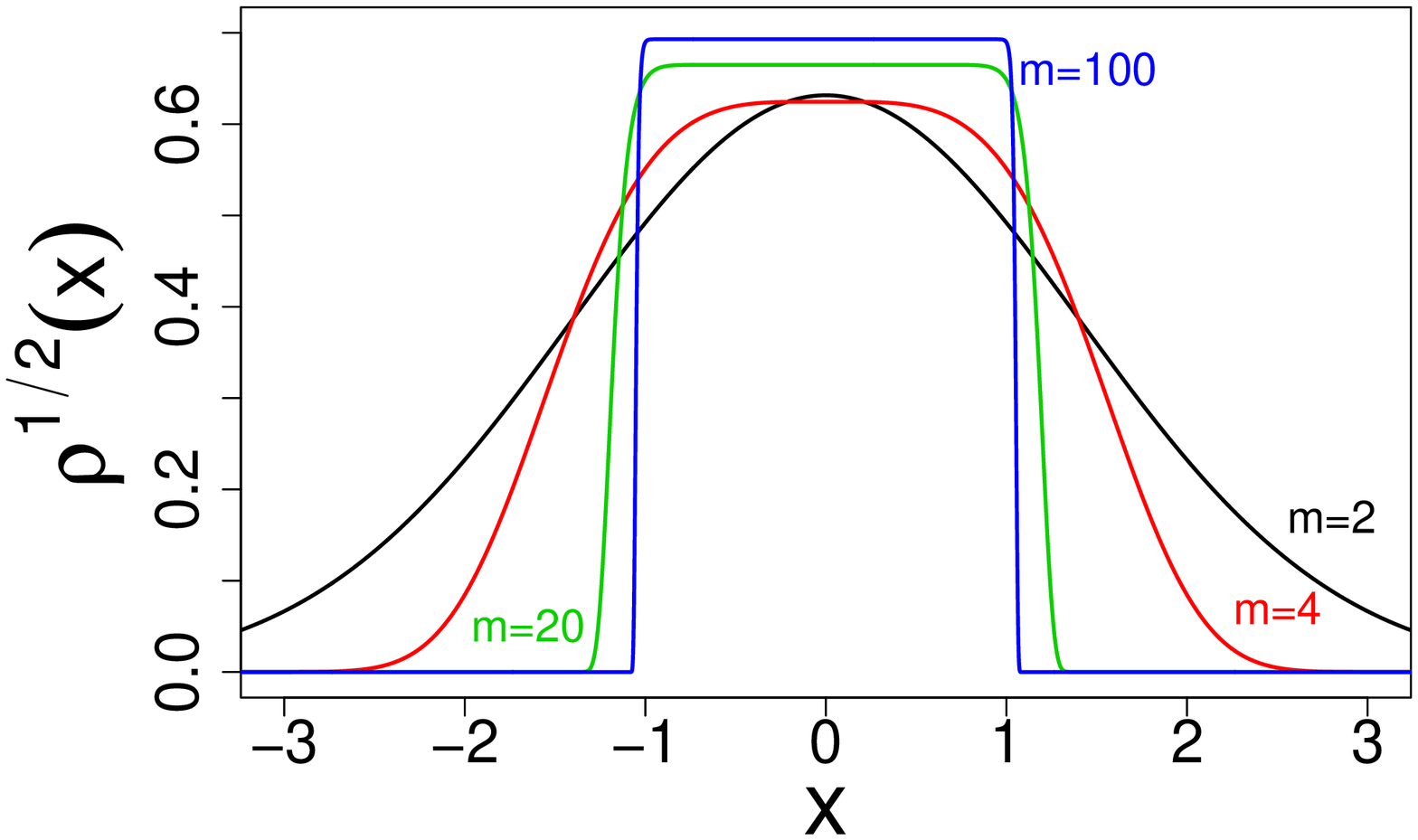}

\caption{$U(x)= x^m/m $, $\rho _*(x)$ and $\rho _*^{1/2}(x)$ for $m=2,4 ,20, 100$. In the left panel,
 the  dashed line depicts the infinite well  potential  contour.  Each of the displayed  $\rho _*^{1/2}(x)$ is an
   eigenfunction  of $\hat{H}= - \Delta  + {\cal{V}}(x)$  corresponding to the eigenvalue zero,  Shapes of
    ${\cal{V}}(x)$ for selected values of $m$ are  depicted  below,  in Fig. 8.}
\end{center}
\end{figure}

While inspired by the targeted stochasticity concept of Ref. \cite{klafter},  we follow a procedure described in
\cite{zero} to infer the   Schr\"{o}dinger semigroup  from a given a priori invariant probability density function
 (strictly speaking, from its  square root). The semigroup   dynamics is to deduced   from the knowledge of
  the  strictly positive {\it  zero energy eigenfunction}     $\rho _*^{1/2}(x) \sim   \exp [- U(x)/2]$, $U(x)=
   x^m/m, \, m=2n \geq 2$  of the sought for  (semigroup)   motion generator, see Fig.7  and  Refs.
   \cite{vilela,zero,klafter}  for rudiments of the reverse engineering concept and this
     of  the  targeted stochasticity.\\

\begin{figure}[h]
\begin{center}
\centering
\includegraphics[width=55mm,height=55mm]{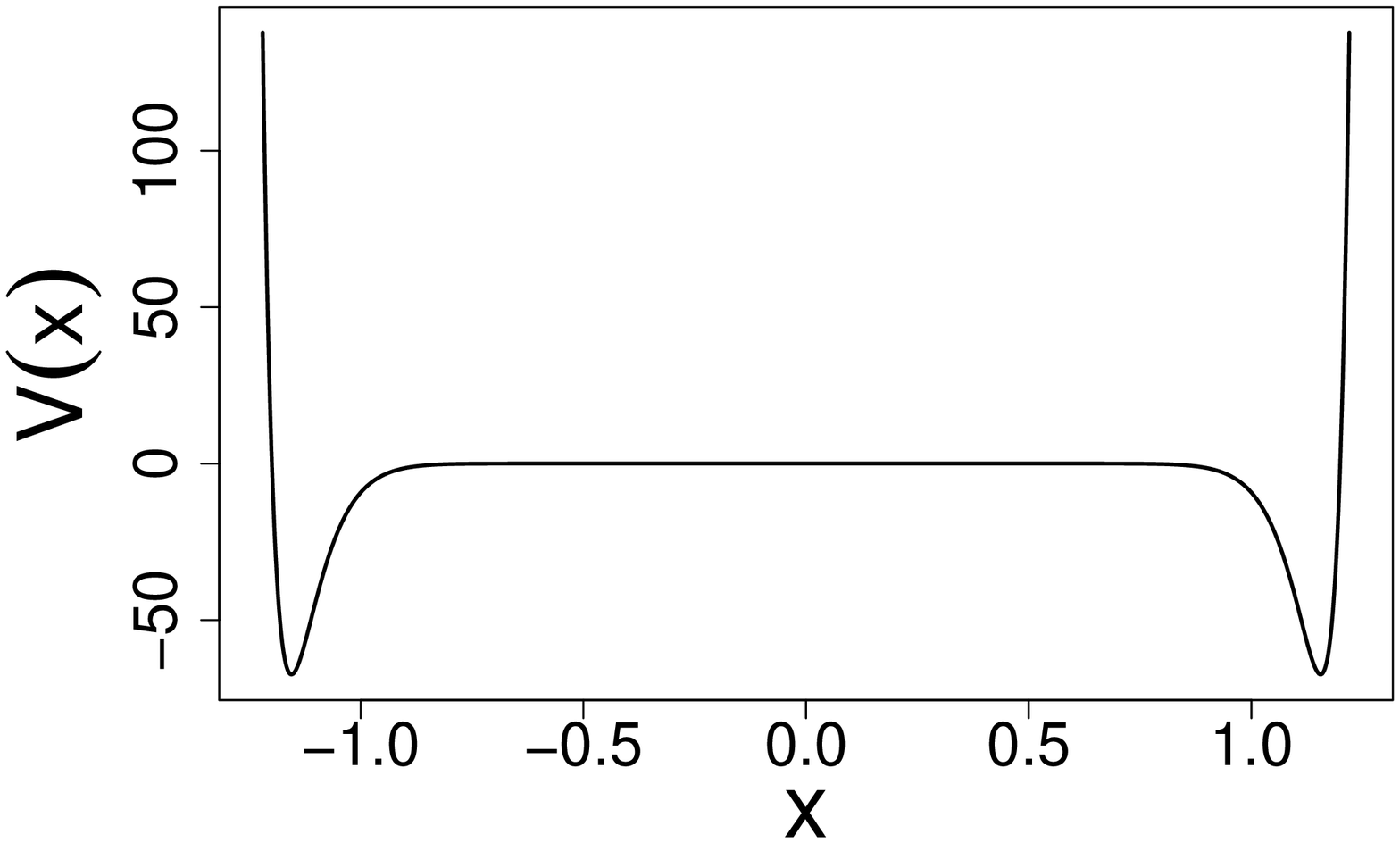}
\includegraphics[width=55mm,height=55mm]{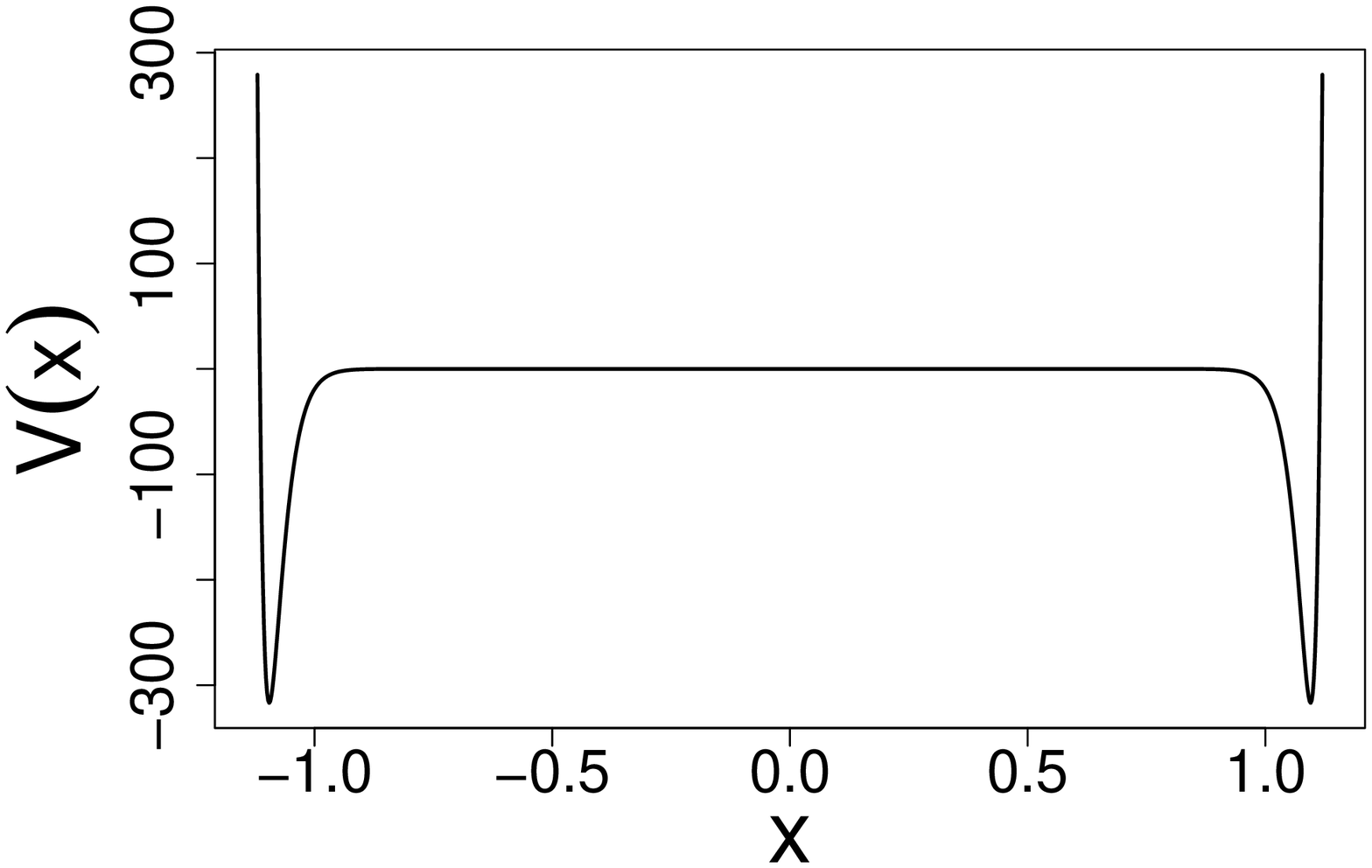}
\includegraphics[width=55mm,height=55mm]{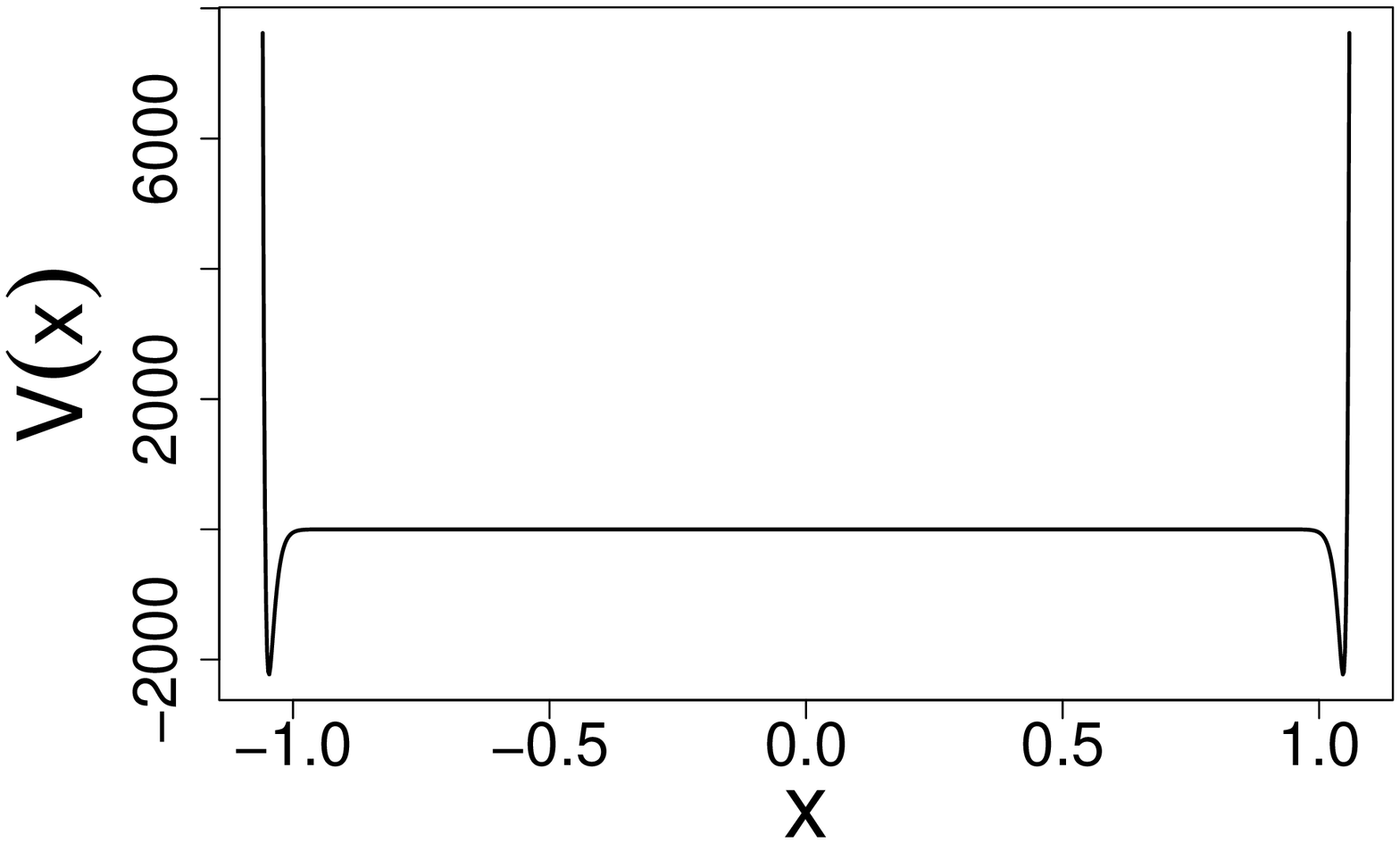}
\caption{$U(x)= x^m/m$: the reconstructed Feynman-Kac potential   ${\cal{V}}(x)= {\cal{V}}_m(x)=
 {\frac{x^{m-2}}{2}}  \left[  {\frac{x^{m}}{2}}+(1-m)\right]$  for $m=20, 40, 100$.
  Note   significant scale differences along the vertical axis. Minima  of  the semigroup potential
  are located  at points $x= \pm [(m-2)/2]^{1/m}$  i.e.   $|x| \sim   m^{1/m} >1$.}
\end{center}
\end{figure}

\begin{figure}[h]
\begin{center}
\centering
\includegraphics[width=55mm,height=55mm]{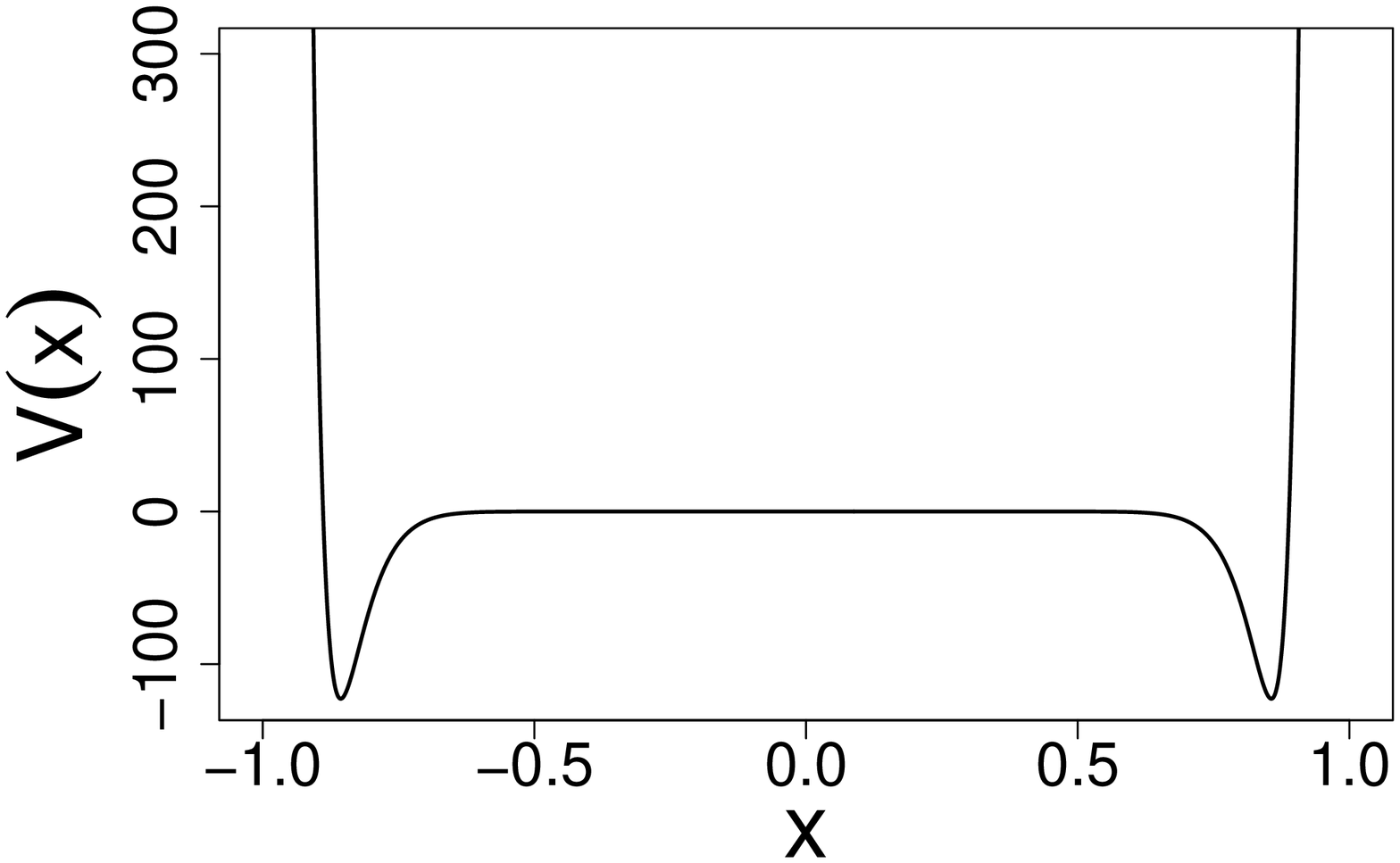}
\includegraphics[width=55mm,height=55mm]{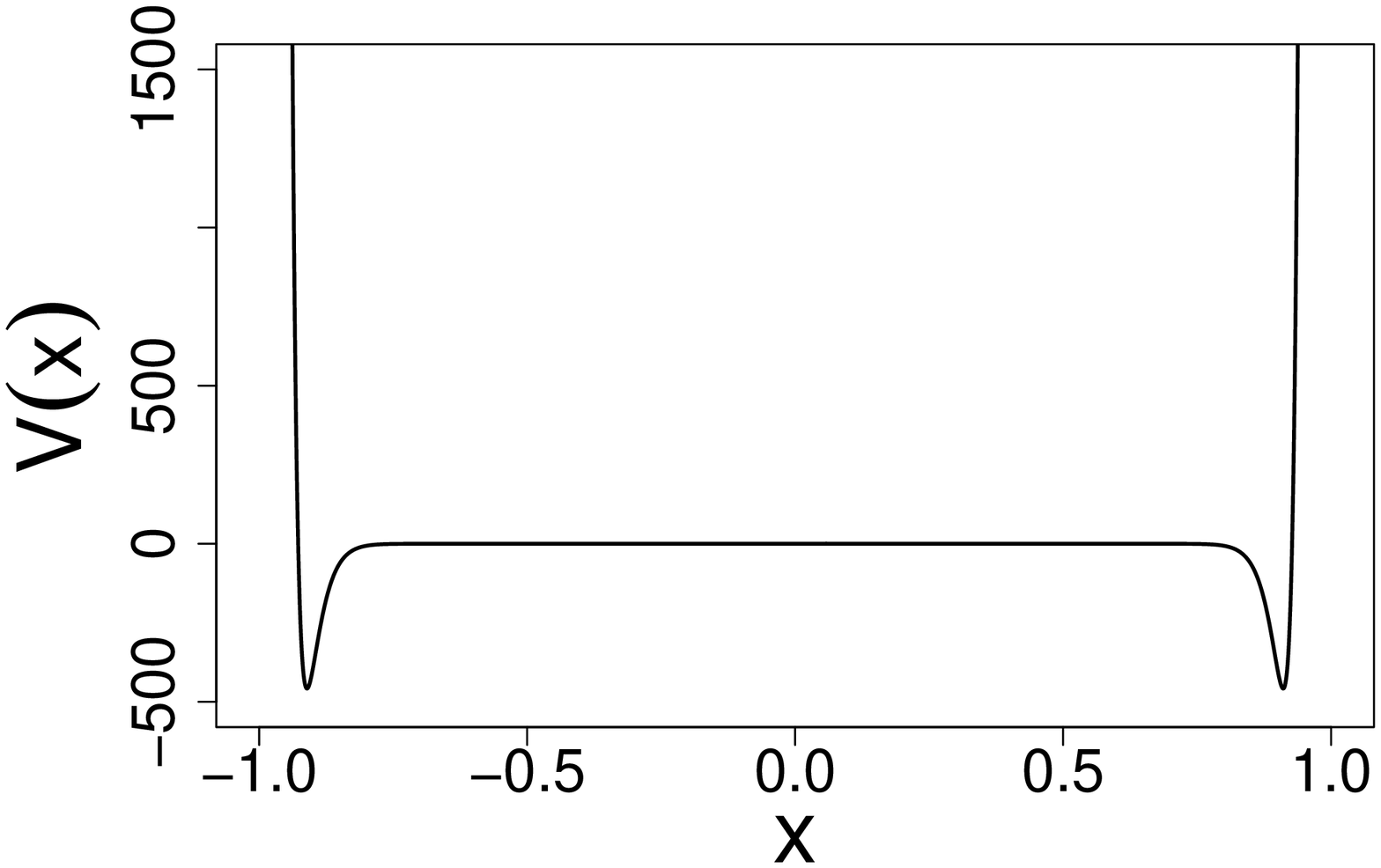}
\includegraphics[width=55mm,height=55mm]{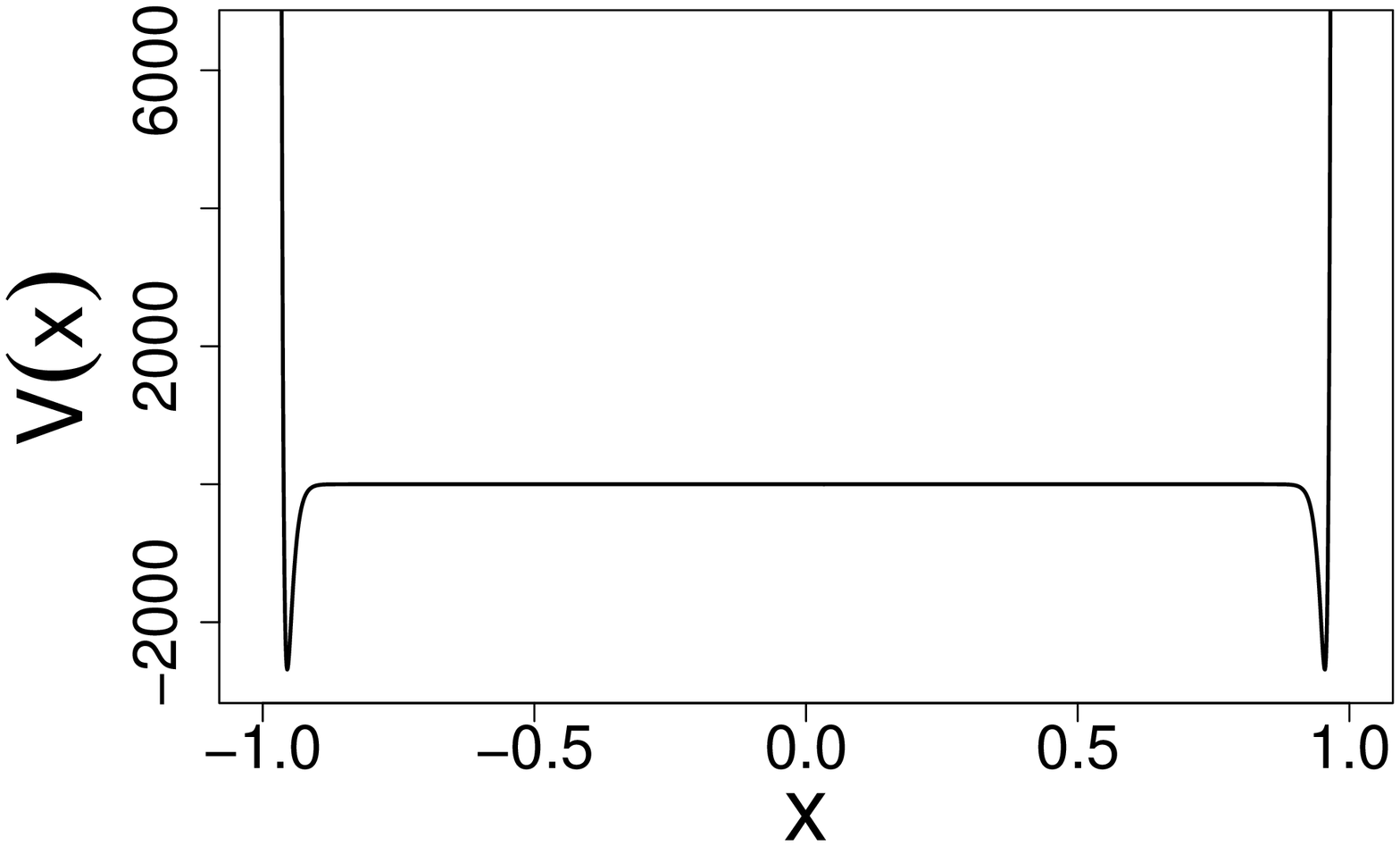}
\caption{A comparative figure for  $U(x)= m x^m$.  The reconstructed Feynman-Kac potential ${\cal{V}}(x)
= {\frac{m^2}2} x^{m-2} [{\frac{m^2}2} x^m +(1-m)]$  is
depicted  for $m=20, 40, 100$.  Note   significant scale differences along the vertical axis.
 Minima  of  the semigroup potential are located  at points $ x=\pm [(m-2)/m^2]^{1/m}  <1 $. They belong to the interior of
 $[-1,1]$,  which  needs to be contrasted with the behavior reported in Fig. 7. While approaching  $\pm 1$,
  the minima escape to minus infinity. In parallel  we have    ${\cal{V}}(\pm 1) \rightarrow + \infty $.}
\end{center}
\end{figure}

   By setting  $D=1$  in Eqs. (7)-(10),     we   interpret  the introduced $m$-family of
     $\rho _*^{1/2}(x) \sim \exp (-x^m/2m)$  as   ground-state  solutions of the generalised diffusion equation
 \be
 \partial_t\Psi= \Delta \Psi - {\cal{V}} \Psi = - \hat{H} \Psi
 \ee
  with
   $\hat{H} =  - \Delta  +{\cal{V}}$, where the  reconstructed (given $\rho _*^{1/2}(x)$)  Feynman-Kac potential
   reads, c.f. Figs. 8 and 9:
\be
{\cal{V}}(x)=   {\frac{\Delta \rho _*^{1/2}(x)}{ \rho _*^{1/2}(x)}} = {\frac{x^{m-2}}{2}}  \left[  {\frac{x^{m}}{2}}+(1-m)\right].
\ee

{\bf Remark:} We recall that the effects of noise  (with an arbitrary  intensity $D>0$)  can be  reintroduced through
the formula  $ {\cal{V}}(x)\equiv  {\cal{V}}_m(x)= ({x^{m-2}}/2)     \left[ (x^{m}/ 4D) +(1-m)\right]$.   A specific choice of the
 the noise intensity  value  becomes  immaterial in the large $m$ limit.\\

 In the literature,  slightly less restrictive reconstruction problems for the Schr\"{o}dinger-type dynamics have been
  considered. Namely, \cite{vilela,turbiner}, one may depart as well from the eigenvalue problem  in $L^2(R)$
\be
[\hat{H} - E ]\psi _0(x) = - \Delta  \psi _0(x) +  [{\cal{V}}(x)  - E ]\psi _0(x) \Rightarrow   {\cal{V}}(x)-E= {\frac{\Delta \psi _0(x)}{\psi _0(x)}}
\ee
with  the given  a priori $L^2(R)$  eigenfunction  $\psi _0(x)$ (typically free of nodes, albeit it is not a must,
see Remark in  below), and ask for the well
 behaved potential function which makes the eigenvalue problem  for $\hat{H}$  properly  defined and solvable.
  This is in fact the main idea behind  the "reconstruction of the dynamics form the
 eigenstate", \cite{vilela,streit}.

It is clearly   an inversion of the standard logic, where one first selects the appropriate   potential and then seeks  solutions (and eigensolutions) of the Schr\"{o}dinger
or Schr\'{o}dinger-type (generalized diffusion one, our case) equation.\\

  {\bf Remark:}  It is customary to reconstruct the  dynamics from the  strictly positive function, which  is interpreted as the  ground state of the sought for motion
  generator. However, it is not a must. One may as well admit  other eigenfunctions, that lead to non-negative probability densities, instead of the strictly positive one.
   It is possible to handle the problem of zeroes. A detailed analysis of the harmonic oscillator case in this regard (evaluation of drifts and related semigroup potentials
   of the form (8) from excited eigenstates)  is provided in the  text between  formulas (46)  and (50) of Ref. \cite{nonegative}.  See also Ref. \cite{non}. \\

The functional form   (43) of ${\cal{V}}(x)$ looks quite intriguing, if  compared with  the  familiar   anharmonic
 expression $|x|^m$,  notoriously   employed in the quantum theory  literature as the  potential function for the $m$-family of
  Schr\"{o}dinger operators,   whose   spectral  properties  in the   $m  \rightarrow \infty $ limit have  been a subject of elaborate
   studies,  \cite{voros}.

    In the present paper, we  are  mostly  interested in the large $m$ behavior of the potential  (43),
    and specifically  whether   the  reconstructed semigroup spectral problem  might be perceived  as  "close" to  that related to
  the   Neumann Laplacian $\Delta _{\cal{N}}$, compare e.g. Section IV.B.

\subsection{On  the spectral affinity  ($m \gg 2$ regime) with the reflecting Brownian motion on the interval.}

Let us consider the eigenvalue problem
\begin{equation}
\hat{H} \psi=-\Delta \psi + \mathcal{V}\psi = \lambda\psi,
\end{equation}
where $\mathcal{V}(x)$ has the  two-well functional form (30).
We are interested in testing whether with the growth of $m$ one may  relate  the
corresponding semigroup  generator  with the Neumann Laplacian $\Delta _{\cal{N}}$   in $L^2([-1,1])$.

Let us choose the Neumann basis on $[-1,1]$, c.f. \cite{bounded1,gar}  and Section IV.B:
  \begin{equation}
\psi_n(x)=\cos\left(\frac{n\pi}{2}(x+1)\right),\qquad n=0,1,\ldots.
\end{equation}
Presuming that it is the reference basis system, we make a hypothesis that any eigenfunction $\psi (x)$  in Eq. (32), for large $m$, should be "close" to the corresponding Neumann eigenfunction. Likewise, we expect the same   "closeness" property for the corresponding eigenvalue (in the large $m$ regime).

If (32) is a valid eigenvalue problem, the eigenvalues should follow from
the  $L^2(R)$  expectation value
\begin{equation}
\lambda_{n,m}=<\psi_n|\hat{H}|\psi_n>,
\end{equation}
that needs to be approximated by the  expectation  value   of $\hat{H}$  in $L^2([-1,1])$  with respect to the Neumann basis.

According to  Fig. 8, for  $|x| >0.99$   the singular behavior of ${\cal{V}}(x)$, in the large $m$ regime, is incompatible with the
 Neumann basis properties at the interval boundaries.  If the boundary parameter $a$  goes to  $1$, we need to take into account the
  term  $(m-1)x^{m-2}$, which does not vanish at $\pm 1$  and gives a large input at the boundary, in the numerically assisted integration procedure.

\begin{figure}[h]
\begin{center}
\centering
\includegraphics[width=65mm,height=65mm]{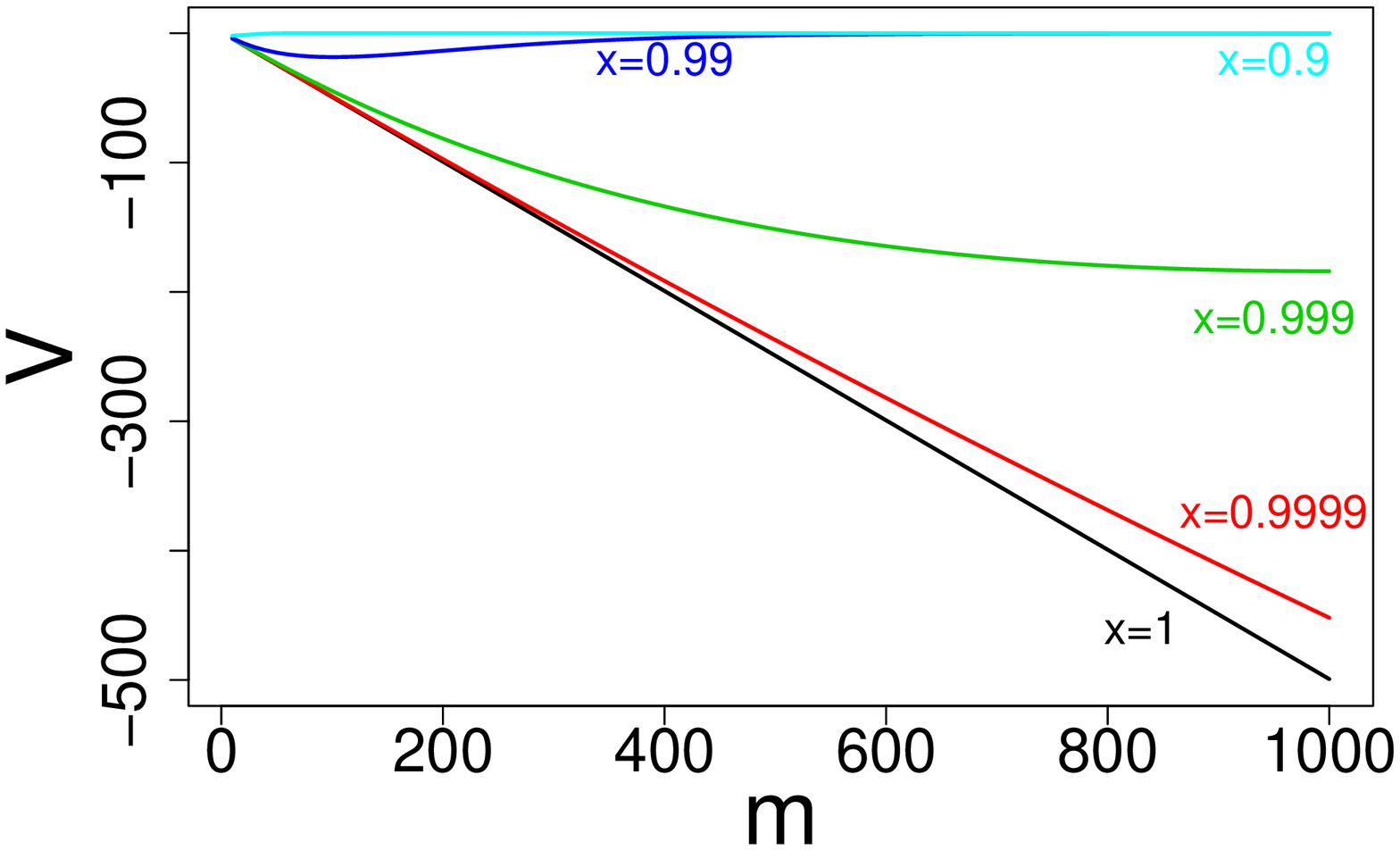}
\includegraphics[width=65mm,height=65mm]{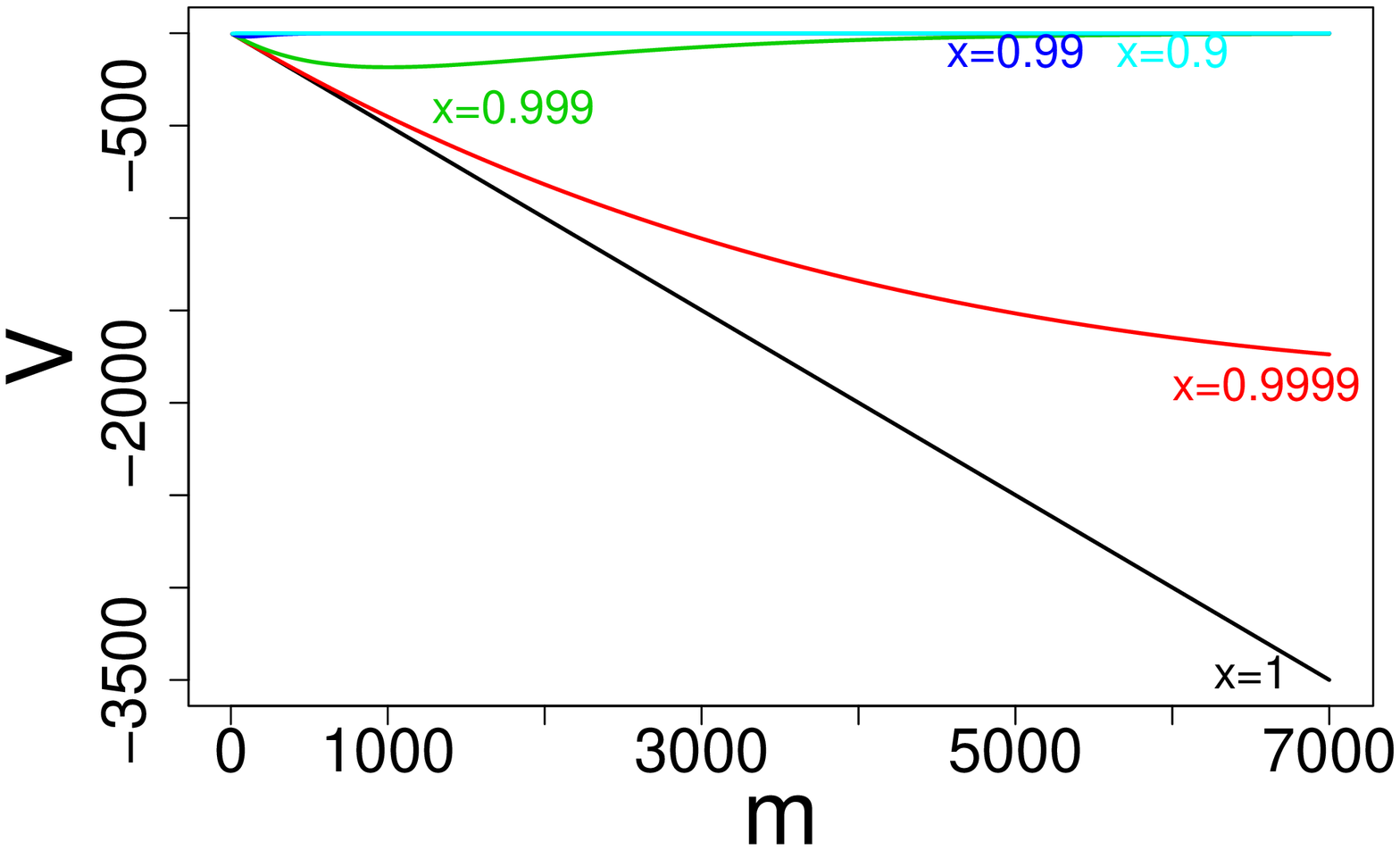}
\caption{The $m$-dependence of ${\cal{V}}(x)= {\cal{V}}_m(x)= {\frac{x^{m-2}}{2}}
 \left[  {\frac{x^{m}}{2}}+(1-m)\right]$ in the close  vicinity of $x=1$,  $x=0.9, 0.99, 0.999, 0.9999$ and  at  $x=1$.
  Two different scales are employed.  The convergence properties for $x=0.9$ and  $x=0.99$ suggest that  the
  link with the reflecting Brownian motion may be considered reliable  within the interval, say,   $[-0.99,+0.99]$.}
\end{center}
\end{figure}

 Therefore, instead of extending the involved integration to the whole of $[-1,1]$, we restrict the interval boundaries to
 $[-a,a], a<1$.
Accordingly, we  disregard  the misbehaving part of the integral  (this  misbehavior is a consequence of
 replacig/apprixmating the "true " eigenfunction by the Neumann one). Moreover, we   isolate a leading term $(n\pi /2)^2, n\geq 1$ which is characteristic
for the Neumann well spectral solution. The remaining integral  expression  is expected to  decay to  $0$ as $m\rightarrow \infty $:
\begin{equation}
\lambda_{n,m}   \sim \left(\frac{n\pi}{2}\right)^2 + \int\limits_{-a}^a\mathcal{V}(x)\cos^2\left(\frac{n\pi}{2}(x+1)\right)dx.
\end{equation}
In Table II we collect  the pertinent  integral values for a couple of $m$ values, for specific choices of the integration parameter $a=0.9, 0.99$ and $a=0.999$,  with reference to the first excited level, i.e. $n=1$.
\begin{table}[h]
\begin{center}
\begin{tabular}{|c||c||c||c|}
\hline
$m$ & $a=0.9$ & $a=0.99$ & $a=0.999$\\
\hline
\hline
$10$ & $-0.345908$ & $-0.849936$ & $-0.923367$ \\
\hline
$20$ & $-0.127473$ & $-0.806033$ & $-0.957354$ \\
\hline
$30$ & $-0.045061$ & $-0.737262$ & $-0.958168$ \\
\hline
$50$ & $-0.005528$ & $-0.607352$ & $-0.945658$ \\
\hline
$100$ & $-2.87\cdot 10^{-5}$ & $-0.368941$ & $-0.903151$ \\
\hline
$200$ & $-7.63\cdot 10^{-10}$ & $-0.135228$ & $-0.818505$ \\
\hline
$300$ & $-2.03\cdot 10^{-14}$ & $-0.049511$ & $-0.740936$ \\
\hline
$500$ & $-1.43\cdot 10^{-23}$ & $-0.006634$ & $-0.606782$ \\
\hline
$1000$ & $-1.89\cdot 10^{-46}$ & $-4.36\cdot 10^{-5}$ & $-0.368025$ \\
\hline
\end{tabular}
\caption{Computed values of the integral term in Eq. (47), for $n=1$, and
selected  choices of $a$ and $m$. The  conspicuous  decay of this  term
with $m \rightarrow \infty $ is confirmed for $a\leq 0.99$.}
\end{center}
\end{table}

\begin{figure}[h]
\begin{center}
\includegraphics[width=10cm,height=7cm]{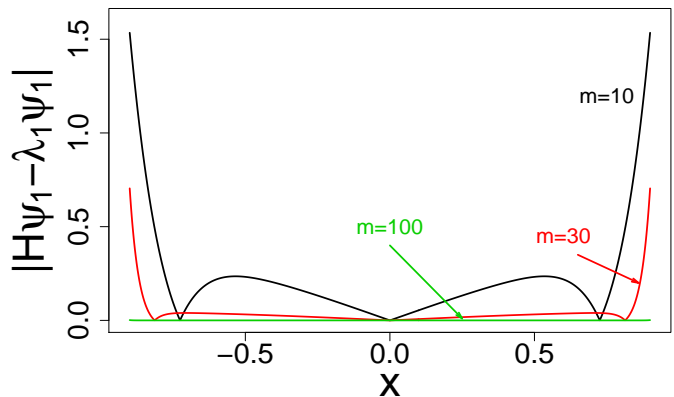}
\caption{The detuning plot  $|H\psi_1-\lambda_{1,m}\psi_1|(x) $   for $n=1$, with $\psi _1(x)$  interpreted as "close" to  $cos(\pi (x+1)/2)$ in the interval $[-0.9,0.9]$. Curves  $m=10,30,100$  are depicted.}
\end{center}
\end{figure}

{\bf Remark:}   The approximation accuracy of the "true" eigenfunction $\psi _1(x)$ by the Neumann eigenfunction
 $cos(\pi (x+1)/2)$ is excellent  up to $a=0.99$ and improves with $m\rightarrow \infty $. The
 approximation  finesse   surely breaks down   for $a>0.99$  if,  while  approaching the boundary value $1$,
   we do not exceed $m=1000$.  See. e.g.   Table II at the column corresponding to $0.999$.
  In fact for any $a<1$,  for each value of $m$  we can evaluate the integral.
  As $m\rightarrow \infty $ the integral will (possibly discouragingly   slowly) converge to $0$.
   For example, if we take $a=0.999$, at $m= 1000$ the integral value is $-0.368025$.  Would we  have evaluated
    the integral for  $m = 2000, 3000, 5000, 10 000$, the respective values  read:
     $-0.135332, -0.0238809, -0.00336391, -0.0000226092$. The convergence to zero is  conspicuous. \\

We can give another argument  supporting  the statement that   signatures of the aforementioned  spectral affinity in the
interval $[-a,a], a\leq 0.99$  have been met, by   depicting (Fig. 13)  the detuning diagram, i.e. the plot of
$|H\psi_1-\lambda_{1,m}\psi_1|$ for  $a=0.9$ and $m=10,30,100$, as a function of  $x$,
 provided $\psi _1(x)\equiv cos(\pi (x+1)/2)$  in $[-a,a]$.
  The detuning can be made arbitrarily small with the growth of $m$, as long as  we stay within $[-a,a]$, $a<1$.

\section{False bistability   of   two-well  Feynman-Kac potentials.}

Rather specific functional form of the deduced semigroup (Feynman-Kac)    potential (30) and its increasingly
 singular behavior at the boundaries of  $[-1,1]$, with the growth of $m$,
makes problematic a direct computer-assisted evaluation of higher  level eigenvalues and eigenfunctions.
 The level of  involved technical  difficulties has  been noted before, \cite{turbiner},  in connection with  an
  example of  the  sextic potential   ${\cal{V}}(x)= ax^6 - bx^2$.
  We note that for $m=4$,    e.g. $U(x) = x^4/4$,  (43)  reduces to the sextic potential with
  $a=1/4$ and $b=  3/2$.
For  $m=6$,  e.g. $U(x) = x^6/6$, (43) takes the form of   the decatic potential   ${\cal{V}}(x)= ax^{10} - bx^4$
with $a=1/4$ and $b=5/2$.

  We mention an extended  discussion of decatic potentials presented in  Refs. \cite{brandon,maiz,turbiner}, which
    can be regarded as a  reference record of various curiosities  concerning the solvability of  spectral problems
     closely related to our $\hat{H}= -\Delta +   {\cal{V}}(x)$. These  curiosities are shared by
   a  broad   class of {\it quasi-exactly solvable} Schr\"{o}dinger equations,   whose solutions are not amenable to standard
      algebraic methods and basically  unknown, except for very few examples. A direct recourse to
       computer assisted arguments is unavoidable.

A related problem is the   {\it false bistability}   of   bistable-looking  two-well  Feynman-Kac potentials.
 Namely, as follows from the previous  discussion, operators $\hat{H}= -\Delta + {\cal{V}}(x)$,   with
 ${\cal{V}}$ defined by Eq. (30),  have  a nonnegative energy spectrum beginning from the eigenvalue zero.
   As noticed in Ref. \cite{turbiner},  in this case  traditional WKB
  approximation methods are    inapplicable to retrieve  spectral data, because the classical zero energy
  level in the considered case refers to an unstable   equilibrium point of the two-well potential.

  Moreover, an insight into computer-generated solutions shows that the system  appears to  ignore  ("does not feel")
    the presence of two potential wells, contrary to expectations rooted in the  folk understanding of diffusion problems
    in double-well   potentials, \cite{risken,pavl}.  (Note  that we make a distinction between two-well and double-well
potentials; the latter have always a quartic form).

    Let us mention the classic   double-well potential: $U(x)= ax^4/4  - bx^2/2 $, with $a>0, b>0$.
       The potential has two  negative-valued   minima (local well depth)     at $x_{1,2} = \pm \sqrt{b/a}$,
     and  a  local maximum equal $0$ at $x_0=0$.  (Note that a modified potential  $U(x)+ b^2/4a  =    a(x^2/2 - b/2a)^2$
           is   nonnegative and has a local maximum  $b^2/4a$, while local minima take  the value zero).
      In the course of the  Langevin-driven  Brownian motion with a drift $b(x) \sim - bx^3 +  ax$, even a  {\it weak} noise causes particles
  not to remain in the same (local) well,  and makes them to "oscillate" forth and back between two wells. That  should possibly result in
 the   bimodal  stationary  pdf,  with two conspicuous maxima and a minimum in between them (possibly at $x_0$).

 One might  justifiably be  tempted to invoke the standard imagery the  escape time  analysis (for a randomly
wandering particle from the well of temporary residence etc.),  \cite{vankampen}-\cite{baner}.
  However, as  follows from our further discussion  (Section VI.B) this temptation is somewhat deceiving and relies on the
   relative balance between parameters $a$ and $b$.

\subsection{When do we encounter   the  bottom eigenvalue zero  for  two-well systems,
  or why  there is no signatures of the potential  bistability ?}

Double-well potentials in the classic quartic  version have received an ample coverage in the literature,
specifically in connection with effects of tunneling through the barrier, which gives rise to a  splitting of
otherwise degenerate negative energy (might be positive, but then necessarily
 with energies   below the top  of the  central  double-well   barrier  i.e. below the  local maximum of the potential).
  A standard procedure, employed   to analyze the low-lying  spectrum, amounts to   a  local aproximation of each well by a suitable harmonic oscillator
   potential, \cite{landau}.

Let us consider  the general  eigenvalue problem for a Hermitian (eventually self-adjoint) Schr\"{o}dinger type operator
\be
-\Delta\psi + \mathcal{V}\psi = \lambda \psi.
\ee
We presume the potential to be bounded from below and symmetric (this assumption does not harm the generality of further arguments).
Let $\mathcal{V}(x)$  has  a  minimum  (in the least a local minimum)   at  $x_0\in R$ and  consider a Taylor expansion of $\mathcal{V}(x)$, which
 in a sufficiently close vicinity of $x_0$ may be reduced to
\be
\mathcal{V}(x)=\mathcal{V}(x_0)+\frac{1}{2}\mathcal{V''}(x_0)(x-x_0)^2 .
\ee

Let us furthermore  assume that $\mathcal{V}(x_0)<0$ in the vicinity of $x_0$ and at $x_0$.

Inserting an  approximate expression  (32)   for $\mathcal{V}(x)$   to the eigenvalue equation  (33), we arrive at an approximate eigenvalue problem:
\be
-\Delta\psi(x) + \frac{1}{2}\mathcal{V''}(x_0)(x-x_0)^2\psi(x) = (\lambda-\mathcal{V}(x_0)) \psi(x)
\ee
Setting $z=x-x_0$, $dz=dx$  we get
\be
-\Delta\psi(z) + \frac{1}{2}\mathcal{V''}(x_0)z^2\psi(z) = (\lambda-\mathcal{V}(x_0)) \psi(z).
\ee
This expression can be readily compared with the standard harmonic oscillator spectral problem. Indeed, since we have:
\be
-\frac{\hbar^2}{2m}\frac{d^2}{dz^2}\psi(z)+\frac{1}{2}m\omega^2 z^2\psi(z) = E\psi(z),
\ee
Eqs.  (52) and (53) become  compatible upon formal rearrangements  of constants and units.
(Note that  we  need $\mathcal{V''}(x_0)> 0$.)  Let us make identifications
\be
\frac{\hbar^2}{2m}\equiv 1,\qquad \frac{m\omega^2}{2}=\frac{\mathcal{V''}(x_0)}{2},\qquad E=\lambda-\mathcal{V}(x_0).
\ee
While keeping in  mind that $\frac{\hbar^2}{2m}\equiv 1$, we get
\be
\frac{\hbar^2}{2m}\cdot\frac{m\omega^2}{2}=\frac{\hbar^2\omega^2}{4}=
\frac{\mathcal{V''}(x_0)}{2}.
\ee
 For the  standard   quantum  harmonic oscillator   the ground state eigenvalue $n=0$ equals
   $E_0=\hbar\omega/2$,    we  can infer a corresponding $\lambda _0$ from the (approximate) identity $E_0=\lambda _0 -\mathcal{V}(x_0)$, so arriving at
\be
\lambda_0 = \frac{\hbar\omega}{2} + \mathcal{V}(x_0).
\ee

An   assumption that in the vicinity of the local minimum ${\cal{V}}(x_0)$,  the  potential function ${\cal{V}}(x)$ is negative,
 implies   a rough {\it  existence  criterion}    for  negative eigenvalues in the two-well  potential case.
  Indeed,  a demand   $\lambda_0<0$  enforces
  \be
\frac{\hbar\omega}{2} < -\mathcal{V}(x_0).
\ee
Both sides of the inequality are positive, hence  after taking the square and employing  (55)  we can
 rewrite the existence criterion  for the negative $\lambda _0$,  as   the restriction  upon the curvature
 of the two well potential $V(x)$  at $x_0$:
  \be
\mathcal{V''}(x_0) < 2[\mathcal{V}(x_0)]^2.
\ee
If  this  inequality   holds true,  the existence of negative eigenvalues  is allowed  in the spectral
solution for $\hat{H}$.

The above argument is   not limited to the double-well (quartic) potentials and remains valid for more general
two-well potentials.
We  work with a special subclass in  the two-parameter family of two-well potentials:
\be
\mathcal{V}(x)=a x^{2m-2}-bx^{m-2},\qquad a,b>0,\quad m>2,
\ee
 for which two local (well) minima are  symmetrically located   at $\pm x_0$, where
\be
x_0=\left(\frac{(m-2)b}{2(m-1)a}\right)^{1/m}.
\ee
Note that assuming $b> 3a$, for all  $m$   we have $x_0 > 1$, while for  $b<2a$ the minima are located in the interior of $[-1,1]$.

Our  specific  case of  the   two-well potential  Eq. (30)  is  identified by setting  $a=1/4$, $b=(m-1)/2$
and assuming $m=2n,  n>1$  (we recall that the case $n=1$ refers to the  standard  harmonic oscillator potential with ground state energy subtracted, \cite{faris,gar}).  This implies   $x_0=(m-2)^{1/m}$.

Further calculation is performed for  our  two-well potential   (30).
 We have:
\be
\mathcal{V''}(x_0)=\frac{m(m-1)}{2}(m-2)^{(2m-4)/m},
\ee
and
\be
[\mathcal{V}(x_0)]^2=\frac{m^2}{16}(m-2)^{(2m-4)/m}.
\ee
This implies a sharp inequality
\be
\frac{m(m-1)}{2}<2\cdot \frac{m^2}{16} \Rightarrow m<4/3.
\ee

Since we assume  $m>2$,  the  condition  $m<4/3$  is violated from the start, which provides a formal explanation to why
negative eigenvalues do not exist  in the spectrum of    $\hat{H}= -\Delta + {\cal{V}}$  with the two-well potential
${\cal{V}}$  in the form  (30).
This conclusion stays in an obvious  conformity  with the  fact that (according to the  derivations of  Section I.B)
$\hat{H}$  has the eigenvalue zero corresponding to the unimodal strictly positive eigenfunction. In turn, the latter property
  validates  our   harmonic oscillator approximation at local minima of the looking-bistable  two-well potential and
   the inferred, Eq. (45) existence criterion  for  negative eigenvalues.\\

\subsection{Bistability versus  false bistability for two-well potentials, or when (why)  negative eigenvalues cease to exist.}

  \subsubsection{Quartic double-well  as the Feynman-Kac potential.}

\begin{figure}[h]
\begin{center}
\centering
\includegraphics[width=65mm,height=65mm]{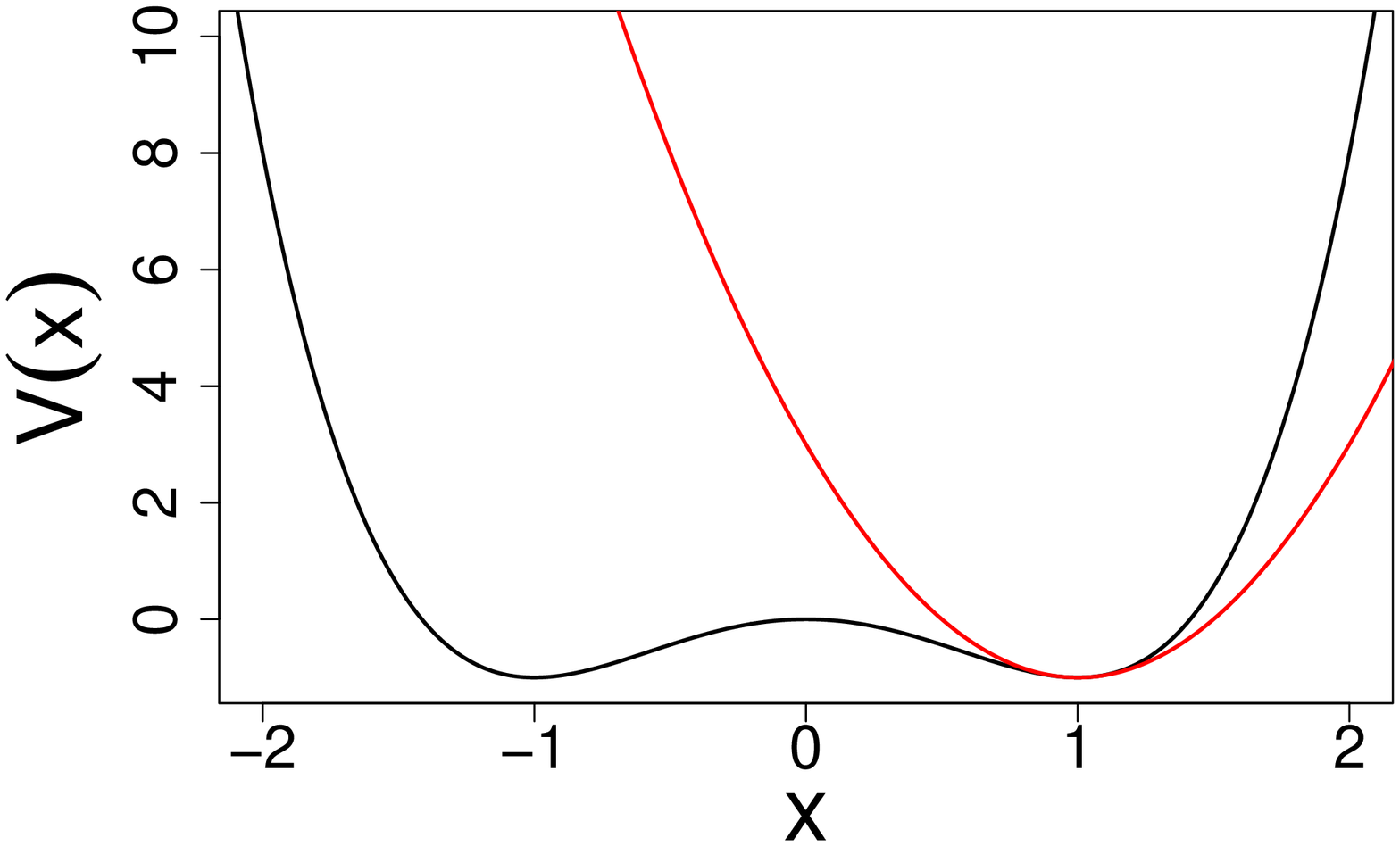}
\includegraphics[width=65mm,height=65mm]{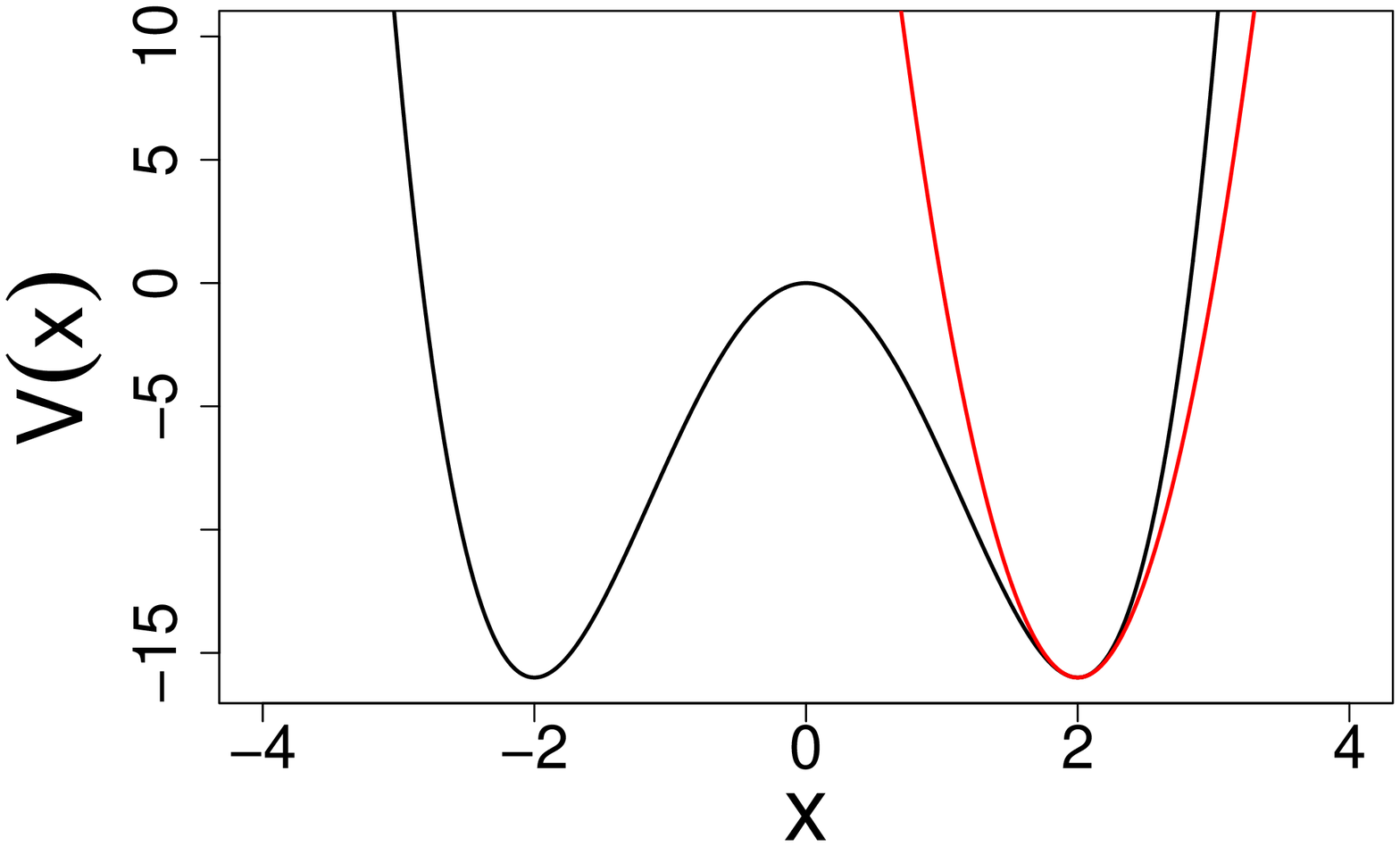}
\caption{Double-well  ${\cal{V}}(x)=  x^4-  2\alpha ^2 x^2$  (black)  for $\alpha =1$ (left panel)
and $\alpha =2$ (right panel) with an approximating parabola $V(x)= 4\alpha ^2(x-\alpha )^2 - \alpha ^4$ (red)
 at the bottom of the $x_0=\alpha$  well.  Note that $2\alpha $ is a distance between two local minima and $\alpha ^4$ stands for the depth of each local well.
 We point out that scales on vertical and horizontal axis  are  different.}
\end{center}
\end{figure}

  \begin{figure}[h]
\begin{center}
\centering
\includegraphics[width=65mm,height=65mm]{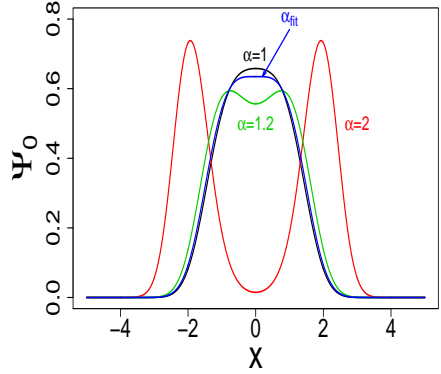}
\caption{Ground state function  $\psi _0(x)= \rho _*^{1/2}(x)$   for $\hat{H}= - \Delta + {\cal{V}}$   with the double-well
 potential ${\cal{V}}(x)=  x^4-  2\alpha ^2 x^2$ chosen a priori. Eigenfunction shapes are depicted for  $\alpha =1, 1.2, 2$
 and a fitted "transitional"   value  $\alpha _{fit} = 1.0564$  around which the eigenfunction topology changes
 from unimodal to bimodal.  The corresponding (numerically computed) ground state eigenvalues read:
   $+ 0.137786$ ($\alpha =1$), $- 0.489604$ ($\alpha =1.2$), $- 12.1363$ ($\alpha =2$).
    For $\alpha =1.1$ we have the eigenvalue $- 0.135576$.
According to an independent reasoning of Ref. \cite{turbiner1}, the  ground state eigenvalue  zero should appear
 for $\alpha _{critical} \sim 1.0534677$. (Eigenvalues were computed by means of the Strang splitting method,
  extensively  tested by us in Ref. \cite{gar2}).}
\end{center}
\end{figure}

Although   the widely studied in the literature   quartic   double-well case  does not belong to our $m=2n$  family of two-well potentials, it may serve as a useful  credibility   test of our  methods.    Arguments  of the previous subsection can be readily adopted to this  familiar  example. Let us consider
 \be
 {\cal{V}}(x) = ax^4 - bx^2
 \ee
and investigate how the  the existence   or non-existence of negative eigenvalues  for $\hat{H} = -\Delta + {\cal{V}}$    depends on the mutual balance of  steering parameters   $a$ and $b$.
 We note  that the local  maximum is located at $0$,  two negative-valued  local  minima  (depth of the wells) equal $-b^2/4a$     at  $x_{1,2}= \pm \sqrt{b/2a}$, and the distance
    between the minima (e.g. an effective width of the central barrier)  equals  $2\sqrt{b/2a}$.\\
Note that a nonnegative   definite version of the   double-well  potential reads
      $ {\cal{V}} + b^2/4a  =    a(x^2/2 - b/2a)^2$.\\

In conformity with our  previous discussion, let us address the existence issue of negative eigenvalues  for the
 operator $\hat{H}= - \Delta + {\cal{V}}$, with  ${\cal{V}}(x)$ given in the general   double-well form (51).
   We employ the  inequality   (45), whose validity means that negative eigenvalues are permitted.
   We  need to evaluate   the  local minimum  value  of ${\cal{V}}(x_0)$  and the curvature of the potential $\mathcal{V''}(x_0)$.
 These read:  $\mathcal{V}(x_0)= - b^2/4a $  and    $\mathcal{V''}(x_0) = 4b$.
Accordingly,  the   negative eigenvalue existence criterion takes  the form
\be
  4b < 2(b^2/4a)^2 \Rightarrow
  b^3>  32a^2.
  \ee

Let us examine this criterion  for  a simple but popular in the literature double well example:
\be
{\cal{V}}(x)=  (x^2-\alpha ^2)^2 -  \alpha ^4  =  x^4-  2\alpha ^2 x^2
\ee
with two minima at  $\pm \alpha,  \alpha >0$  (the effective width of the cetral barrier equals $2\alpha $).
The subtracted term  $\alpha ^4$ actually sets the depth  of each well (alternatively - height of the central barrier whose  local maximum at $x=0$  equals zero).
 A passage from  the  previous notation  involves substitutions:  $a=1, b=2\alpha ^2$.

We have ${\cal{V}}(\pm \alpha )=- \alpha ^4$ and ${\cal{V''}}(\pm \alpha ) = 8 \alpha ^2$ and to have undoubtedly
 accomodated  the  negative eigenvalues we need  $\alpha > 2^{1/3} \sim 1.25992$.
  In terms of  ${\cal{V}}(x)= ax^4 - bx^2$  this amounts to $a=1$  and $b> 3.1747$.

Accordingly, the bistability of   two-well potentials  (and the double-well in this number)  may not
be reflected in the topological properties of ground state eigenfunctions. These functions  may be  unimodal
or bimodal, depending on the value of $\alpha $,  and there is a transitional value $\alpha _{critical}$
 which results in the eigenvalue zero of the  operator $-\Delta + {\cal{V}}$.

For $\alpha <\alpha _{critical}$ the ground state eigenvalues are positive, while for   $\alpha > \alpha _{critical}$
negative  ground state eigenvalues are admitted. Thus, the  topology change of the ground state function form unimodal to
 bimodal shape is reflected in the sign change of related eigenvalues.

Zero energy eigenvalue corresponds to the   "transitional shape", where the associated $\alpha _{critical}$
  stands for a bifurcation point: the local maximum of the unimodal function degenerates and  bifurcates into twin
   local maxima  of the bimodal one. The local maxima existence  is the signature of bistability and the presence of
   negative eigenvalues  for $\hat{H}= -\Delta + {\cal{V}}$.

\begin{figure}[H]
\begin{center}
\centering
\includegraphics[width=75mm,height=75mm]{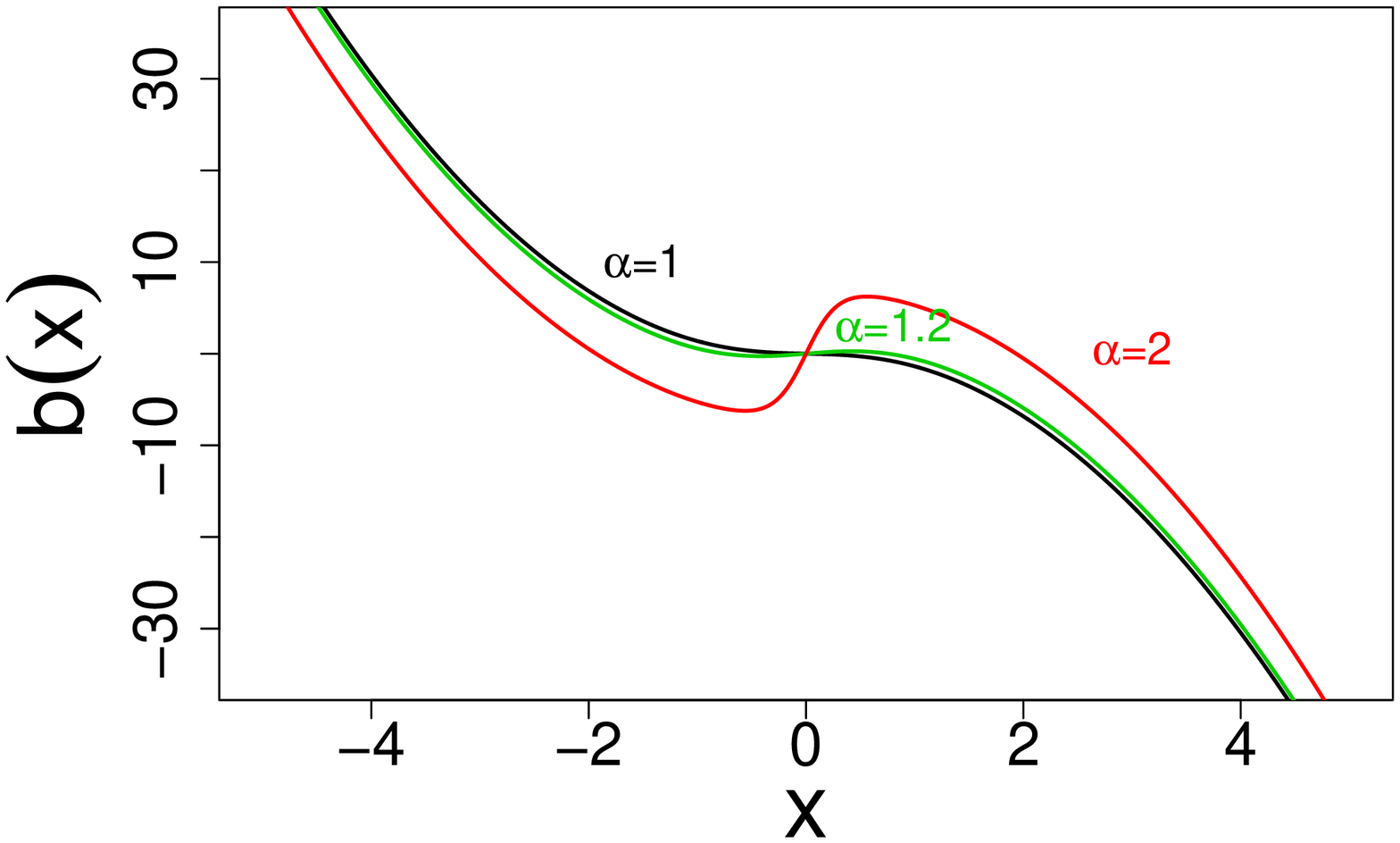}
\includegraphics[width=75mm,height=75mm]{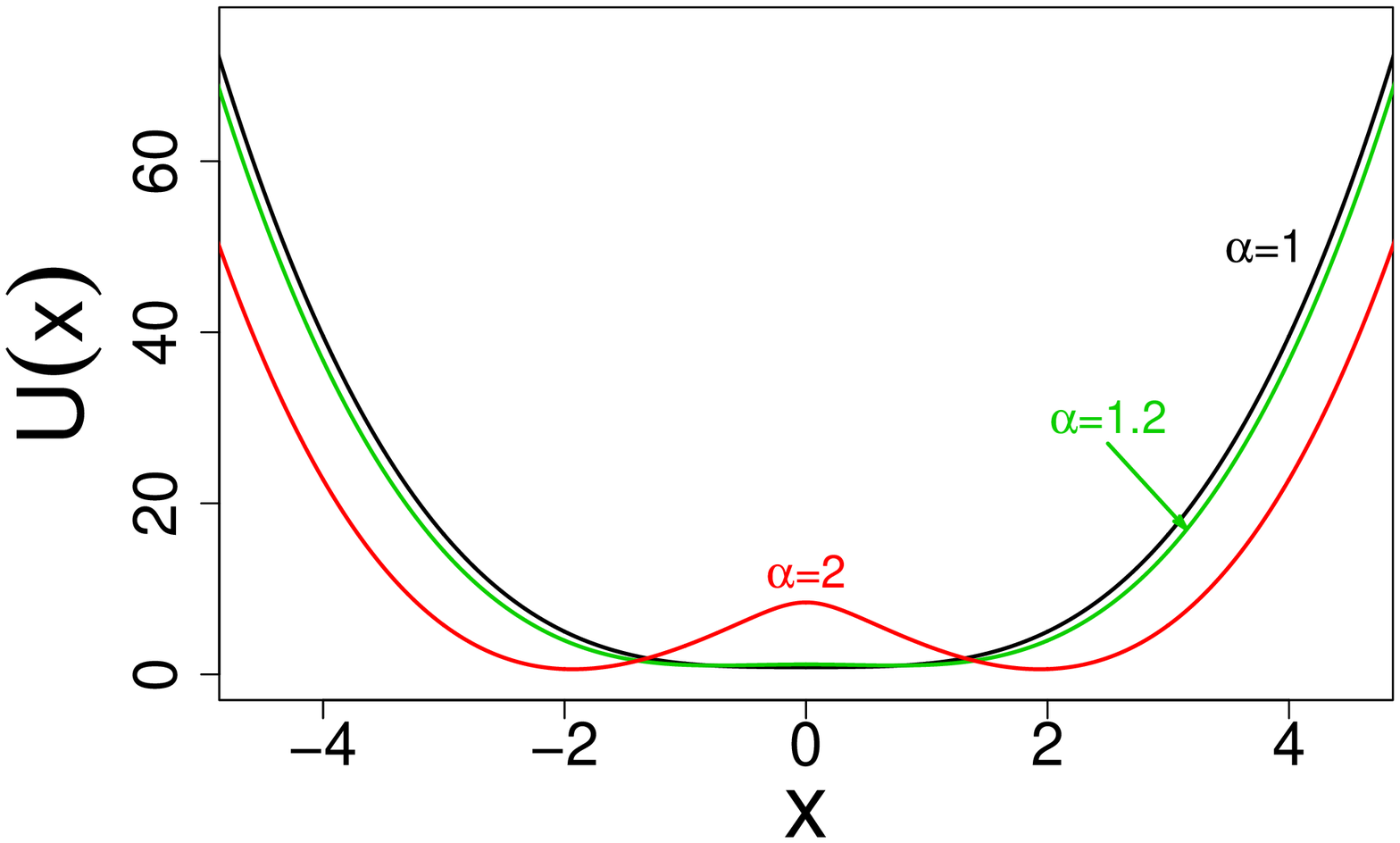}
\caption{Langevin-driven Brownian motion: invariant density and the drift for $\partial _t\rho = \Delta \rho - \nabla (b\rho )$
are reconstructed from the  Feynman-Kac potential  ${\cal{V}} = x^4 - 2\alpha ^2 x^2$ and the spectral solution for $\hat{H}$ (see Fig. 15).
The Newtonian (Boltzmann, $U= - \ln \rho _*$) potential  $U(x)$, such that $b= - \nabla U$,  is depicted as well.}
\end{center}
\end{figure}

\subsubsection{Reverse engineering for the Feynman-Kac  quartic double well: Reconstruction of the associated
 Langevin-driven Brownian motion.}

It is Fig. (15) which, as a byproduct of the discussion of the  double-well  ${\cal{V}}(x)$, Eq. (51),  shape  impact upon the  functional form (e.g. shape variability from  modal to   bimodal)  of the ground state function and the
 bottom eigenvalue of $\hat{H}$,  is our departure point.  We have in hands all  numerical data  necessary to  retrieve the
  affiliated Langevin-driven Brownian motion.
   We effectively follow the so-called reverse engineering procedure, \cite{klafter,zero,stef}.

 For each  value of $\alpha $, we employ numerical data specifying the related ground state function $\rho _*^{1/2}(x)$.
  Accordingly, we have $\rho _*(x)$ in hands.  This in turn allows to deduce (numerically as well) the corresponding
  drift field  $b(x) = - \nabla \ln \rho _*(x)$   of the affiliated Langevin-Fokker-Planck driven dynamics. Given $b(x)$,
  the  functional form of  nonegative valued (an additive constant input  is  here  irrelevant)
potentials   $U(x)$ have been reconstructed.

  Accordingly, Langevin-driven Brownian motions with drifts  $b(x)$ depicted in Fig. (16), follow a
  spectral relaxation pattern to an invariant pdf $\rho _*(x)$,    with time rate determined by the
   bottom eigenvalue of $hat{H}$ which is shared with  actual $L$ and $L^*$, see e.g. Section II.

   \subsubsection{Specifying ${\cal{V}}(x)$ if    $U(x)$ is quartic.}

   We have mentioned before that the Langevin-driven process  with   $b(x) = - x^3$ , i.e.  $U(x)= x^4/4$,   is intimately related with the
   Brownian motion "in a potential landscape" set by the Feynman-Kac potential  in the sextic
    form ${\cal{V}}(x) = x^6/4  - 3x^2/2$.
   It is interesting to know that  a thorough dicussion of  the mapping of the the Fokker-Planck evolution,  with
    $b(x)$ inferred from $U(x)= ax^4 - bx^2$,     to the  corresponding  pseudo-Schr\"{o}dinger one
     (incorporating effects of  arbitrary  noise intensity $D$,    has been accomplished in Ref. \cite{okopinska}.
  Namely, for $U(x)= ax^4 - bx^2$, $a,b>0$ and $D>0$, we get  the sextic Feynman-Kac potential:
  \be
  {\cal{V}}(x) =  g_0 + g_2x^2 +g_4x^4 + g_6x^6,
  \ee
 where:  $g_0=b$, $g_2= b^2/D - 6a$, $ g_4= - 4ab/D$, $g_6= 8b^2/D$.  We note that the mutual balance  of
  parameters $a$ and $b$ is reflected in the uni- or bimodality of the Fokker-Planck invariant pdf $\rho _*(x)$.
   Accounting for the presence of noise  (e.g. accepting any   $D>0$) implies that the associated  sextic
   Feynman-Kac potential  may  topologically vary from two-well  to  the  three-well shape (this
    to be compared with an intuitive reasoning  of  Ref. \cite{vankampen}).

\subsubsection{Sextic two-well  potential.}

In contrast to the quartic double-well,  the sextic two-well potential
\be
{\cal{V}}(x) = a x^6 - bx^2,
\ee
upon identifications  $a=1/4$ and $b=3/2$, becomes a member of our family (30).

Coming back to the notation (55), we realize that  the extrema  are  located at $0$  and  $\pm x_0$ where $x_0 = (b/3a)^{1/4}$.   Accordingly we have
${\cal{V}}(x_0)= - {\frac{2b}{3}} \left(\frac{b}{3a}\right)^{1/2}$
 and   ${\cal{V''}}(x_0)= 8b$.   Accordingly,  a  criterion for  the existence of negative eigenvalues takes
  the form  $27a < b^2$.

 In passing  we note that in the    case  which corresponds to the bottom eigenvalue zero,
  the pertinent inequality does not hold true.

\begin{figure}[H]
\begin{center}
\centering
\includegraphics[width=75mm,height=75mm]{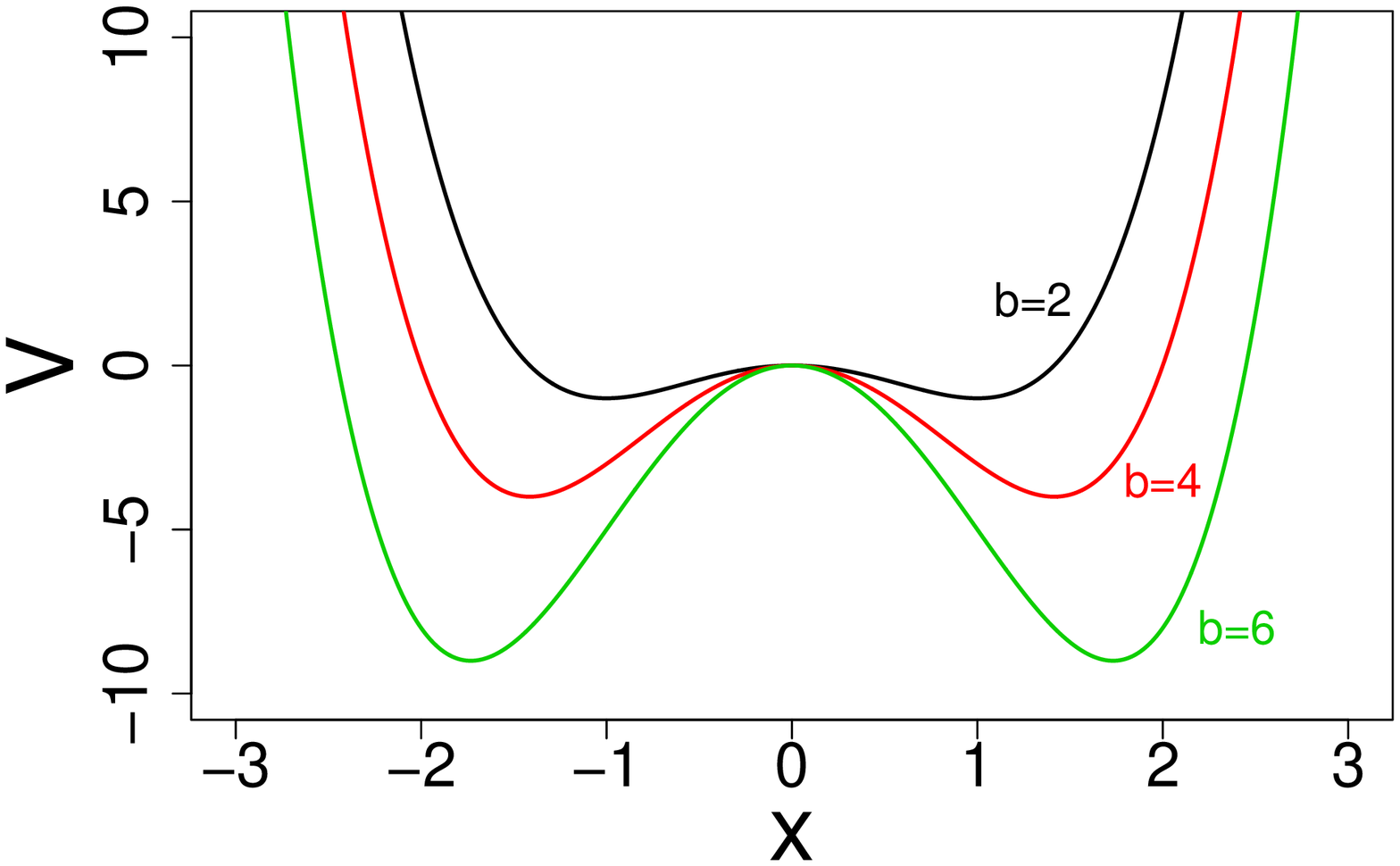}
\includegraphics[width=75mm,height=75mm]{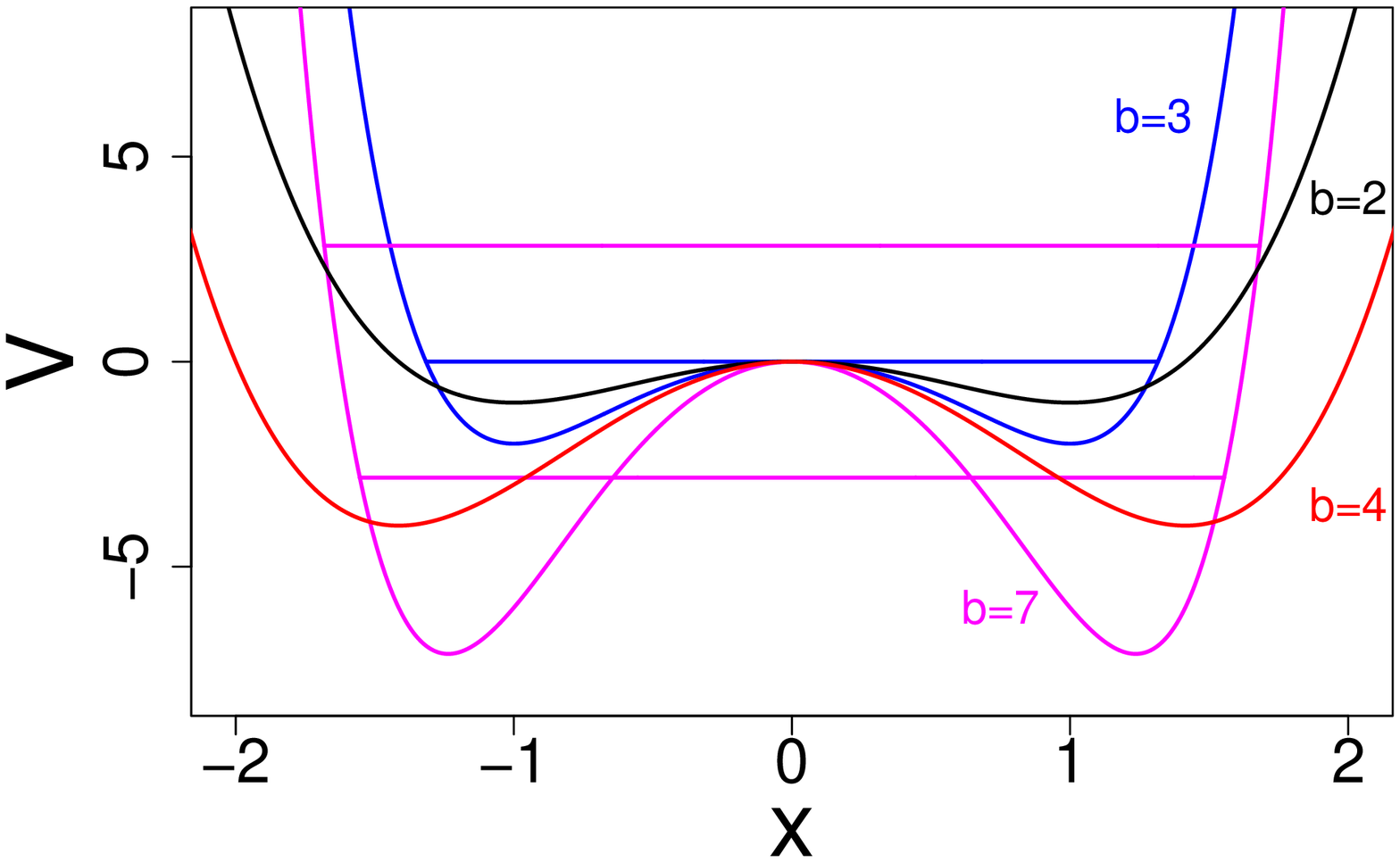}
\caption{Left panel: quartic double-well  ${\cal{V}}(x)=  x^4- b x^2$, b=2  (black) - only positive eigenvalues,
 b=4 (red) - negative eigenvalue, b=6 (green) negative eigenvalue.
 Right panel: sextic double well ${\cal{V}}(x)=  x^6- b x^2$, depicted for $b=3$ (blue)  and $b=7$ (pink), the corresponding eigenvalues  are displayed in terms  of horizontal    lines.  For comparison, on the same panel we depict quartic double well potentials: $b=2$ (black) and $b=4$ (red).}
\end{center}
\end{figure}

In Ref. \cite{turbiner}  a one-parameter  family of sextic potentials
\be
{\cal{V}}(x)= \alpha ^2 x^6 - 3\alpha  x^2
\ee
has been introduced ($\alpha >0$ is presumed). By inspection one can verify that for  each member of this  family the operator $ -\Delta + {\cal{V}}$ has the energy zero eigensolution  of the form $\psi _0(x) \sim \exp (- \alpha x^4/4)$.

We can readily verify  that (identify $a= \alpha ^2 $ and $b=3\alpha $) the sufficient condition for the existence of negative eigenvalues $27a < b^2$   does not hold true.

Plugging $\alpha =1$ we get ${\cal{V}}(x) = x^6-  3x^2$ with the related $\psi _0(x) \sim  \exp (-  x^4/4)$. Note that
the exponent $x^4/4$   is twice the exponent   $x^4/8$  (inferred from $\rho ^{1/2}_*$), which
 we have associated with the potential function (30) i.e. ${\cal{V}}(x) = x^6/4 - 3x^2/2 $.

In Ref. \cite{turbiner}, and example is given of the sextic potential ${\cal{V}}(x) = x^6 - 7x^2$, for which by inspection we can verify that $-\Delta + {\cal{V}}$ has  two explicitly known eigenfunctions  and eigenvalues.
Namely:  the ground state eigenfunction (yet unnormalized)  $\psi _0(x)= (2x^2 - \sqrt{2}) \exp (-x^2/4)$ corresponds to the negative eigenvalue $E_0= - 2\sqrt{2}$, while  $\psi _2(x)= (2x^2 + \sqrt{2}) \exp (-x^2/4)$ corresponds to the  positive  eigenvalue $E_0= + 2\sqrt{2}$ (it is not the first,  but the second excited eigenfunction, has the same parity as the ground state).  Plugging  $a=1$ and $b=7$ we readily  verify  that the condition  $27a<b^2$  in the present case    holds  true, as expected.

\section{Outlook}

The  transformation of the Fokker-Plack dynamics  to the affiliated   Schr\"{o}dinger-type   evolution  has been analysed in detail for a sequence of superharmonically confined  systems, with a special emphasis on spectral properties of the Schr\"{o}dinger semigroup and the  role  of its (Feynman-Kac)  integral  kernel  in the construction of  stochastic  models of relaxation, stemming from the knowledge of the ground state function of the Schr\"{o}dinger Hamiltonian  and the integral (Feynman-Kac) kernel of the related Schr\"{o}dinger semigroup.

The expected outcome should have been a possibly "smooth"   limiting behavior  (this  proved not to be  the case,  pointwise-limits do not reproduce the standard "sharp" Neumann condition)    of  Schr\"{o}dinger-type operators $\hat{H}= -\Delta + {\cal{V}}(m)$ as $ m\rightarrow \infty $,   that would justify a reliable  approximation of the reflected Brownian motion with the generator $(-\Delta )_{\cal{N}}$ by the  Brownian motion  in extremally steep anharmonic potentials. This  (reliable approximation) property is normally expected to be transferable to the path-wise behavior as well.

 In view of rather unusual two-well   shape of pertinent Feynman-Kac potentials, and their singular at the boundaries $ m \to \infty $  limiting behavior, there is no smooth passage between two formalisms (e.g. an approximate and sharp model of reflection).  Nonetheless, the approximation by superharmonic models can be justified, if we admit a disregard of the narrow  zone of a finite "thickness"  (c.f. our considerations  in sections III and IV, and especially  Tables and Figures therein),  in the vicinity of the boundary. The   optimal   "thickness"   depends on how  large  $m$   is.

We find   worth emphasizing  an analysis  of Section V, where the concept of false bistability has been introduced. It is really amusing, that in the statistical physics literature nobody has ever   paid attention to the  generic  issue of the eigenvalue zero  of the diffusion generator (which, together with the Fokker-Planck operator is transformed into the affiliated  Hamiltonian $\hat{H}$),  and non-existence of negative eigenvalues of $\hat{H}$ for a broad family of two-well potentials, and the classic  quartic double-well problem in particular.  In this regard, we  have deduced a non-existence  criterion, which extends earlier perturbative studies  of the  (quantum)  double-well, \cite{turbiner1}.

A special role in the paper is played by  Appendices A, B and C. To enhance the coherence of our arguments, we have moved     to Appendix A a description of all  consecutive  steps being executed  in the construction of  transition pdfs of the  conditioned Brownian  motion (killing, its taming, and recovering Brownian relaxation for eg the Ornstein-Uhlenbeck process and the likes).  Motion   in the interval with  Dirichlet boundaries is described in detail as well.  In Appendix B, we demonstrate the workings of specific sequential approximations , whose limiting case is the  Brownian motion in the   interval with   inaccessible  Dirichlet boundaries.
Appendix C is of utmost importance, because there we demonstrate how  the relaxation properties    of conditioned Brownian motions can be deduced  within the classic, but seldom  revived  in the literature, Schr\"{o}dinger's boundary data and interpolation problem. That includes a derivation "for nothing" of the  basic  functional  form (C.11) (C.12),   shared by  various transition probability densities   introduced   in the present paper,  and   derived by other methods (albeit with the   Schr\"{o}dinger problem reference)  in Ref. \cite{gar}.

The present  Brownian endeavour in  has been   planned as a detour, which actually  turned over to an extended preparatory  step,  in the attempt  to  tackle theoretical and computer-assisted   problems arising   in the  approximate  description of L\'{e}vy-type processes in bounded domains with reflection. Partial results in existence,  were obtained along  Brownian-inspired  lines  (Langevin-L\'{e}vy  motion in steep anharmonic  potentials  versus fractional Schr\'{o}dinger spectral problems), \cite{smol}.  In passing, we note   that  the mapping of the fractional Fokker-Planck dynamics into the affiliated fractional Schr\'{o}dinger-type problem, has been  investigated in the past, \cite{vilela}-\cite{belik} and \cite{lorinczi,lorinczi1}, in connection with "L\'{e}vy  motion in energy landscapes".

Leaving aside an approximation issue, reflected L\'{e}vy processes  proper   are  nowadays  an active research topic,   albeit  physical and mathematical  papers on the subject  show  quite  different, often incongruent,   attitudes  and  proposals  towards the  proper  description of  (i)  the  involved path-wise behavior and   (ii) a proper formulation of Neumann boundary condition for  nonlocal operators, \cite{bounded1,smol}.

Once more, we indicate that in the present paper  we have  addressed problems that are    casually   considered as "obvious" and taken for granted, or possibly irrelevant in the considered contexts  (this refers  to the validity of the $m\to \infty $ limit, the related approximation accuracy control, and  to the nonexistence of negative eigenvalues for a broad class of two-well potentials,  not forgetting about a profound problem of the quasi-exact solvability \cite{turbiner} of related Hamiltonian systems).   The Brownian route has been chosen as a playground for checking jeopardies and  possible   inadequacies of the "obvious"  strategy, and   to   "clean the stage"  prior to passing to  an analysis of  technically more involved   issue of  reflected  L\'{e}vy  flights  and their superharmonic approximations, along the lines indicated in Ref. \cite{smol}.

\begin{appendix}

\section{Killing  versus spectral  relaxation.}
 \subsection{Ornstein-Uhlenbeck process vs  harmonic oscillator semigroup: dimensional issues.}

 Since, the explicit presence of dimensional constants somewhat blurs a connection between the Langevin equation-induced  Fokker-Planck) dynamics (1)-(5),
  the inferred   semigroup one (6)-(9) and the  emergent spectral problem (10),  we shall discuss  in some  detail the  dimensional issues in the  case of the standard harmonic attraction $U(x)= k x^2/2$.

 The drift $b(x) = -(k/m\beta)x =  - \kappa x $ defines  the   Ornstein-Uhlenbeck process, for which the  semigroup  (Feynman-Kac)  potential (8) takes the form:
\be
{\cal{V}}(x)= {\frac{\kappa ^2 x^2}{4D}}  -  {\frac{\kappa }2}.
\ee
 A multiplication by $2mD$ restores the standard dimensional version (Joule as the  energy unit)  of the potential, so  we realize that  upon setting $\omega = \kappa $, the  potential
\be
2mD\, {\cal{V}}(x)= {\frac{m\omega ^2 x^2}2}  - 2mD\,  {\frac{\kappa }2}
\ee
can be interpreted as  that of a   harmonic oscillator  system,   with an $mD\omega $ subtracted.
One should not confuse this $\omega $ with a natural frequency  $\sqrt{k/m}$  associated with  the  harmonic
 potential  $U(x)=k x^2/2$.

 By means of a formal identification  $D\equiv \hbar /2m$ where $\hbar $ is   the  reduced Planck constant, we give $2mD\, \hat{H}$   the familiar form:
 \be
 2mD\, \hat{H}  \equiv - {\frac{ \hbar ^2}{2m}}\, \Delta  +  {\frac{m\omega ^2 x^2}2}   -  {\frac{\hbar \omega}2}
 \ee
of the quantum harmonic oscillator Hamiltonian with a ground state energy renormalization.  More general discussion of this  subtraction issue can be found in Refs.
\cite{gar,vilela,faris,streit}, see also Section 3 in Ref. \cite{spectral}  for spectral aspects of the relationship between the  OU process and the harmonic oscillator semigroups.

We know that the  eigenvalue problem (9),  while extended   to the operator (A.3),  has the spectral solution   in $L^2(R)$   with eigenvalues $(2mD)\lambda _n  =  E_n - E_0= \hbar \omega n$, $n\geq 0$. The  bottom  eigenvalue equals  zero.
 The  corresponding   $L^2(R)$-normalized  (ground state) eigenfunction reads:
\be
\phi_0(x) = \left({\frac{m\omega}{\pi \hbar }}\right)^{1/4}  \exp \left[ -  {\frac{m\omega x^2}{2\hbar }}\right] \equiv \left( {\frac{k}{2\pi k_BT}}\right)^{1/4}
 \exp \left[- {\frac{U(x)}{2k_BT}}\right] = \rho _*^{1/2}(x),
\ee
where $U(x)= kx^2/2$, and   to  recover   the functional form   (5)  of   the invariant density  $\rho _*(x)$,   we have reintroduced the "thermal" notation, e.g. $\hbar \equiv  2mD$, $D=k_BT/m\beta $ and $\omega = \kappa = k/m\beta $.  Accordingly  $F_* =  (1/2) \ln  ( k/2\pi k_BT)$.\\

\subsection{Eigenfunction expansions: Killing can be tamed.}

 We note that if the energy is measured in units of $\hbar \omega $, while the distance in units of $\sqrt{\hbar /m\omega }$, the rescaled energy operator with subtraction takes the form
\be
\hat{H}  = {\frac{1}2} (- \Delta + x^2  -1)= -{\frac{1}2} \Delta   + {\cal{V}}(x)
\ee
with the spectrum $E_{n} = n $, $n\geq 0$ and the ground state function  $\phi _0(x) = \pi ^{-1/4} \exp (-x^2/2)$.   Note that the spectrum of $2 \hat{H} = - \Delta + x^2 - 1$   coincides with  $2 E_{n} = 2n$ and begins from the eigenvalue zero.\\

    Consider  the  (rescaled, without subtraction) harmonic oscillator problem denoted  $\hat{H}_0= (1/2)(-\Delta + x^2)$. Its spectral solution comprises  a sequence of eigenvalues $\epsilon _n= n +  {\frac{1}2}$  and corresponding eigenfunctions   $\phi _n(x) = [4^n (n!)^2 \pi ]^{-1/4} \exp(-x^2/2)\,  H_n(x)$  which are $L^2(R)$ normalized.   Here  $H_n(x)$ is the n-th Hermite polynomial
  $H_n(x) =(-1)^n (\exp x^2) \, {\frac{d^n}{dx^n}} \exp(-x^2) $.   Consequently  $\phi _0(x)=\pi ^{-1/4} \exp(-x^2/2)$ and  $\epsilon _0 = 1/2$.   The
integral kernel of $\exp(-t\hat{H}_0)$  is a symmetric function of space variables:
\be
 k_0(t,x,y)=  \sum_j
\exp(- \epsilon _j t) \, \phi _j(y) \phi _j(x) = k_0(t,y,x)
 \ee
 and stands for a transition  density of the diffusion-type process with  killing, with a well defined probabilistic
  interpretation  including e.g. the survival time and its  exponential decay in the large time asymptotic,
   c.f. Refs. \cite{gar,faris}.   The kernel induces  the semigroup
propagation:
$\Psi (x,t) = [\exp (-\hat{H}_0t)\, \Psi ](x)= \int_R k_0(t,x,y)\Psi (y) dy$.\\
The integral kernel of the "renormalized" energy operator $  \hat{H}=  \hat{H}_0 - 1/2  = (1/2)(-\Delta + x^2 -1)=  - (1/2)\Delta + {\cal{V}}$ has the form:
\be
k(t,x,y) = \exp (+ \epsilon _0\,  t)\,  k_0(t,x,y) = \phi _0(x)\phi _0(y)  +  \sum_{j=1}^{\infty} \exp[- (\epsilon _j - \epsilon_0) t] \, \phi _j(y) \phi _j(x),
\ee
with a conspicuously time-independent   ground state contribution.  We note that any suitable   $\Psi (x) =\sum_{j} \alpha _j \phi _j(x)$  evolves in time  according to $\Psi (x,t)= \alpha _0 \phi _0(x) +  \sum_{j=1}^{\infty} \exp[- (\epsilon _j - \epsilon_0) t] \,   \alpha _j\,  \phi _j(x)$. \\
  The asymptotic demand $\Psi (x,t) \rightarrow \phi _0 = \rho _*^{1/2}$  can be met only if $\alpha _0=1$.   Hence,   a proper form for $\Psi (x)$ of interest is
\be
\Psi (x)= \phi _0(x) + \sum_{j=1}^{\infty} \alpha _j \, \phi _j(x).
 \ee

In principle,  one can work with any   $\Psi \in L^2(R)$.  The form (A.8) can be introduced,  if  $\Psi $ is not
 orthogonal to $\phi _0$. Then,   $(\Psi, \phi_0) = \alpha _0 \neq 0$  entails the replacement of
   $\Psi $ by $(1/ \alpha _0) \Psi $, which  does the job.\\

\subsection{Transition densities: Elimination of killing.}

 The link with the Fokker-Planck dynamics of $\rho (x,t)$ for  the Ornstein-Uhlenbeck process,
 while  departing from the present spectral  notation, can be restored   as follows.
 To  comply with   the factorisation (6) and the  Schr\"{o}dinger-type  dynamics (7),  we  identify   $\rho _*(x) = [\phi _0(x)]^2$  and  set  $\Psi (x)\,  \phi _0(x)  = \rho (x,0) $.
 The Fokker-Planck operator takes the form $L^* = (1/2)\Delta -
\nabla   [b(x) \, \cdot ]$   and $b(x)= -x$, (the diffusion constant $D$ is here replaced by $1/2$).

 The asymptotic (invariant, stationary) probability density of the
 pertinent process reads
$ \rho _* (x)= \phi _0^2(x)=(1/\pi )^{1/2}\, \exp(-x^2)$.
We recall that the stationary density  $\rho _*(x)$   of the F-P equation
 $\partial _t \rho = D\Delta \rho - \nabla (b \rho )$, where $b= -  \nabla U = D \nabla \ln \rho _*$,
  has the form $ \rho _* \sim \exp (- U(x)/D)$.

  It is the choice of $D=1/2$ and  $U(x)= x^2/2$, which gives rise to the above $\rho _*(x) \sim \exp (-x^2)$.
   Then, we  have  $b(x)=  (1/2) \nabla \rho _*(x) =  \nabla \ln \phi_0(x)= -x $ and  ${\cal{V}}= (1/2)(x^2-1)$ appears in the
   appropriate version of Eq. (7): $ \partial_t\Psi= (1/2) \Psi - {\cal{V}} \Psi $, whose stationary solution $\rho _*^{1/2}(x)$ actually is.

     An alternative choice of $D=1$, while keeping   $U=x^2/2$,  implies $\rho _*(x)\sim \exp (-x^2/2)$. Then,
   we have $b(x) = \nabla \ln \rho _*(x)= -x$ and ${\cal{V}} = (1/2) (x^2/2  -1)$. With such scaling,  Eq. (7) takes the
    form $ \partial_t\Psi= \Delta \Psi - {\cal{V}} \Psi $, \cite{zero,stef,geisel,brockmann}.

We can give   a   concise summary of   of relationships  between    transition densities  $k_0(t,y,x)$, $k(t,y,x)$ and $p(t,y,x)$,   introduced   in subsection II.B.
  We   begin from the killed process,  for which the  integral kernel of $\exp(-t\hat{H}_0)$,  with
  $\hat{H}_0= (1/2)(-\Delta + x^2)$, reads  (the order of space variables is important):
\be k_0(x,y,t) =   k_0(t,y,x) =  [\exp(-t\hat{H}_0)](y,x)=
 \exp(-t/2)\, (\pi [1-\exp(-2t)])^{-1/2} \exp \left[{\frac{1}2} (x^2-y^2) -
   {\frac{(x- e^{- t}y)^2}{(1- e^{-2 t})}}\right].
\ee
Note a conspicuous presence of the time-dependent factor $\exp (-t/2)$, comprising the contribution from the
lowest eigenvalue $1/2$ of $\hat{H}_0$, which is directly responsible for the exponential decay of the transition pdf.

The transition probability density of the   relaxation  process governed by Eqs. (1), (2), (12),  in  the present case reads:
\be
p(t,y,x) =
k(t,y,x)\, {\frac{\phi _0(x)}{ \phi _0(y)}} = e^{+ t/2} k_0(t,y,x)\, {\frac{\phi _0(x)}{ \phi _0(y)}},
 \ee
where  $\phi _0(x)= \rho _*^{1/2}(x)$.   Eq. (30) reproduces the
transition density of the  Ornstein-Uhlenbeck process in $R$.   Here,   $k_0(t,y,x)$ refers to a process with killing,
 \cite{faris,gar},   while $k(t,y,x) $ refers to the  spectrally   relaxing process described by (7) and  (12).
  The presence of the ratio $\phi _0(x)/ \phi _0(y)$  is a signature of the  (emergent, see Ref. \cite{gar})
   Doob-type conditioning which  transforms $k(t,y,x)$   into a legitimate (probability measure preserving)   transition
    probability density  $p(t,y,x)$
   of a Markovian diffusion process  with an invariant probability density, \cite{gar,pinsky, pinsky1}.\\

\subsection{Interval with absorbing  endpoints  versus  non-absorbing   (Dirichlet)   infinite well.}

   For the process  with  killing (absorption)  at  the boundaries of the interval $[-1,1]$, the   generator of motion
   is the  ordinary  Laplacian  $\Delta _{\cal{D}}$    with Dirichlet boundary conditions    imposed as  its domain restriction ($L^2([-1,1])$
      functions, vanishing at   $ x = \pm  1$,   \cite{karw}).
A physically legitimate step  is here to   analyze the dynamics   in terms of  a    potential  function $U(x)$, which
 mimics the  infinite square well enclosure and directly refers  to  quantum theory routines.
Formally, one is inclined to   set  $U(x)=0$  for $x\in (-1,1)\subset R$ and  $U(x) \equiv \infty $ for $R\setminus [-1,1]$ and
introduce the operator $\hat{H}_0= - \Delta + U(x)$   (see  \cite{robinett,karw,diaz}).
This  may be interpreted as a "physical"  encoding     of  $-\Delta _{\cal{D}}$  (we recall our introductory  mention of
 the case of a constant potential plus the Dirichlet boundary data imposed upon the operator $\hat{H}$,
  see also \cite{pinsky}).

 For the operator $ \hat{H}_0 = -\Delta _{\cal{D}}$, the orthonormal
eigenbasis in $L^2([-1,1])$ consists of functions $\psi _n(x)= cos(n\pi x/2)$ for $n$ even and $sin(n\pi x/2)$ for $n$ odd,
while the  respective eigenvalues read $E_n = (n\pi /2)^2$.
 The lowest eigenvalue equals  $\pi ^2 /4$ and the  $L^2([-1,1])$ normalized  ground state function is  $\psi _1(x)=
 \cos(\pi x/2)$.

It is clear that any $\psi \in L^2([-1,1])$, in the domain of the
infinite well Hamiltonian,   may be represented as $\psi  (x) =
\sum_{n=1}^{\infty } c_n \psi _n(x)$.  Its time evolution follows
the Schr\"{o}dinger semigroup pattern $\psi (x)  \rightarrow \Psi
(x,t)=  [\exp(-\hat{H}_0 t) \psi ](x) = \sum_{n=1}^{\infty } c_n \exp (-E_n
t)\, \psi _n(x)$  and asymptotically sets at the value $0$, with the  exponential  rate of convergence $E_1= \pi ^2/4$.

Let us consider $\hat{H} = -\Delta _{\cal{D}} -  E_1$ instead of $\hat{H}_0=-\Delta _{\cal{D}}$  proper.
The    positive-definite  ground state $\psi
_1(x) \doteq \rho _*^{1/2}(x)$ corresponds to the zero eigenvalue of
$\hat{H}$. The "renormalized" semigroup evolution implies the relaxation pattern $\Psi
(x,t) = \exp(+E_1 t) \sum_{n=1}^{\infty } c_n \exp (-E_n t)\, \psi
_n(x) \rightarrow \psi _1 (x) =\rho _*^{1/2}(x)$.  The present (asymptotic)  relaxation rate equals $E_2- E_1= 3\pi ^2/4$.

 The semigroup kernel $k(t,x,y)=\exp(-t \hat{H})(x,y)$,  defines  a time homogeneous random process in the interval.

 In view of: $k_0(t,x,y) =  \sum_{n=1}^{\infty } \exp [-(n\pi /2)^2\,  t]\,
  \psi _n(x)  \,  \psi _n(y)$, its   spectral representation reads  (compare e.g. (26)):
 \be
 k(t,x,y)=  \exp (+ \pi^2t /4) k_0(t,x,y)=
\sum_{n=1}^{\infty } \exp [(1- n^2) \pi ^2\, t/4]  \sin[n\pi (x+1)/2]\, \sin[n\pi (y+1)/2]=k(t,y,x).
 \ee
 By construction, $\Psi (x,t) = \int k(t,x,y) \Psi _0(y) \, dy$, if not orthogonal to $\psi _1(x)$, is bound to    relax  to  $\psi_1 (x)$.

 The   probability density
function  (pdf) $\rho _*(x) = [\psi _1(x)]^2=  cos ^2 (\pi x/2)$  is an  equilibrium solution  of the Fokker-Planck equation
 $\partial _t \rho =    \Delta _{\cal{D}}\rho  - (\nabla b\rho )$,  with   the forward drift  $b(x) =\nabla \ln  \rho _*$,
   \cite{gar,risken}.   The resultant   diffusion process in the well (equivalently interval with inaccessible endpoints)
    has the Lagevin (and thence Fokker-Planck) representation  and   is determined by means of  the
    transition probability density   of the form (29) \cite{risken,gar,bounded1}
\be
p(t,y,x) = k(t,y,x) {\frac{\rho _*^{1/2}(x)}{\rho _*^{1/2}(y)}},
\ee
so that  the evolution  $\rho (x,t)= \int p(t,y,x)\, \rho _0(y)\, dy$ proceeds   entirely within the interval $D\subset R$ and never leaves $D$.

\begin{figure}[h]
\begin{center}
\centering
\includegraphics[width=60mm,height=60mm]{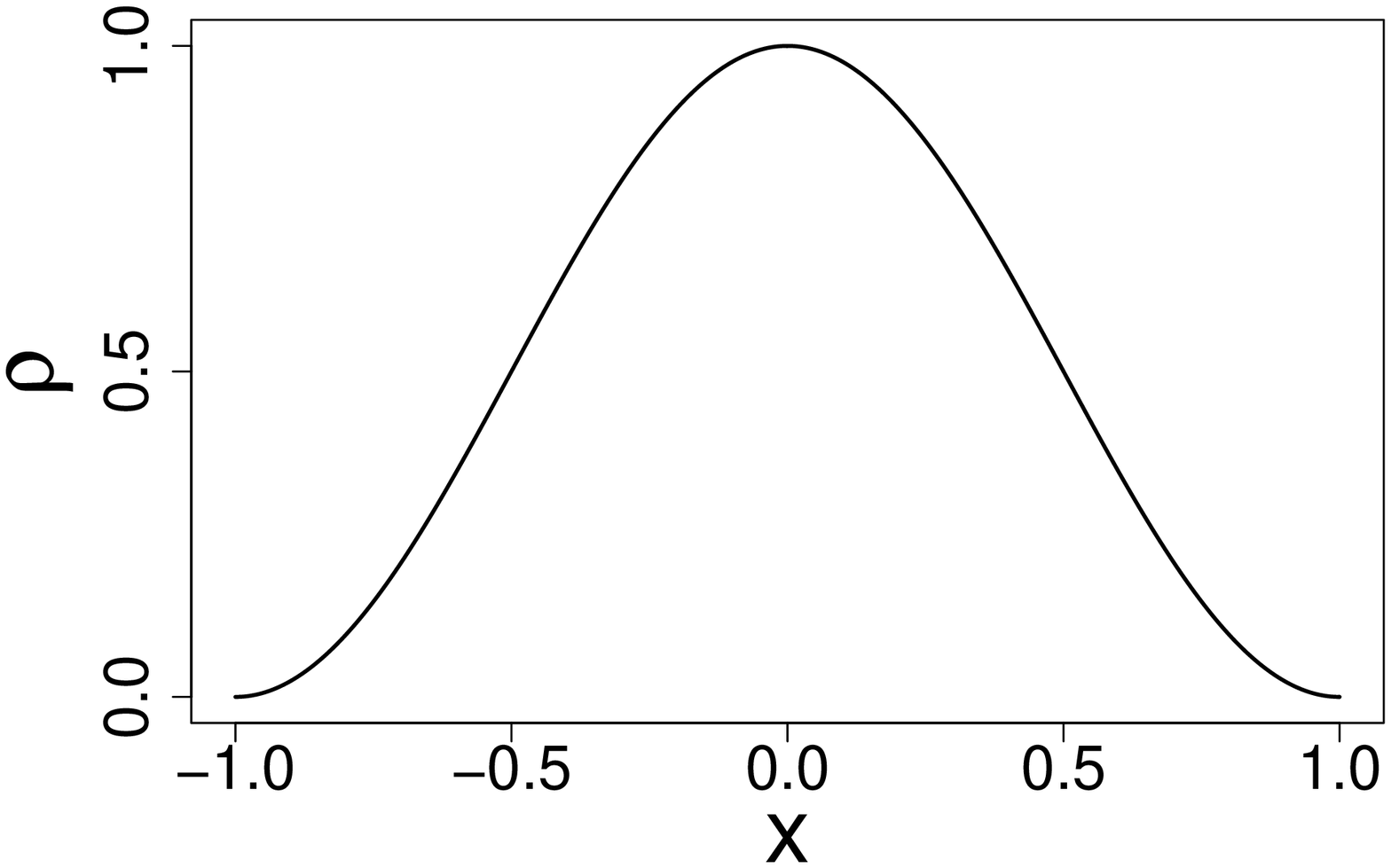}
\includegraphics[width=60mm,height=60mm]{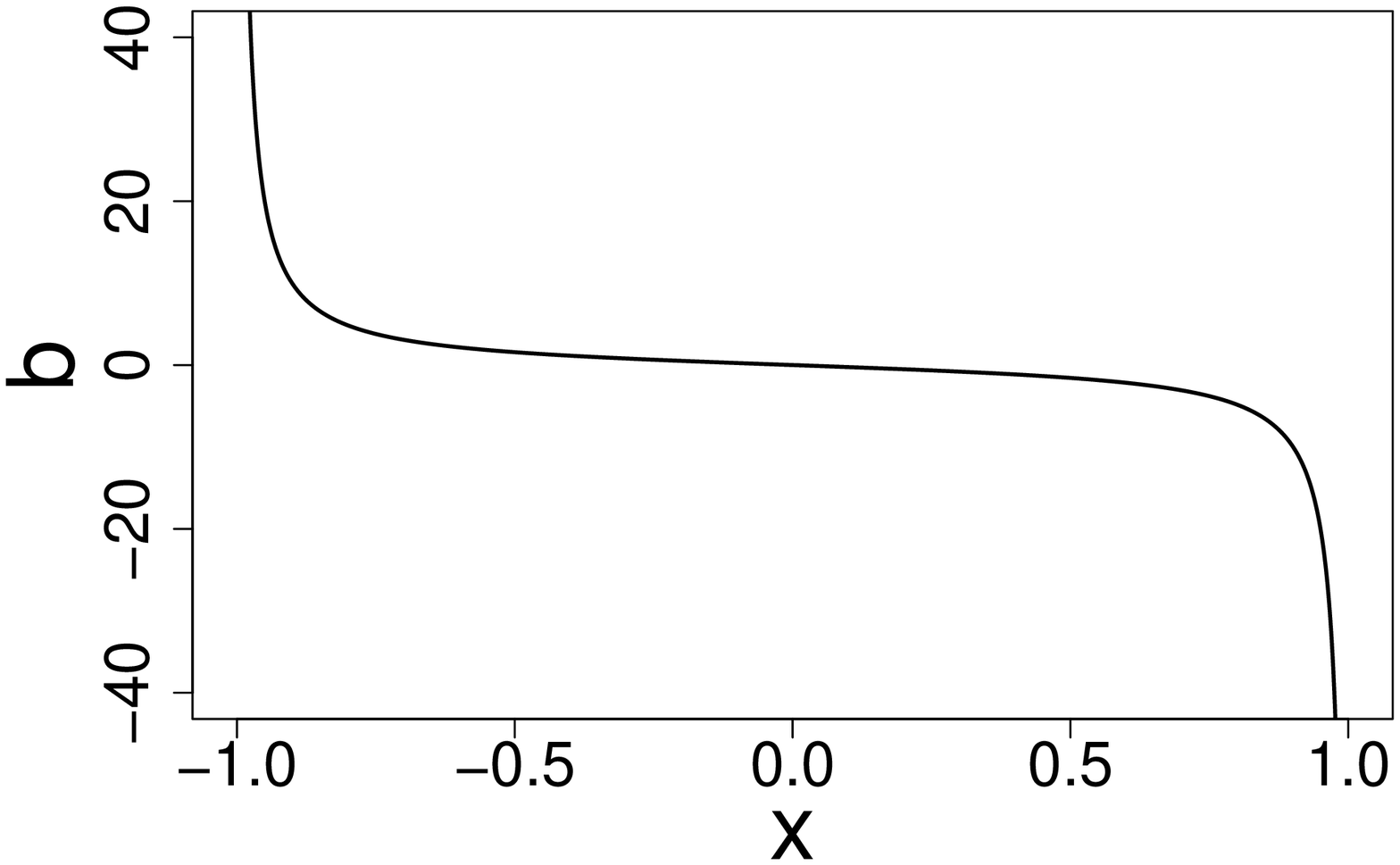}
\caption{Boltzmann-type equilibrium in the interval $[-1,1]$. Left, the  stationary  pdf
$\rho _*(x)= cos^2(\pi x/2)$. Right,  the forward drift  $b(x) = \nabla \ln \rho _*(x)=   - (\pi /2)\,  \tan (\pi x /2) $  of the Fokker-Planck
equation $\partial _t \rho = {\frac{1}2}   \Delta _{\cal{D}}\rho  - (\nabla b\rho )$  in $(-1,1) \subset R$.
It is $\rho _*(x)$ to which the solution $\rho (x,t)$  of the F-P equation relaxes  (spectral relaxation pattern)
 in the large time asymptotic.}
\end{center}
\end{figure}

 The role of the  noise  intensity parameter $D$ needs to be carefully observed.  By  choosing  $D=1$  we arrive at
  the eigenvalues $(n\pi )^2/4$, while   $D=1/2$  implies   $(n\pi )^2/8$.
The drift functions read   $b(x)= -  \pi \tan (\pi x/2)$ or $b(x)=  - (\pi /2) \tan (\pi x/2)$  respectively.
Recalling that we  interpret $b(x)$ as $- \nabla U(x)$, one recovers   $\rho _*(x) = \exp[- U(x)]$, where
$U(x)= - 2 \ln cos(\pi x/2)$  ($D=1$), c.f.  \cite{risken}

In accordance with Eqs  (6)- (8), we can establish  the relationship between the forward drift $b(x)$ and the functional form of the Feynman-Kac potential
  ${\cal{V}}(x)$ in question.   By substituting  $b(x)=  - (\pi /2) \tan (\pi x/2)$ to the $D=1/2$ expression  (8),
   ${\cal{V}} = {\frac{1}2} (b^2 + \nabla b)$,
   we readily verify that the Feynman-Kac potential actually is a constant:   ${\cal{V}} = - \pi ^2/8 $
   for all $x \in (-1,1)$, see e.g. also \cite{risken,pinsky}.  We encounter here  a shift of
   the   well bottom  along  the energy axis  from the value $0$ to $-E_1= - \pi ^2/8$, compare e.g.  Fig. 5.2
    in Ref. \cite{risken}, see also \cite{gar,faris}.

    This well conforms with the general statement, \cite{pinsky}, that the Schr\"{o}dinger operator  with
     constant potential {\it and } Dirichlet boundary   condition corresponds to the Brownian motion
      that is conditioned   never to exit $(-1,1)$, see also \cite{gar}.

   Recalling the path integration  (Feynman-Kac)  formula (11),  and setting there ${\cal{V}} = - \pi ^2/4 $   ($D=1$ case), we
       realize that  $-\int_0^t {\cal{V}}(\omega(\tau  )) d\tau   = +t \pi ^2 /4$  identically. Hence the transition
       density  $k(t,x,y)$ takes the form
\be
 k(t,x,y) = e^{(+ t\pi ^2/4)}    \int d\mu _{(x,0,y,t)}(\omega) =   e^{(+ t\pi ^2/4)}  \exp [t \Delta _{\cal{D}}] (x,y),
\ee
where $  \exp [t \Delta _{\cal{D}}] (x,y) = k_0(t,x,y)$, c.f. Eq. (30).

\section{Steep potential wells and   Brownian motion in  the Dirichlet well/interval.}

\subsection{$\hat{H}= - {\frac{1}2} \Delta + x^m$: $m$-dependence of ground state functions  and eigenvalues,
 signatures of  spectral  "closeness"  to  $-  {\frac{1}2}\Delta _{\cal{D}}$.}

The operator $\hat{H} = - D\Delta + x^m$, with  $m$ even  (and typically $D=1$ or $D=1/2$),       is known to provide a  fairly accurate
   spectral approximation  of the standard  infinite well problem with Dirichlet boundaries, whose  eigenvalues
    and eigenfunctions can be  reproduced approximately in  the form of of $1/m$ expansions, see e.g.
    \cite{froman,bender} and \cite{gar1}.

\begin{figure}[h]
\begin{center}
\centering
\includegraphics[width=88mm,height=88mm]{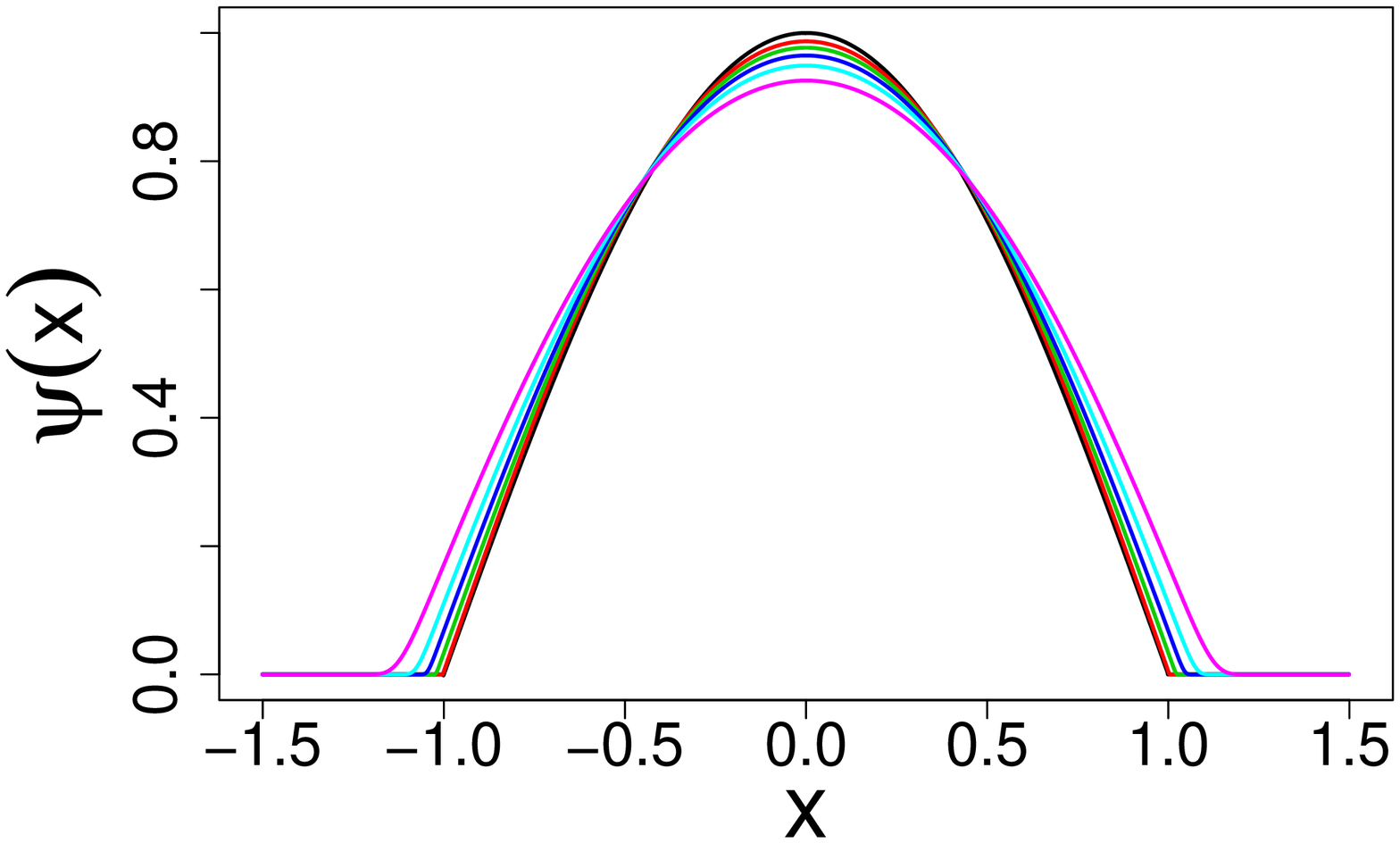}
\includegraphics[width=88mm,height=88mm]{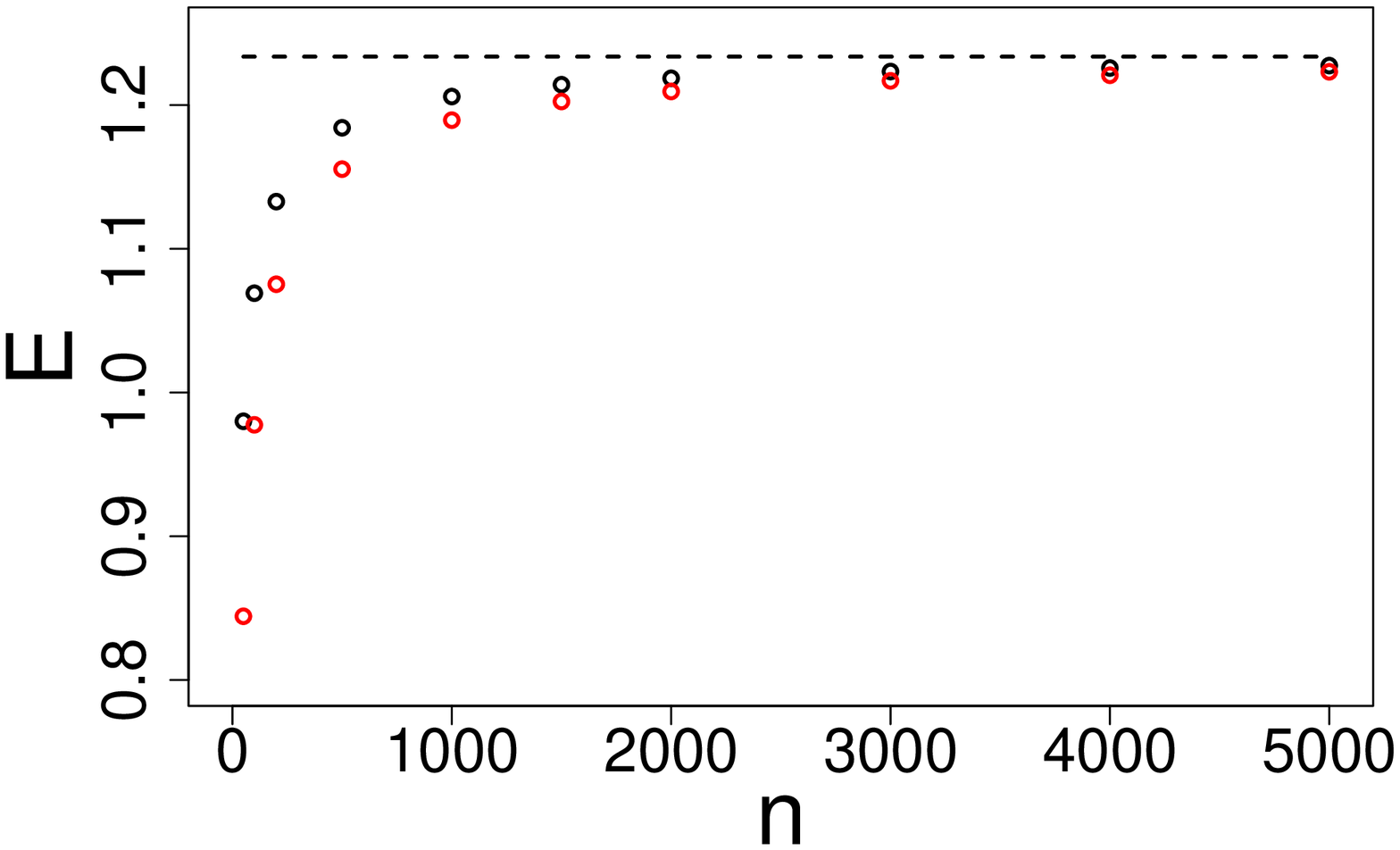}
\caption{Left panel: ground state eigenfunctions  $\rho _*^{1/2}(x)$  of $\hat{H}= - {\frac{1}2} \Delta + x^m$,
$m=50, 100, 200, 500, 5000$; one may assign $m$ to a concrete curve by following the maxima in the  increasing order, the  top maximum (black) curve
 corresponds to the  asymptotic $\rho _*^{1/2}(x) = \cos(\pi x/2)$. Right panel: The $m$-dependence of the ground state
  eigenvalue $E=E_1(m)$  is depicted both for the potential  $x^m$ (black)   and  $x^m/m$ (red).
  The convergence to the asymptotic  infinite  Dirichlet  well value   $E_1=\pi ^2/4 \simeq  1.2326$
   is undisputable.  The reported eigenvalues  read (case of $x^m$, listed
    in the growing $m$-order):  $E_1(m) = 0.980021,  1.06912, 1.13285, 1.20595, 1.21421, 1.21865, 1.22335, 1.2258, 1.22748$.}
\end{center}
\end{figure}

Since analytic solutions for moderately sized values of $m$ are unavailable, we have adopted a computer assisted  route
 towards the approximate   evaluation of  eigenfunctions and eigenvalues of $\hat{H}$, with a focus on their
 convergence properties with the growth of $m$.  Its is based on    the   Strang splitting  method,
 originally  devised  in another (L\'{e}vy processes and fractional Schr\"{o}dinger operators)  context,  see e.g.
 \cite{gar2,gar3}.

We reproduce  final computer-assisted results,  obtained for the case $D=1/2$ i.e.  for the operator $\hat{H} = - {\frac{1}2} \Delta  + x^m$.
Guided by the "reconstruction of the dynamics  from the ground state"  idea, we confine our  attention to the   $m$-dependence of the
 ground state eigenvalue   (in the present case it is not  equal zero)  and the related eigenfunction.
 The ground state functions are depicted for  $m = 50, 100, 200, 500, 5000$, two  latter curves are  indistinguishable from
 the asymptotic $\cos (\pi x/2)$.

 The  ground state  eigenvalues (comparatively for the case $x^m$ and $x^m/m$)  are numerically evaluated for
 $m=50, 100, 200, 1000, 1500, 2000, 3000, 4000, 5000$. The convergence to $\pi ^2/4\simeq 1.2326$ is graphically
 confirmed.

 The case of $x^m/m$ has been covered in a parallel computation, but made comparatively explicit in Fig. 10,  only on the
  level of eigenvalues.
 As far as the  eigenfunctions  are concerned, their  qualitative behavior  for  the potential  $x^m/m$
  is  the same as that reported   for $x^m$    in Fig. 10.

  The presented analysis demonstrates that for   large  even  $m$, the spectral problem for $\hat{H}= - D \Delta + x^m$ well approximates
   the  spectral problem for the Dirichlet operator $-\Delta _{\cal{D}}$ on the interval (equivalently, in the infinite well).

  \subsection{$\hat{H} = - \Delta  + {\cal{V}}$, with ${\cal{V}}(x)  = x^m - E_1(m)$:
reconstruction of confined   Brownian motions (reverse engineering).}

Since  for each $m=2n\geq 2 $, in the least numerically, we have in hands the ground state function $\rho _*^{1/2}(x)$,
which  corresponds to the eigenvalue zero: $\hat{H} \rho _*^{1/2} = (- \Delta  + {\cal{V}}])\rho _*^{1/2}=0$,
 we can  readily  follow  the reverse engineering idea (originally devised for L\'{e}vy flights) \cite{klafter,zero,stef}.

\begin{figure}[h]
\begin{center}
\centering
\includegraphics[width=55mm,height=55mm]{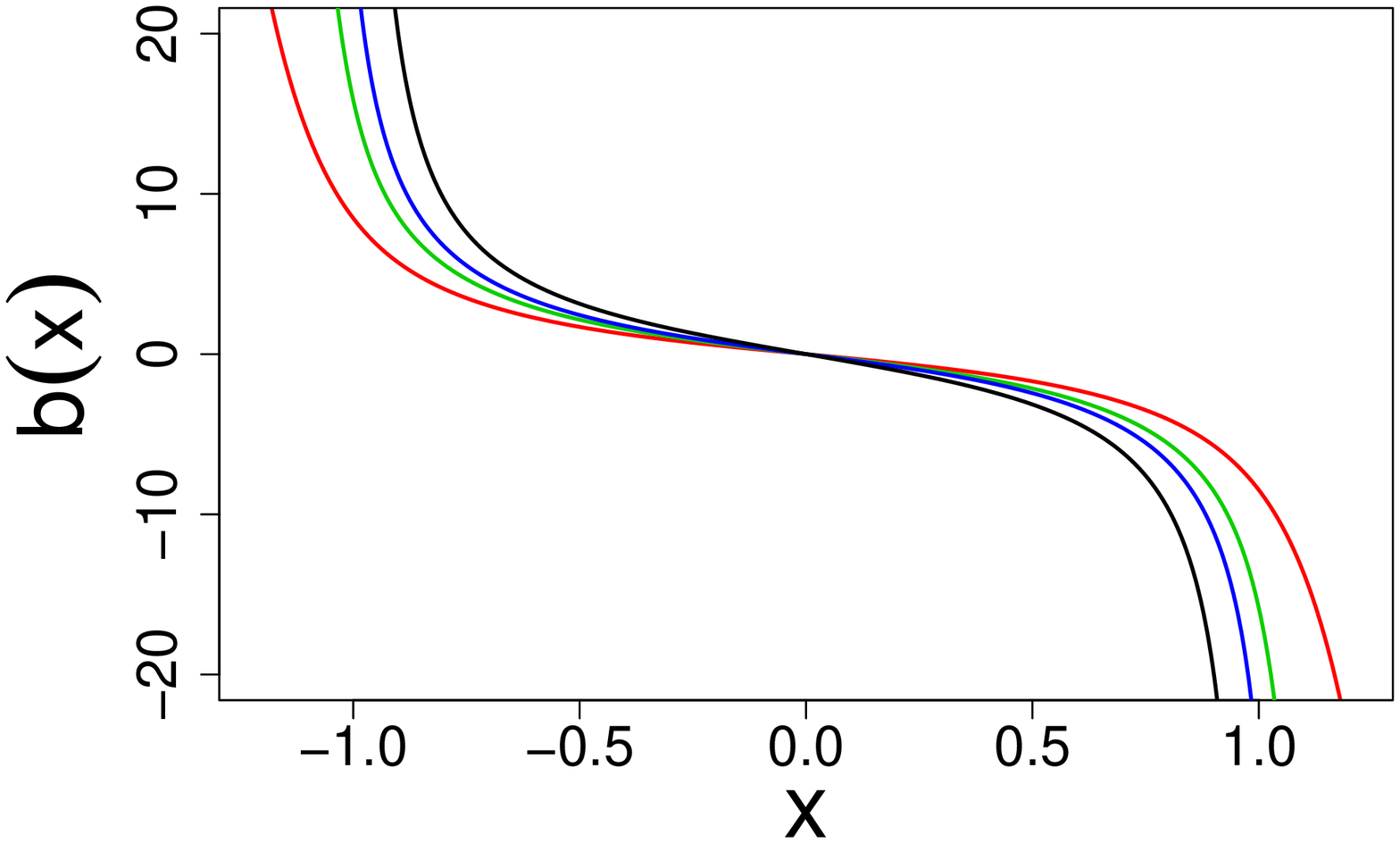}
\includegraphics[width=55mm,height=55mm]{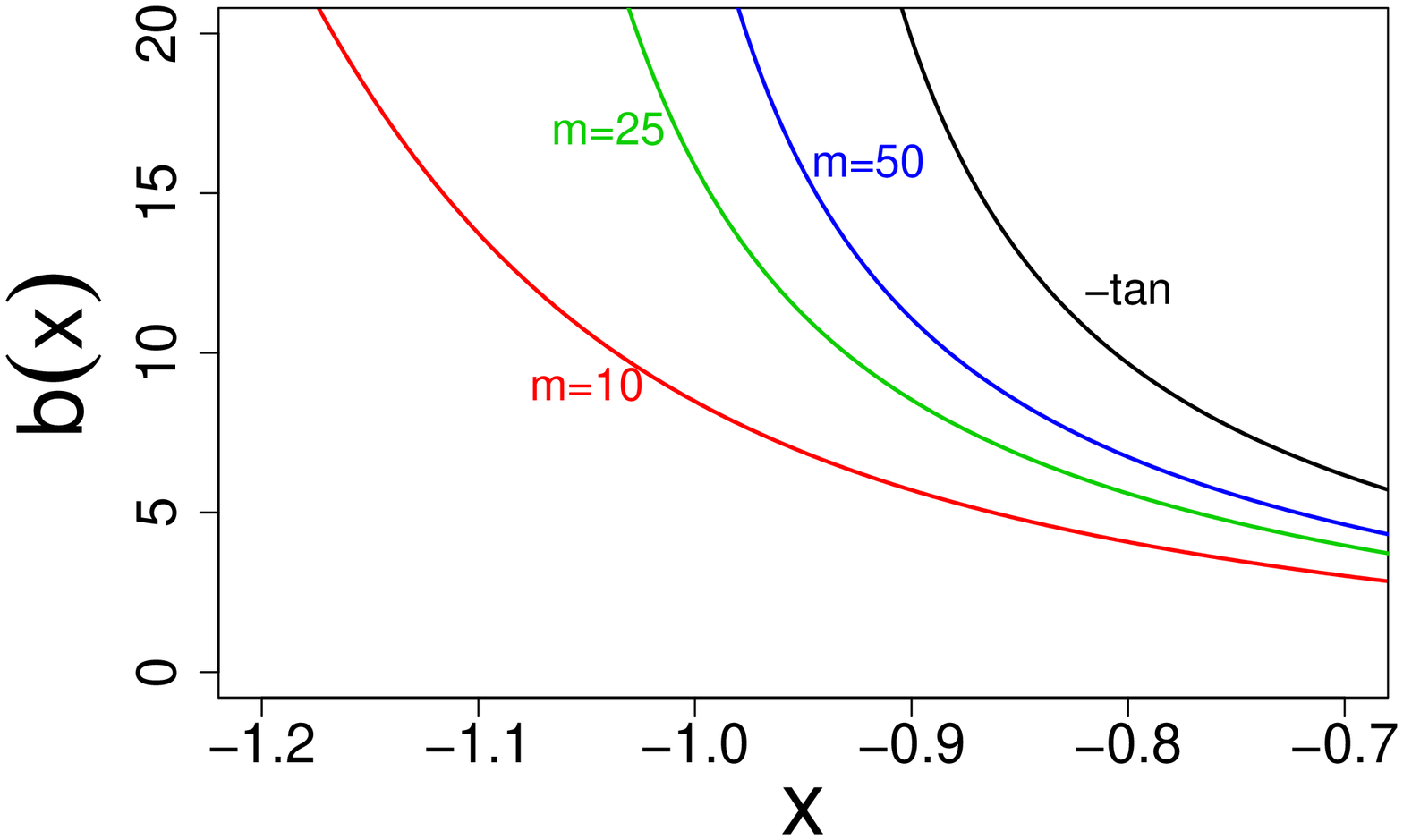}
\includegraphics[width=55mm, height=55mm]{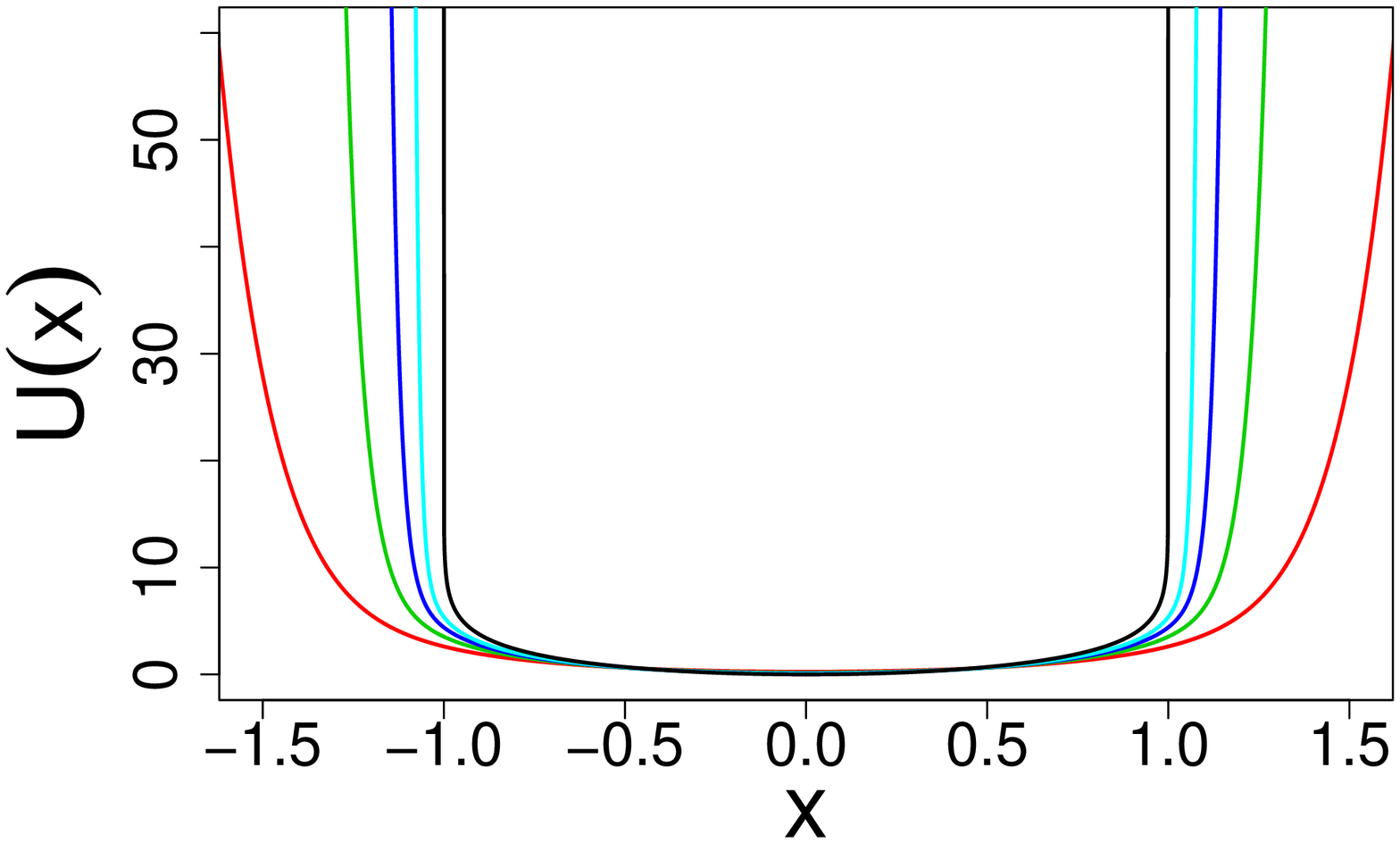}
\caption{Reverse engineering:   we depict the reconstructed drift $b(x)$  for $m=10$ (red), $m=25$ (green), $m=50$ (dark blue) and the asymptotic
$b(x)= -\pi \tan (\pi x/2)$. Middle  panel: the enlargement in the vicinity of $x= -1$.    Right panel:
the reconstructed Newtonian (actually Boltzmann, c.f. section II)  potential  $U(x)= - \ln \rho _*(x)$, black curve  corresponds to the
 asymptotic potential  $U(x)= - 2\ln \cos (\pi x/2)$,  an  additional blue  curve refers to   $m= 100$.  Note that
 $b =\nabla \ln \rho _* =  - \nabla U$.}
\end{center}
\end{figure}

 Its basic aim has been to reconstruct the Langevin (Fokker-Planck)-driven diffusion process, for which
  an invariant probability density  is given  a priori.  This  is precisely
  our   point of departure,  since  for a couple of $m$   we have in hands the
   pertinent $\rho _*(x)$.

   Our  reconstruction procedure goes as follows. Given $\rho ^{1/2}_*(x)$, we take its square.
    Next we adopt the numerical algorithm
   allowing to retrieve $b(x)$, through a direct evaluation of $b(x)= \nabla \ln \rho_*  = [\nabla \rho _*(x)]/\rho _*(x)$.
   The results are depicted in Fig. 11  and show convincingly  why (drift spatial behavior)  the Langevin-driven
    Brownian motion in $R$  sets down at an equilibrium pdf.    We indeed sequentially   approximate the
    taboo process (\cite{gar}) in the interval, while  keeping under  control   how  the  asymptotic pdf, drift and  Newtonian (Boltzmann, see section II)
    potential are  approached   with the growth of even $m$.  In  the  interval (alternatively, in the Dirichlet infinite well),
     the invariant pdf is  $\rho _*(x)=\cos^2(\pi x/2)$. The related Fokker-Planck equation is  reproduced by  evaluating
       $b(x)= \nabla \ln \rho _*(x)  = - \nabla U(x)$,  with  $U(x)= - 2\ln \cos (\pi x/2)$   (this corrresponds to D=1/2)
        playing the role of the Newtonian (and Boltzmann, see section II) potential.

\section{Schr\"{o}dinger's boundary data and interpolation problem: Arena for conditioned Brownian motions.}

The previous discussion  of   diversly   confined Brownian motions, have a natural affinity with the classic  Schr\"{o}dinger's boundary data and interpolation problem. Its  roots range back to  the original paper by E. Schr\"{o}dinger, \cite{schr}, the theory  being revived by mathematicians and physicists, see e.g. \cite{zambrini},   \cite{gar4,nonegative,non} and references therein.

The Schr\"{o}dinger boundary data problem  is known to provide  a unique Markovian interpolation between any two strictly positive probability densities,  designed to form the input–output statistics data  for a  random process process bound to   run  in a  finite (observation)   time interval.  The key input, if one attempts to reconstruct  the   pertinent Markovian dynamics, is to select the jointly continuous in space variables, positive and contractive semigroup kernel. Its  choice  is arbitrary, except for the strict positivity  (not a must, but we keep this restriction in the present paper) and continuity demand.

The semigroup dynamics in question we  infer  from the  classic notion of  the  Schr\"{o}dinger  semigroup   $\exp(-t \hat{H})$, with a proviso that the semigroup generator $\hat{H}$  actually  stands  for  a  legitimate
(up to scaled away physical constants)  Hamiltonian operator  of the form   (dimensional constants are scaled away)    $\hat{H}=   - \Delta +  {\cal{V}}(x)$.

Roughly, the essence of the Schr\"{o}dinger  boundary data problem,  \cite{zambrini},  goes as follows.
We consider Markovian propagation scenarios running in a finite time interval $[0,T]$,  with the input–output statistics data  provided in terms of two strictly positive boundary densities $\rho (x,0)$ and  $\rho (x,T)$,  $T>0$, \cite{zambrini,gar3}.  We demand the existence of the  bivariate    transition probability $m(A,B)$, $A, B \subset R$, admitting the  probability density
   \be
 m(x,y)= f(x) k(x,0,y,T) g(y)
 \ee
 with marginals  $\int _R m(x,y)dy = \rho (x,0)$ and $\int _R  m(x,y)dx  = \rho (y,T)$   that may be constrained to (integrated over) some Borel sets $A$ and $B$  contained in $R$ to yield $\rho _0(A)$  and $\rho (B,T)$ as marginals for $m(A,B)$.

In Eq. (C.1), $f (x)$ and $g(y)$  are the a priori unknown   strictly positive  functions,   that need to be deduced  from  the imposed boundary data (i.e. marginals that are presumed to be known a priori).  To this end,
 in addition to the density boundary data we  select  a  strictly positive, jointly continuous in space variables   kernel function $k(x,0,y,T)$, which we consider as  a restriction (to endpoints of the considered time interval $[0,T]$) of  a certain strongly continuous dynamical semigroup kernel      $k(y,s,x,t),\,   0\leq s<t \leq T$. This assumption will secure  the Markov property of the associated   interpolating   process.
Actually, we shall consider time homogeneous processes generated by the semigroup $\exp[- (t-s)\hat{H})$, with the  kernel $k(t-s,y,x)$.

Under those circumstances,  \cite{zambrini} and \cite{gar3,nonegative,non},  once we define functions
\be
\theta (x,t) = \{ \exp[-(T-t)\hat{H}]\, g \} (x)  = \int  k(x,t,y,T) g(y) dy
 \ee
 and
 \be
  \theta _*(y,t)   =    \{ \exp( -t\hat{H}) \, f \}(y) =  \int k(x,0,y,t) f(x) dx
\ee
one can demonstrate  the existence of  a transition  probability  density  (note that even if $k(t,y,x)=k(t,x,y)$ the symmetry property is not respected by $p(t,x,y)$  in below)
\be
p(y,s,x,t) = k(y,s,x,t) {\frac{\theta (x,t)}{\theta (y,s)}},
\ee
which implements a Markovian propagation of the probability density
\be
\rho (x,t) =  \theta (x,t) \theta _*(x,t),
\ee
according to  the pattern
\be
 \rho (x,t) = \int  p(y,s,x,t) \rho (y,s) dy= \theta (x,t) \int k(y,s,x,t) \theta _* (y,s) dy = \theta (x,t) \theta _*(x,t),
\ee
 providing an interpolation between the prescribed boundary data in the time  interval $[0,T]$.

Here we note the exploitation of the semigroup property in propagation formulas (C.2), (C.3).
Namely,   for $0\leq s<t \leq T$    we have, \cite{zambrini}:
\be
\theta (x,s)=  \int  k(x,s,y,T)g(y) dy = \int \int k(x,s,z,t) k(z,t,y,T) g(y) dy dz =
\int k(x,s,z,t) \theta (z,t) dz
\ee
and likewise for:
\be
\theta _*(y,t) = \int k(x,0,y,t)f(x)dx = \int \int k(x,0,z,s) k(z,s,y,t)f(x) dx dz =
 \int  k(z,s,y,t) \theta _*(z,s) dz
\ee

For a given semigroup    $\exp (-t\hat{H})$ the specific  Hamiltonian  generator $\hat{H}$,  the kernel $\exp(-t\hat{H})(y,x) = k(t,y,x)$ and the emerging transition probability density  $p(t,y,x)$ of the time homogeneous stochastic process  are unique in view of the
uniqueness of solutions $f (x), g(y)$ of the Schr\"{o}dinger boundary data problem, \cite{zambrini}.
  In  the  case of Markov processes, the knowledge of the transition probability density $p(y,s,x,t)$ (here $p(t-s,y,x)$) for all intermediate times $0\leq s<t \leq T$ suffices for the derivation of all other relevant characteristics of the pertinent random motion.

Further exploiting the Schr\"{o}dinger semigroup lore,  we can  write evolution equations  (C2.), (C.3)  in the   form  displaying an intimate link with   Schr\"{o}dinger-type  equations.   Namely, while staying  in  a  finite  interval $[0,T]$,   we have   $\partial _t\theta _*=  \Delta  \theta _*  -  {\cal{V}}\theta _*$ and
 $\partial _t\theta = -\Delta  \theta  +  {\cal{V}}\theta $ respectively.

We have not yet   specified any restrictions upon the properties  of the potential function ${\cal{V}}(X)$, nor its  concrete functional form.  In the present paper, the potential is expected to be a  continuous function and show up definite confining properties, which we  set   by demanding  $\lim_{|x | \rightarrow \infty}  {\cal{V }}(x) = \infty $.  Then, the Hamiltonian operator $\hat{H}$ generically admits a   positive  ground state  function  $\varphi _0(x)$  with an isolated bottom eigenvalue  $\lambda _0$ (typically a  fully discrete spectrum is admitted).

At this point, we shall narrow the generality of the  addressed  Schr\"{o}dinger boundary data and interpolation problem, by assuming that  actually for all  $t\in [0,T]$ we have
\be
\theta (x,t)= \exp (t \lambda _0)\,  \varphi _0(x)
\ee
where $\hat{H} \varphi _0(x)  = \lambda _0 \varphi _0(x)  $.  Accordingly, we recover  a canonical link of the preselected potential profile ${\cal{V}}(x)$  with the  profile of the  ground state function of    $\hat{H}$:
\be
{\cal{V}}(x) - \lambda _0 =  {\frac{\Delta \varphi _0 (x)}{\varphi _0 (x)}}.
\ee

Let us indicate that the subtraction of $\lambda _0$ from the potential is a standard way to assign the eigenvalue zero to  the ground state function  $\varphi _0(x)$, c.f.   \cite{gar3,vilela,faris}.  Indeed, trivially  there holds    $(\hat{H}_{\alpha } - \lambda _0)  \varphi _0(x) =0$.    In this connection see e.g. our discussion of how killing can be tamed and subsequently eliminated in the Appendix. See also the formula (A.10) referring to the Ornstein-Uhlenbeck process.

Interestingly, a substitution of (C.9) to  Eq.  (C.5) implies
\be
p(y,s,x,t) = \exp[ \lambda _0 (t-s)]\,    k(y,s,x,t) {\frac{\varphi _0 (x)}{\varphi _0(y)}}.
\ee
This is a prototypical  functional form (\cite{gar}, see also (A.10) and (A.12), (A.13))  of the transition probability density   of  the  ground state  transformed  process, whose pdf $\rho (x,t)$  asymptotically relaxes to
\be
\rho _*(x) =  {\frac{\varphi ^2_0 (x)}{ \int \varphi ^2_0(y) \, dy }}.
\ee
We point out that in the present paper we have  predominantly  employed  the  notation   $\rho _*^{1/2}(x)  =  \varphi _0 (x)/\sqrt{\int \varphi ^2_0(y)dy}$.

For completeness, let us rewrite the defining formula (C.5) for $\rho (x,t)$in conformity with Eq. (6), i.e.
in the form  $\rho (x,t)= \rho _*^{1/2} (x) \Psi (x,t)$, where
\be
\Psi (x,t) = \exp(\lambda _0 t)  \theta _*(x,t)   \sqrt{\int \varphi ^2_0(y)dy}.
 \ee
 In virtue of $\partial _t\theta _*= - \hat{H} \theta _* $, where  $\hat{H}=   - \Delta +  {\cal{V}}$,      we  realize that  $\partial _t \Psi = -(\hat{H}  - \lambda _0) \Psi $.

Consequently, the Fokker-Planck equation (14)  takes the form:
  \be
\partial _t \rho =   L^* \rho (x,t)=     \rho _*^{1/2}\, \partial _t \Psi  =  - \rho _*^{1/2}(\hat{H}  - \lambda _0) \rho _*^{-1/2}\, \rho (x,t)
\ee
encoding the  similarity  (strictly speaking, unitary)  transformation relating   the Fokker-Planck operator  $L^*$ and $\hat{H}- \lambda _0$:  $L^* \equiv   - \rho _*^{1/2}(\hat{H}  - \lambda _0) \rho _*^{-1/2}$.
 Because of (17), (18) and  $L = \rho _*^{-1} L^* \rho _*$, we ultimately  arrive at the   diffusion generator  $L$ in the form:   $L\equiv  -  \rho _*^{-1/2}(\hat{H}  - \lambda _0) \rho _*^{1/2}$.

Processes  governed by transition pdfs of the form (C.11)  we have discussed in the main body of the present paper  and in the   preceding  Appendices A and B.
A detailed analysis of  a number of  other  exemplary cases can be found in Refs.  \cite{gar,mazzolo} and  \cite{gar3,gar4,gar,nonegative,non, vilela,streit,klauder,faris}.   For more advanced mathematical background  see.e.g.  \cite{lorinczi1,pavl}.

\end{appendix}


\begin{thebibliography}{99}
  \bibitem{bounded1} P. Garbaczewski and V. Stephanovich, "Fractional Laplacians  in bounded domains: Killed, reflected, censored, and  taboo  L\'{e}vy flights", Phys. Rev. E {\bf 99}, 042126,   (2019).
 \bibitem{asmussen}  L.N. Andersen et al., Lévy Matters V, Lecture Notes in Mathematics 2149, (Springer International Publishing,  Switzerland,  2015).
   \bibitem{dybiec2}  B. Dybiec, E. Gudowska-Nowak, E. Barkai and A. A. Dubkov, "L\'{e}vy flights  versus L\'{e}vy walks in   bounded domains", Phys. Rev. E {\bf 95}, 052102, (2017).
 \bibitem{metzler} T.  Guggenberger et al. , "Fractional Brownian motion in a finite interval: correlations effect depletion or accretion zones of particles near boundaries, New J. Phys. {\bf 21}, 022002, (2019).
\bibitem{appr} A. Malsgov and M. Mandjes,   "Approximations for reflected fractional Brownian motion", Phys. Rev. E {\bf 100}, 032120, (2019).
 \bibitem{denisov} S. I. Denisov, W. Horsthemke and P. H\"{a}nggi, "Steady-state L\'{e}vy flights in a confined domain", Phys. Rev. E {\bf 77}, 061112, (2008).
 \bibitem{dubkov} A. Dubkov and B. Spagnolo, "Langevin approach to L\'{e}vy flights in fixed potentials: Exact results for stationary probability distributions", Acta Phys Pol. B {\bf 38}, 1745, (2007).
 \bibitem{dubkov1}  A. A. Kharcheva et al, "Spectral characteristics of steady-state L\'{e}vy flights in confinement potential  profiles", J. Stat. Mech.  (2016)  054039.
   \bibitem{spectral}  R. Toenjes, I. M. Sokolov and E. B. Postnikov, "Spectral properties of the fractional Fokker-Planck operator for the L\'{e}vy flight in a harmonic potential", Eur. Phys. J. B {\bf 87}, 287, (2014).
\bibitem{fogedby} S. Jespersen, R. Metzler and H. C. Fogedby,  "L\'{e}vy flights in external force fields: Langevin and fractional Fokker-Planck equations and their solutions",  Phys. Rev. E {\bf 59}, 2736, (1999).
\bibitem{gar2}  M. \.{Z}aba  and  P. Garbaczewski, "Solving fractional Schr\"{o}dinger spectral problems: Cauchy oscillator and Cauchy well",
J. Math. Phys. {\bf 55}, 092103, (2014).
\bibitem{gar3} M. \.{Z}aba  and  P. Garbaczewski,  "Nonocally induced (fractional) bound states: Shape analysis in the infinite Cauchy well",
J. Math. Phys. {\bf 56}, 123502, (2015).
 \bibitem{chechkin}  A. Chechkin et al., "Stationary states of non-linear oscillators driven by L\'{e}vy noise", Chemical Physics, {\bf 284}, 233, (2002).
\bibitem{chechkin1} A. V. Chechkin et al.,  "L\'{e}vy flights in a steep potential well",  J. Stat. Phys., {\bf 115}, 1505, (2004).
\bibitem{dybiec} B. Dybiec, E. Gudowska-Nowak and P. H\"{a}nggi, "L\'{e}vy-Brownian motion on finite intervals: Mean first passage analysis", Phys. Rev. E {\bf 73}, 046104, (2006).
\bibitem{dybiec1} B. Dybiec, I. M. Sokolov and A. V. Chechkin, "Stationary states in single-well  potentials under symmetric L\'{e}vy noises", J. Stat. Mech.  (2010)  P07008.
  \bibitem{lorinczi}  K. Kaleta and J. L\H{o}rinczi, "Transition in the decay rates  of stationary distributions of L\'{e}vy motion in an energy  landscape", Phys. Rev. E {\bf 93}, 022135, (2016).
\bibitem{lorinczi1} J. L\H{o}rinczi, F. Hiroshima and V. Betz, "Feynman-Kac-Type Theorems and Gibbs Measures on Path Space", (De Gruyter Studies in Mathematics  vol. 34, De Gruyter, Berlin, 2020)
\bibitem{vilela} S. Eleuterio and  R. Vilela Mendes, "Stochastic ground-state process", Phys. Rev. B {\bf 50}, 5035, (1994).
\bibitem{zero} P. Garbaczewski and V. A. Stephanovich,"L\'{e}vy targeting and the principle of detailed balance", Phys. Rev. E {\bf 84}, 011142, (2011).
 \bibitem{stef}  P. Garbaczewski and V. Stephanovich, "L\'{e}vy flights in inhomogeneous environments", Physica A {\bf 389}, 4419, (2010).
\bibitem{brockmann} D. Brockmann and I. M. Sokolov, "L\'{e}vy flights in external force fields: From models to equations", Chem. Phys. {\bf 284}, 409, (2002).
\bibitem{geisel} D. Brockmann and T. Geisel, "L\'{e}vy flights in inhomogeneous media", Phys. Rev. Lett. {\bf 90}, 170601, (2003).
 \bibitem{belik} V. V. Belik and D. Brockmann, "Accelerating random walks by disorder", New J. Phys., {\bf 9}, 54, (2007).
  \bibitem{risken} H. Risken, {\it The Fokker-Planck equation}, (Springer, Berlin, 1992).
    \bibitem{pavl} G. A. Pavliotis, {\it Stochastic processes and applications}, (Springer, Berlin, 2014).
 \bibitem{linetsky}  V. Linetsky,  "On the transition densities for reflected diffusions", Adv. App. Prob. {\bf 37}, 435-460, (2005).
\bibitem{bickel}  T. Bickel, "A note on confined diffusion", Physica A {\bf 377}, 24-32, (2007).
\bibitem{pilipenko}  A. Pilipenko, {\it An introduction to stochastic differential equations with reflection}, (Potsdam University Press, Potsdam, 2014).

 \bibitem{gar} P. Garbaczewski, "Killing (absorption) versus survival in random motion",  Phys. Rev. E {\bf 96}, 032104, (2017).
   \bibitem{mazzolo} A. Mazzolo, "Sweetest taboo processes", J. Stat. Mech. 073204, (2018).

\bibitem{pinsky} R. G. Pinsky, "Comparison theorems for the spectral gap of diffusion processes and Schr\"{o}dinger operators on an interval", J. London Math. Soc.
{\bf 72}, 621, (2005).
\bibitem{pinsky1}  R. G. Pinsky, "Spectral gap and rate of convergence to equilibrium for a class of conditioned Brownian   motions",  Stochastic Proc. Appl. {\bf 115}, 875, (2005).
\bibitem{gar1} P. Garbaczewski  and M. \.{Z}aba, "Nonlocal random motions and the trapping problem", Acta Phys. Pol. B {\bf 46}(2),  231,  (2015).
\bibitem{entropy} P. Garbaczewski, "Entropy and the thermodynamics of diffusion processes", Acta Phys. Pol. B {\bf 39}, 1087, (2008).
 \bibitem{froman}N. Fr\"{o}man and P. O. Fr\"{o}man, "On the application of the generalized quantal Bohr - Sommerfeld quantization condition
to single-well potentials with very steep walls", J. Math. Phys. {\bf 19}, 1823, (1978).
\bibitem{bender} S. Boettcher and C. M. Bender, "Nonperturbative square-well approximation to a quantum theory", J.  Math. Phys. {\bf 31}, 2579, (1990).
\bibitem{voros} A. Voros, "Exact anharmonic quantization condition", J. Phys. A {\bf 27}, 4653, (1994).
\bibitem{robinett} M. Belloni and R. W. Robinett, "The infinite well and Dirac delta function potentials as pedagogical, mathematical and physical models in quantum mechanics", Physics Reports, {\bf 540}, 24, (2014).
\bibitem{karw} P. Garbaczewski and. Karwowski, "Impenetrable barriers and canonical quantization", Am. J. Phys. {\bf 72}, 924, (2004).
\bibitem{diaz}  J. I. Diaz,  "On the ambiguous treatment of the Schr\"{o}dinger equation for the infinite potential well and an alternative via flat solutions: The one-dimensional case", Interfaces and Free Boundaries 17(3),333, (2015)
  \bibitem{klafter} I. Eliazar and J. Klafter, "Lévy-Driven Langevin systems: Targeted stochasticity", J. Stat. Phys, {\bf 111}, 739, (2003).
\bibitem{sokolov} R. Toenjes, I. M. Sokolov and E. B. Postnikov, " Nonspectral relaxation in one-dimensional Ornstein-Uhlenbeck process", Phys. Rev.
Lett., {\bf 110}, 150602, (2013).
    \bibitem{turbiner} A. V. Turbiner,  "One-dimensional quasi-exactly solvable Schr\"{o}dinger equations", Physics Reports, {\bf 642}, 1, (2016).
    \bibitem{turbiner1} A. Turbiner, "Double Well Potential: Perturbation Theory, Tunneling, WKB (beyond instantons)", Int. J. Mod. Phys. A  {\bf 25}, 647, (2010).
 \bibitem{brandon} D. Brandon and N. Saad, "Exact and approximate solutions to Schr\"{o}dinger’s  equation with decatic potentials", Open Physics, {\bf 11}(3), 279, (2013).
\bibitem{maiz} F. Maiz et al., "Sextic and decatic anharmonic oscillator potentials: Polynomial solutions", Physica B {\bf 530}, 101, (2018).
 \bibitem{nieto} J. Daboul and M. M. Nieto, "Quantum bound states with zero binding energy", Phys. Lett. A {\bf 190}, 357, (1994).
 \bibitem{makowski}  A. J. Makowski, "Exact, zero-energy, square-integrable solutions of a model related to the Maxwell's fish-eye problem", Ann. Phys. {\bf 324}, 2465, (2009).
\bibitem{landau}  L. D. Landau and E. M. Lifshitz, {\it Quantum Mechanics: Nonrelativistic   Theory},  (Pergamon, Oxford, 1995).
 \bibitem{gar4} P. Garbaczewski and R. Olkiewicz, "Feynman–Kac kernels in Markovian representations
of the Schro\"{o}dinger interpolating dynamics", J. Math. Phys. {bf 37}(2), 732, (1996).
\bibitem{nonegative} Ph. Blanchard and P. Garbaczewski, "Natural boundaries for the Smoluchowski equations and affiliated diffusion processes", Phys. Rev. E {\bf 49}(5), 3815, (1994).
 \bibitem{non}  Ph. Blanchard,  P. Gabaczewski and R.Olkiewicz, "Non-negative Feynman-Kac kernels in Schr\"{o}dinger's interpolation   problem", J. Math. Phys. {\bf 38}, 1, (1997).
    \bibitem{streit}  S. Albeverio, R. H{\o}egh-Krohn  and L. Streit," Energy forms, Hamiltonians, and distorted Brownian paths", J. Math. Phys. {\bf 18}, 907,  (1977).
 \bibitem{klauder} H. Ezawa, J. R. Klauder, and L. A. Shepp, "A path space picture for Feynman-Kac averages", Ann. Phys. (NY) {\bf 88}, 588 (1974).
 \bibitem{faris} W. G. Faris, "Diffusive motion and where it leads", in {\it  Diffusion, Quantum Theory and Radically Elementary Mathematics}, edited
by W. G. Faris (Princeton University Press, Princeton, 2006),  pp. 1–43.

\bibitem{kuczma} M. Kuczma, {\it An introduction  to the theory of functional equations and inequalities}, (Birkh\"{a}user, Basel, 2009).

\bibitem{vankampen} N. G. van Kampen, "A soluble model for diffusion in a bistable potential", J. Stat. Phys. {\bf 17}, 71, (1977).
\bibitem{larson} R. S. Larson and M. D, Kostin, "Kramers theory of chemical kinetics: Eigenvalues and eigenfunction analysis", J. Chem.
 Phys, {\bf 69},4821, (1978).
 \bibitem{risken1} M. M\"{o}rsch, H. Risken and H. D. Vollmer, "One-dimensional diffusion in  a soluble model potential",
  Z. Physik {\bf B 32}, 245, (1979).
  \bibitem{liu} F. So and K. L. Liu, "A study of the Fokker-Planck equation of bistable systems by the method of state-dependent diagonalization",
  Physica {\bf A 277}, 335, (2000).
\bibitem{okopinska} A. Okopi\'{n}ska, "The Fokker-Planck equation  for bistable potential in the optimized expansion",
 Phys. Rev. E {\bf 65},  062101, (2002).
\bibitem{baner} K. Banerjee and J. K. Bhattacharjee, "Anharmonic oscillators and double wells: Closed-form global approximants for eigenvalues",
 Phys. Rev. {\bf D 29}, 1111, (1984).
\bibitem{smol} P. Garbaczewski, “L\'{e}vy flights in steep potential wells: Langevin modeling versus direct response to energy  landscapes”,  talk at the  32nd  Marian Smoluchowski Symposium  on Statistical Physics, Cracow, Poland,   Sept. 2019, available through the Symposium page   https://zakopane.if.uj.edu.pl/event/9/contributions/312/

\bibitem{schr}  E. Schr\"{o}dinger,  "Sur la théorie relativiste de l' \'{e}lectron et l'  interpr\'{e}tation  de la m\'{e}canique quantique",  Ann. Inst. Henri Poincare {\bf 2}(4), 269, (1932).
  \bibitem{zambrini} J. C. Zambrini,  "Stochastic mechanics according to E. Schr\"{o}dinger", Phys. Rev. A {\bf 33}(3), 1532, (1986).

\end{thebibliography}
\end{document}